\documentclass{emulateapj}
\usepackage{natbib}
\usepackage{lscape}
\usepackage{amsmath}
\usepackage{enumerate}

\slugcomment{\footnotesize \today}

\newcommand{\degree}{\mbox{$^{\circ}$}}

\newcommand{\am}{\mbox{\arcmin}}
\newcommand{\as}{\mbox{\arcsec}}

\newcommand{\kms}{\mbox{km s$^{-1}$}}
\newcommand\cmv{\mbox{cm$^{-3}$}}
\newcommand\cmc{\mbox{cm$^{-2}$}}

\newcommand{\um}{$\mu$m}

\newcommand{\lsun}{\mbox{L$_\odot$}}
\newcommand{\msun}{\mbox{M$_\odot$}}
\newcommand{\ta}{{$T_A^*$}}
\newcommand{\tex}{\mbox{$T_{\rm ex}$}}
\newcommand{\tmb}{\mbox{$T_{\rm mb}$}}
\newcommand{\tpk}{\mbox{$T_{\rm pk}$}}

\newcommand{\tk}{\mbox{$T_{\rm K}$}}
\newcommand{\td}{\mbox{$T_{\rm D}$}}

\newcommand{\mean}[1]{\mbox{$\langle#1\rangle$}} 

\newcommand{\miso}{\mbox{$M_{\rm iso}$}} 
\newcommand{\mvir}{\mbox{$M_{\rm vir}$}} 
\newcommand{\rgal}{\mbox{$R_{\rm Gal}$}} 

\newcommand{\ammonia}{\mbox{{\rm NH}$_3$}}

\newcommand{\Snu}{\mbox{$S_{\nu}$}}

\begin{document}

\title {\bf The Bolocam Galactic Plane Survey. VII. Characterizing the Properties of Massive Star-Forming Regions}
\author{Miranda K. Dunham\altaffilmark{1,2,3},
Erik Rosolowsky\altaffilmark{4},
Neal J. Evans II\altaffilmark{2},
Claudia Cyganowski\altaffilmark{5},
James S.~Urquhart\altaffilmark{6}
}
\altaffiltext{1}{Department of Astronomy, Yale University, P.O. Box 208101, New Haven, CT 06520-8101}
\altaffiltext{2}{Department of Astronomy, The University of Texas at Austin,
       1 University Station C1400, Austin, Texas 78712--0259}
\altaffiltext{3}{Email:  miranda.dunham@yale.edu}
\altaffiltext{4}{University of British Columbia, Okanagan,
     3333 University Way, Kelowna BC V1V 1V7 Canada}
\altaffiltext{5}{NSF Astronomy and Astrophysics Postdoctoral Fellow, Harvard-Smithsonian Center for Astrophysics, Cambridge, MA 02138}
\altaffiltext{6}{CSIRO Astronomy and Space Science, P.~O.~Box 76, Epping, NSW 1710, Australia}

\begin{abstract}
We present the results of a GBT survey of \ammonia(1,1), (2,2), (3,3) lines
towards 631 Bolocam Galactic Plane Survey (BGPS) 
sources at a range of Galactic longitudes in the inner Galaxy.  
We have detected the \ammonia(1,1) line towards 72\%\ of 
our targets (456), demonstrating that the high column density features
identified in the BGPS and other continuum surveys accurately predict the 
presence of dense gas.  We have determined kinematic distances and 
resolved the distance ambiguity for all BGPS sources detected in \ammonia.
The BGPS sources trace the locations of the Scutum and Sagittarius 
spiral arms, with the number of sources peaking between \rgal$\sim4-5$ kpc. We 
measure the physical properties of each source and find that 
depending on the distance, BGPS sources are primarily clumps, with some cores 
and clouds.  We have examined the physical properties as a function of 
Galactocentric distance, and find a mean gas kinetic temperature of 15.6 K, and
that the \ammonia\ column density and abundance decrease by nearly an order of 
magnitude  between $\rgal\sim 3 - 11$ kpc.  Comparing sources
at similar distances demonstrates that 
the physical properties are indistinguishable, which suggests a similarity in
clump structure across the Galactic disk.  We have also compared the
BGPS sources to criteria for efficient star formation 
presented independently by Heiderman et al.~and Lada et al.,  and massive star formation
presented by Kauffmann et al.  48\%\ of our sample
should be forming stars (including massive stars) with high efficiency,
and 87\%\ contain subregions that should be efficiently forming stars.
Indeed, we find that 67\%\ of the sample exhibit signs of star formation
activity based on an association with a mid-IR source.

\end{abstract}


\section{Introduction}\label{intro}
Massive stars ($M>$ 8\msun) play a significant role in the evolution of
their immediate environment, as well as evolution on a galaxy-wide 
scale via their many feedback processes including strong UV radiation
that creates HII regions, outflows and winds which inject energy back into
the surrounding medium, and supernova at the end of their lives 
(Kennicutt 2005).  Given the importance of massive stars in galactic 
evolution, it is imperative that we understand how and where massive 
stars form within the Milky Way in order to intrepret observations of
other galaxies.  

Large sample sizes
are beneficial in this task, and many large samples have been chosen
in various ways.  For example, 
Plume, Jaffe \& Evans (1992) selected 179 H$_2$O masers
from Cesaroni et al.~(1988) and followed up with observations of dense
gas tracers (CS $J=7-6$ and CO $J=3-2$).  This H$_2$O maser sample was followed up with 
lower J transitions of CS (Plume et al.~1997; Shirley et al.~2003) as
well as submillimeter dust continuum (Mueller et al.~2002).  Shirley
et al.~(2003) thoroughly characterized the physical properties based on 
CS $J=5-4$ maps of a subsample of 64 sources from Plume et al.~(1992).
They found the following median properties:  $R=0.32$ pc, \mvir$=920$ \msun,
$\Sigma=0.60$ g \cmc.  Plume et al.~(1997) found a mean volume density
of \mean{\log(n/\cmv)}$=5.9$ based on multiple transitions of CS.  
Recently, Wu et al.~(2010) have characterized properties for a subsample of 
50 massive, dense clumps from Plume et al.~(1992) based on HCN and 
CS line emission.  They found a range for the mean of each property
based on the different molecular line tracers used:  
\mean{\mvir}$=1300 - 5300$ \msun, \mean{\Sigma}$=0.29 - 1.09$ g \cmc, and 
\mean{n}$=3.2\times10^{4} - 2.5\times10^{5}$ \cmv.
Other samples were defined based on the presence of ultra-compact HII
(UCHII) regions and IRAS colors.  
Wood \& Churchwell (1989b) complied a catalog of 
massive star-forming regions
from the IRAS All Sky Survey based on typical FIR colors of the cool dust
surrounding UCHII regions (Wood \& Churchwell 1989a).  Beuther et al.~(2002b)
compared 1.2 mm continuum observations and CS molecular line studies toward
69 UCHII regions based on the sample defined by Sridharan et al.~(2002).
Beuther et al.~determined physical properties similar to those of
Shirley et al.~(2003).   

These initial studies of massive star formation
were important first steps toward characterizing large samples, although the 
methods employed in identifying the sources resulted in highly biased
samples.  Requiring the presence of H$_2$O masers or UCHII regions
biases the samples to a particular stage in star formation.
Both phenomena require a significant source of energy, and therefore 
will bias the samples to later evoluntionary states when a
protostar has already formed.  Therefore, the properties determined by these initial 
studies may not be characteristic
of massive star forming regions at all evolutionary stages because of these biases.

The many large-scale, blind Galactic plane surveys recently completed and 
currently underway provide the data sets needed to study massive star formation
in a Galactic context.  In particular, three large scale surveys have recently
done so:  The Boston-University Five College Radio Astronomy Observatory
$^{13}$CO Galactic Ring Survey (GRS; Jackson et al.~2006), the Red
MSX Source (RMS; Hoare et al.~2004; Urquhart et al.~2008) Survey, and
the Bolocam Galactic Plane Survey (BGPS; Aguirre et al.~2011).

Rathborne et al.~(2009) identified 829 clouds and 6124 clumps within the 
$^{13}$CO emission in the GRS.  They compared properties of clouds inside
and outside of the 5 kpc molecular ring, a Galactic structure identified
as an overdensity in molecular gas located at a Galactocentric radius of 5 kpc
that contains the majority of star formation in the Milky Way (e.g.~Burton et al.~
1975; Scoville \& Solomon 1975; Cohen \& Thaddeus 1977; Robinson et al.~1984; 
Clemens et al.~1988; Kolpak et al.~2002).  They found that clouds within
the 5 kpc molecular ring have warmer temperatures, higher column densities,
larger areas, and contained more clumps than clouds located outside of the
molecular ring.  Roman-Duval et al.~(2009) resolved the distance ambiguity
to 750 of the GRS clouds and found that their positions are consistent
with a four arm model of the Galaxy and trace the positions of the 
Scutum-Crux and Perseus spiral arms.  Roman-Duval et al.~(2010)
derived physical properties of 580 of the GRS clouds and demonstrated
that the gas surface density peaks within the 5 kpc molecular ring.  They
found cloud sizes ranging from 0.1 to 40 pc, masses ranging from 10 to $10^6$
\msun, a mean volume density of $230\pm21$ cm$^{-3}$ with the highest density
being $\sim10^3$ \cmv, and a mean surface density of $144\pm3$ \msun\ pc$^{-2}$.

The RMS team has identified a candidate list of $\sim$2000
massive young stellar objects (MYSOs) and ultra-compact HII (UCHII) regions 
based on MSX and IRAS colors (Lumsden et al.~2002).  Urquhart et al.~(2011) 
have recently compared the radial velocities of the MYSOs to the clouds 
identified in the GRS and adopted the kinematic distances determined by
Roman-Duval et al.~(2009) thereby obtaining a complete sample of 196 RMS sources
with luminosities above their completeness limit of $\sim10^{4}$ \lsun.  
The clouds containing a RMS source were found to be larger, more massive, and 
more turbulent than the GRS clouds without a RMS source.  The subsample of
GRS clouds containing RMS sources are also well correlated with the Galactic
spiral arm structure.

The BGPS (Aguirre et al.~2011)
has detected 1.1 mm thermal dust emission from the dense regions
closely associated with star formation.  Dunham et al.~(2010; hereafter refered to as D10) found mean
H$_2$ column densities of $1.2\times10^{22}$ \cmc\ for BGPS sources located within
the Gemini OB1 Molecular cloud, and Schlingman et al.~(2011) found a mean
H$_2$ column density of $6.8\times10^{21}$ \cmc\ for BGPS sources detected
in HCO$^{+}$(3-2).  In contrast to previous large-scale studies
of massive star-forming regions, the BGPS is identifying high column density 
regions regardless of star formation signposts (such as H$_2$O maser emission).  
By identifying sources based on dense gas and dust, the bias toward a 
particular evolutionary stage is avoided. 

In this paper, we present the results of a targeted survey of \ammonia\ toward
631 BGPS sources in the first Galactic quadrant. 
The goals of this paper are to determine the location of a large sample of BGPS 
sources within the Galaxy, characterize the physical properties of a large relatively
unbiased sample of star-forming regions, study their physical properties as a function of
environment using Galactocentric radius as a proxy, and to characterize their
star formation potential and mid-IR source content.  
Section \ref{BGPS} briefly describes the BGPS and source selection, and Section
\ref{nh3survey} describes the \ammonia\ set up and observations.  We present
basic results of the \ammonia\ survey in Section \ref{results}.  In Section 
\ref{distances} we determine kinematic distances and break the kinematic distance 
ambiguity.  Section \ref{analysis} includes a description of each physical 
quantity derived from the BGPS and \ammonia\ surveys.  Section \ref{galactictrends}
explores the trends in physical properties as a function of Galactocentric radius, and 
Section \ref{discussion}
includes a discussion of the star formation activity, comparison to
other large-scale studies, and a discussion of whether BGPS sources are
forming massive stars.  Finally, Section \ref{summary} provides a 
summary of this work.

\section{The Bolocam 1.1 mm Galactic Plane Survey}\label{BGPS}
The BGPS\footnote{See http://irsa.ipac.caltech.edu/data/BOLOCAM\_GPS/} has 
observed approximately 170 square degrees of the northern 
Galactic plane in 1.1 mm continuum emission using Bolocam
at the Caltech Submillimeter Observatory\footnote{The Caltech 
Submillimeter Observatory is supported by the NSF.} (CSO).
The survey area consists of a continuous region spanning
$-10\degree < \ell < 90\degree$ and $|b| \leq 0.5\degree$
as well as select regions of known star formation in the
outer Galaxy: the Perseus tangent, the W3/4/5 region, and 
the Gemini OB1 molecular cloud.  The survey methods and data reduction are thoroughly 
described in Aguirre et al.~(2011), and the source extraction
algorithm and catalog are described in Rosolowsky et al.~(2010).
The BGPS has identified
8,358 continuum emission sources over the entire coverage
area (Rosolowsky et al.~2010), most of which were previously 
unknown.  Here we discuss details relevant to this work and refer the 
reader to these papers for further details.  

\begin{deluxetable*}{clccccccccrr}
\tabletypesize{\tiny}
\tablewidth{0pt}
\tablecaption{\label{mm_props}Observed 1.1 mm Properties}
\tablehead{
\colhead{ID} & \colhead{} & \colhead{RA} & \colhead{Dec} & \colhead{peak RA} & \colhead{peak Dec} & \colhead{$R_{major}$} & \colhead{$R_{minor}$} & \colhead{PA} & \colhead{$R_{obj}$} & \colhead{\Snu(120\as)} & \colhead{\Snu(int)} \\
\colhead{number} & \colhead{Source} & \colhead{(J2000)} & \colhead{(J2000)} & \colhead{(J2000)} & \colhead{(J2000)} & \colhead{(\as)} & \colhead{(\as)} & \colhead{(\degree)} & \colhead{(\as)} & \colhead{(mJy)} & \colhead{(mJy)}
}
\startdata
1307 & G007.501+00.001 & 18 02 30.6 & -22 28 00.3 & 18 02 30.0 & -22 28 07.0 &  23.4 &  18.2 & 173 &  35.41 &     399 (110) &     356 (94) \\
1309 & G007.509+00.403 & 18 01 00.2 & -22 16 03.4 & 18 01 00.4 & -22 15 46.3 &  27.1 &  15.3 & 154 &  28.73 &     550 (140) &     588 (110) \\
1310 & G007.564-00.042 & 18 02 46.4 & -22 26 12.5 & 18 02 47.8 & -22 26 05.6 &  28.2 &  21.6 &  34 &  48.07 &     629 (170) &     533 (170) \\
1311 & G007.600-00.142 & 18 03 14.7 & -22 27 07.6 & 18 03 15.0 & -22 27 10.1 &  25.5 &  15.0 & 142 &  25.78 &     717 (160) &     681 (120) \\
1313 & G007.622-00.000 & 18 02 46.8 & -22 21 46.5 & 18 02 45.8 & -22 21 49.4 &  18.7 &   8.9 & 105 & $<$ 16.5 &     494 (170) &     159 (90) \\
1314 & G007.632-00.110 & 18 03 12.8 & -22 24 28.3 & 18 03 11.9 & -22 24 33.1 &  29.7 &  15.8 &  40 &  33.26 &    1340 (190) &    1450 (150) \\
1316 & G007.636-00.194 & 18 03 31.5 & -22 26 45.8 & 18 03 31.4 & -22 26 49.4 &  18.0 &  14.9 &  77 &  18.37 &     617 (140) &     595 (94) \\
1317 & G007.650-00.166 & 18 03 27.6 & -22 25 13.3 & 18 03 26.8 & -22 25 15.9 &  24.3 &  17.6 &  35 &  34.95 &     735 (150) &     562 (120) \\
1320 & G007.762+00.076 & 18 02 48.1 & -22 12 09.8 & 18 02 46.5 & -22 12 16.0 &  29.1 &  19.6 &  15 &  44.81 &     584 (180) &     565 (170) \\
1322 & G007.898-00.012 & 18 03 23.9 & -22 07 42.2 & 18 03 23.7 & -22 07 45.9 &  29.1 &  14.7 & 173 &  25.12 &     615 (170) &     552 (130)
\enddata \\
\tablecomments{Errors are given in parentheses.}
\tablecomments{The full table is available in the online journal.}
\end{deluxetable*}

The effective FWHM beam size of the BGPS is 33\as, slightly larger
than the nominal Bolocam beam at 1.1 mm due to combining multiple 
observations of each field.  The FWHM beam size corresponds to
a solid angle of $2.9\times10^{-8}$ sr, which is equivalent to 
a tophat function with a 40\as\ diameter ($\Omega=2.95\times10^{-8}$ sr).
The BGPS catalog algorithm, bolocat, presents aperture flux densities within 
40\as, 80\as, and 120\as\ diameter apertures representing
the flux density within one (\Snu(40\as)), two (\Snu(80\as)), 
and three (\Snu(120\as)) times the beam,
respectively.  Bolocat additionally provides an integrated 
flux density (\Snu(int)), which is simply the sum of all pixels assigned
to a given BGPS source and more accurately represents the 
flux densities of irregularly shaped sources.  The aperture-based
flux densities of point sources require correction factors (1.46 for \Snu(40\as),
1.04 for \Snu(80\as), and 1.01 for \Snu(120\as)) to account for
power falling outside of the aperture due to the sidelobes of
the beam.  \Snu(int) does not require a correction since it
is not aperture-based.  

In addition to the aperture corrections, a flux calibration factor
of $1.5\pm0.15$ is required.   This 
factor is based on a comparison of BGPS data with 1.2 mm data
(Rathborne et al.~2006; Motte et al.~2003, 2007; Matthews et al.~2009)
acquired at different facilities, which showed that the BGPS flux 
densities were systematically lower (Aguirre et al.~2011).  
We apply the flux calibration 
factor in order to bring the BGPS flux densities into agreement with
the 1.2 mm surveys.  The aperture corrections and flux calibration
factor are included in the flux densities presented here.

Bolocat includes coordinates of both a geometric centroid and the 
peak 1.1 mm emission.  The major ($\sigma_{maj}$) and minor ($\sigma_{min}$) 
axes of each source are calculated based on emission weighted moments,
and the effective radius ($R$) of each source is calculated as the geometric
mean of the deconvolved major and minor axes:
\begin{equation} R=\eta [(\sigma_{maj}^2 - \sigma_{beam}^2)(\sigma_{min}^2 - \sigma_{beam}^2)]^{1/4},
\label{radiuseqn}
\end{equation}
where $\sigma_{beam} = \Theta_{FWHM}/\sqrt{8 \ln 2}=14\as$, $\Theta_{FWHM}=33\as$, and $\eta$ is a
factor relating the axis dispersions to the true size of the source.
$\eta=2.4$ is adopted as the median value
derived from measuring the observed major and minor axes compared with the
true radius for a variety of simulated emission profiles spanning a
range of density distributions, sizes relative to the beam and
signal-to-noise ratios (Rosolowsky et al.~2010).  Including $\eta$
provides an effective radius that corresponds to the full extent of the 
emission detected by the BGPS.  Specifically, 
$R$ describes the radius over which \Snu(int) 
was calculated.  Due to the cleaning methods
employed (PCA cleaning; see Aguirre et al.~2011), the BGPS
resolves out uniform structure on size scales larger than 
5.9\am.  Thus, the BGPS is sensitive to structures with
diameters between 33\as\ and 5.9\am.  

In order to sample source properties across a large range of 
Galactocentric radius (\rgal), we have obtained 
\ammonia\ and H$_2$O spectral line observations toward 631
BGPS sources within four distinct Galactic longitude ranges:
$7.5\degree \leq \ell \leq 10.0\degree$, $19.5\degree \leq \ell
\leq 22.5\degree$, $31.3\degree \leq \ell \leq 34.5\degree$, and 
$52.5\degree \leq \ell \leq 55.5\degree$.  Sources near $\ell= 33\degree$
were targeted because of the peak in mm continuum sources along that line of sight,
which corresponds to the end of the 
Galactic bar and tangent of the Scutum spiral arm.  The $\ell\sim 54\degree$
field was chosen to sample the tangent of the Sagittarius Arm.
The $\ell\sim 9\degree$ and $\ell\sim 20\degree$ fields were chosen
to sample sources with small \rgal.
Table \ref{mm_props} presents the 631 BGPS sources that lie within
our Galactic longitude ranges:  BGPS source catalog number (column 1), 
source name (column 2), centroid RA and DEC (columns 3 and 4), 
RA and DEC of the peak emission (columns 5 and 6), major and minor
axes (columns 7 and 8), position angle north of $\ell=0\degree$ 
(column 9), radius (column 10), \Snu(120\as) and \Snu(int) (columns
11 and 12).

We have chosen to use \Snu(120\as) as well as \Snu(int) due to their 
complementary nature.   
\Snu(120\as) is a well-defined quantity that is easily compared with 
other data sets but may not represent the true source flux due to 
source crowding and/or irregularly shaped, large sources.  \Snu(int)
is difficult to reproduce and compare with other data sets since it is a
product of the BGPS source extraction algorithm.  However, \Snu(int) will more
accurately represent the source flux density.

\begin{figure}
\begin{center}
\plotone{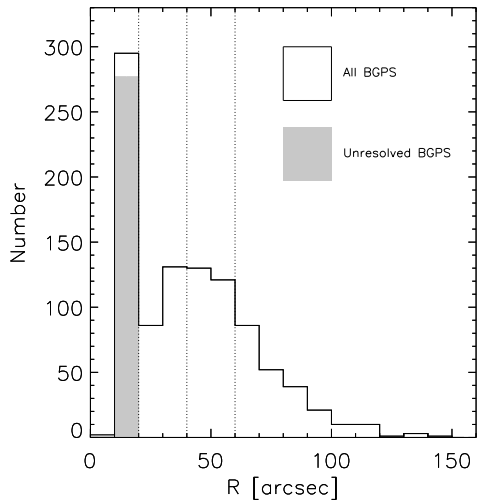}
\figcaption{\label{histr}
Distribution of angular source sizes ($R$) for the 631 BGPS sources that fall within the selected ranges (see section \ref{BGPS}).  The white histogram denotes all source sizes, the gray histogram denotes sources that are unresolved compared to the BGPS effective beam size (FWHM of 33\as, $R=16.5$\as), and the dotted lines mark the radii of the apertures presented in the BGPS catalog. 
}
\end{center}
\end{figure}

\begin{figure}
\begin{center}
\plotone{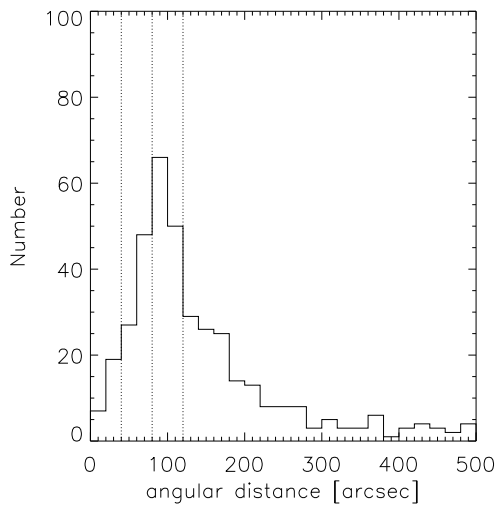}
\figcaption{\label{nearestneighbor}
Distribution of angular distance to the nearest BGPS source.  The dotted lines represent the diameters of the apertures presented in the BGPS catalog.
}
\end{center}
\end{figure}

When choosing which 
aperture size to use, we considered the distribution of source sizes
as well as the distribution of angular separation between BGPS sources.  The 
distribution of angular source sizes is shown in Figure \ref{histr}
where the white histogram represents the angular source size of all
631 BGPS sources, and the gray histogram denotes sources unresolved
compared to the beam.  The dotted lines mark the 3 aperture sizes 
included in bolocat.  The 120\as\ aperture is larger than most sources 
and will therefore underestimate the source flux in the smallest number
of BGPS sources, although the number of sources larger than the 120\as\ 
aperture is not negligible.  Figure \ref{nearestneighbor} shows the distribution of
the angular distance to the nearest BGPS source.  This distribution favors the
smaller aperture sizes since they will suffer the least from source 
contamination.  We must balance the effects of missing flux and
including additional flux when choosing an aperture size.  As demonstrated
in Figure \ref{nearestneighbor}, 217 BGPS sources have at least one source
closer than 120\as.  We have chosen
to use the 120\as\ aperture even though the flux will be overestimated
in 217 of the 631 BGPS sources within our longitude ranges.  \Snu(120\as) 
will, on average, overestimate \Snu(int) by 10\%; the mean 
ratio of \Snu(120\as) to \Snu(int) is 1.09 with a standard deviation of 
0.50.  Therefore, we present \Snu(120\as) as a comparison point for other studies,
while our analysis is based on \Snu(int), which will not suffer from the two problems
discussed.  We stress that \Snu(120\as) will
overestimate the fluxes to approximately one third of BGPS sources included in this 
paper.  

\begin{deluxetable}{ccc}
\tabletypesize{\scriptsize}
\tablewidth{0pt}
\tablecaption{\label{obsdates}\ammonia\ Observation Dates and Atmospheric Opacities}
\tablehead{
\colhead{Session} & \colhead{Date} & \colhead{$\tau_{23}$\tablenotemark{1}}
}
\startdata
1 & 14 February 2008 & 0.047 \\
2 & 16 February 2008 & 0.085 \\
3 & 23 March 2008 & 0.046 \\
4 & 10 January 2010 & 0.031 \\
5 & 14 January 2010 & 0.043 \\
6 & 31 January 2010 & 0.036 \\
7 & 7 February 2010 & 0.039 \\
8 & 11 February 2010 & 0.052 \\
9 & 20 February 2010 & 0.055 \\
10 & 6 March 2010 & 0.041 \\
11 & 9 March 2010 & 0.059 \\
12 & 27 March 2010 & 0.039\\
RMS & 13 February 2010 & 0.054 \\
RMS & 25 September 2010 & 0.120
\enddata\\
\tablenotetext{1}{Atmospheric opacity at 23 GHz.}
\end{deluxetable}

We additionally consider \Snu(40\as) as a representative of the mm
flux per beam since the solid angle of the 40\as\ aperture 
($2.95\times10^{-8}$ sr) is 
equivalent to the solid angle subtended by the effective Gaussian
beam of the BGPS ($2.9\times10^{-8}$ sr).  
\Snu(40\as) is only employed when calculating
the H$_2$ column density per beam and \ammonia\ abundance
(see Section \ref{coldenabund}).

\section{\ammonia\ Survey of BGPS Sources}\label{nh3survey}
We have conducted spectroscopic follow-up observations of the 631 
BGPS sources selected as described in the previous section.   
We observed all sources in the lowest inversion transitions of
\ammonia\ and the 22 GHz water maser transition using the
Robert F. Byrd Green Bank Telescope\footnote{The GBT is operated by
the National Radio Astronomy Observatory, which is a facility of the
National Science Foundation, operated under cooperative agreement by
Associated Universities, Inc.} (GBT).  The dates of observations and 
23 GHz zenith opacities are listed in Table \ref{obsdates}.  
The observations consisted of two minute integrations on each source,
with pointing and focus checks approximately every 1.5 hrs.  
We additionally checked the flux calibration by periodically injecting
a noise signal  through observations of 3C309.1, an NRAO flux calibrator
with a known flux.  We repeated this flux calibration observation
during all sessions except for sessions lasting two hours or less
(sessions 4 and 11).  We did not observe any significant variation
throughout the measurements ($<10$\%), and use a mean calibration
for all observations.

The GBT spectrometer served as the backend and was configured to
observe 4 intermediate frequencies (IFs).  The majority of 
sources were observed with the following four lines:  22 GHz H$_2$O,
\ammonia(1,1), \ammonia(2,2), and \ammonia(3,3).  
Thirty of the spectra were observed with an identical
setup by the Red MSX Source team (RMS), 
who were kind enough to share their raw data.  The sources observed
during 2008 were observed with the \ammonia(4,4) in place of 
\ammonia(3,3).  The frequencies of each line are listed in Table 
\ref{specline}.  We utilized frequency switching with a 5 MHz 
throw, and the RMS team used a throw of 8 MHz.  At these
frequencies, the GBT FWHM is 31\as\ and is well matched to the
effective BGPS FWHM beam size of 33\as.

We reduced the spectral line data using the GBTIDL reduction 
package.  We extracted each scan, and folded it to account for
the effects of frequency switching.  In order to place the data
on the $T_A$ scale, we calibrated the noise diodes using the observations
of sources with known fluxes.  We then subtracted a first order
baseline (second order for the H$_2$O spectra), and converted the
spectra to the \ta\ scale by accounting for the atmospheric opacity.
We then divided by the GBT main beam efficiency, $\eta\sim 0.81$ 
at 23 GHz, to place the spectra on the \tmb\ scale.  Each polarization 
was calibrated separately then averaged together, as were multiple 
observations when applicable.  The mean rms and standard deviation 
of the rms for each possible IF are listed in Table \ref{specline}.

\begin{deluxetable}{lccc}
\tabletypesize{\scriptsize}
\tablewidth{0pt}
\tablecaption{\label{specline}Spectral Line Observations}
\tablehead{
\colhead{} & \colhead{$\nu$} & \colhead{mean RMS} & \colhead{Stdev RMS}   \\ 
\colhead{Line} & \colhead{(GHz)} & \colhead{(mK)} & \colhead{(mK)}  
}
\startdata
\ammonia(1,1) & 23.6944961 &    98 &    28 \\
\ammonia(2,2) & 23.7226328 &    93 &    27 \\
\ammonia(3,3) & 23.8701270 &    96 &    26 \\
\ammonia(4,4) & 24.1394160 &   141 &    33 \\
H$_2$O & 22.2350801 &    87 &    32 
\enddata 
\end{deluxetable}

\begin{figure}
\begin{center}
\includegraphics[angle=90,width=3.5in]{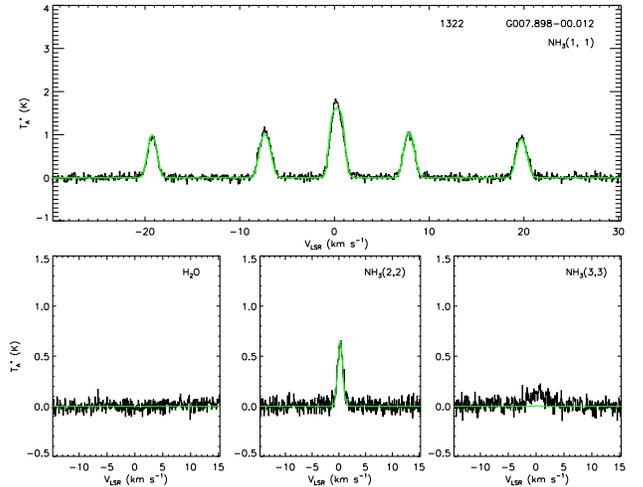}
\caption{\label{nh3spectra} Spectra toward BGPS source 1322.  Upper panel:  \ammonia(1,1). Lower panel from left to right:  22 GHz H$_2$O maser transition, \ammonia(2,2), and \ammonia(3,3).  The green lines show the model fit to the observed spectra.  Similar figures for the 514 \ammonia\ pointings detected in the (1,1) line (see section \ref{results}) are included as an online only figure.}
\end{center}
\end{figure}

Figure \ref{nh3spectra} shows the observed spectra for BGPS source 1322,
including \ammonia(1,1), (2,2), (3,3) and the 22 GHz H$_2$O maser transition.
The green line overplotted in each panel displays the model fit described in
Section \ref{nh3survey}.  

The parameter estimation follows the philosophy developed by
Rosolowsky et al.~(2008) and adapted for use in D10.  The method models the
emission as a beam-filling slab of ammonia with a variable column
density ($N_{\mathrm{NH3}}$), kinetic temperature (\tk), excitation
temperature (\tex), Gaussian line width ($\sigma_{Vlsr}$), and LSR
velocity ($v_{\mathrm{LSR}}$).  Previous versions of the pipeline used
the optical depth of the (1,1) line as a free parameter and then
calculated the column density.  This model reverses the direction of
the calculation, assuming the molecules are in thermodynamic
equilibrium using an ortho-to-para ratio of 1:1, which is the high
temperature formation limit (Takano et al.~2002).  Hence, the ammonia
molecules are partitioned among the energy levels as
\begin{eqnarray}
Z_O=& \notag\\
1+ \sum_{J,K,i}& 2(2J+1)\exp\left\{-\frac{h[BJ(J+1)+(C-B)J^2]+\Delta E(J,K,i)}{kT_k}\right\}\notag\\
\qquad & \qquad \mbox{for } J=3,6,9,\dots; i=0,1,  \label{partitionfunctionO}\\
Z_P=& \notag\\
\sum_{J,K,i}&(2J+1)\exp\left\{-\frac{h[BJ(J+1)+(C-B)J^2]+\Delta E(J,K,i)}{kT_k}\right\}\notag\\
\qquad & \qquad \mbox{for } J=1,2,4,5,\dots; i=0,1,\label{partitionfunctionP}
\end{eqnarray}  
where $Z_O$ and $Z_P$ correspond to ortho and para \ammonia, respectively.
Here, $\Delta E(J,K,i)$ is the energy difference due to the splitting of
the symmetric and anti-symmetric states.  $\Delta E(J,K,1)$ corresponds
to the antisymmetric state, which is $\Delta E/k\sim 1.1$~K above the
symmetric state ($\Delta E(J,K,0)=0$).  The column density of the
molecules in the $N_{\mathrm{NH3}}(J,K,i)$ state is thus
$N_{\mathrm{NH3}} Z_O(J,K,i)/(2Z_O)$ and $N_{\mathrm{NH3}}
Z_P(J,K,i)/(2Z_P)$, where the factor of 2 in the denominator accounts for
our assumption of ortho:para$=$1:1.  From the column densities in the individual
states, we calculate the optical depths in the individual transitions.
From here, the method follows that of previous work: we use the
optical depth, hyperfine structure, the velocity information and the
excitation conditions to model the individual spectra.  The free
parameters are optimized using the MPFIT least-squares minimization
routine including parameter bounds (Markwardt 2009).

These small changes do produce differences in the derived parameters,
reflecting optimizing sightly different functions.  In addition, the
optimization weights the lines slightly differently under the
different schemes.  The variation largely reflects the approximate
nature of the assumed model, namely that of a homogeneous slab with
uniform properties.  While more sophisticated models could produce
better fits, they would require additional information such as mapping
to constrain the distribution of the individual lines.  As an
example, consider the fit to an optically thin object which has a
temperature that decreases with radius.  In this case, the the model
would underestimates the (3,3) line emission by a noticeable margin.
This is because the temperature fit is largely driven by the bright
(1,1) and (2,2) lines, which can be excited throughout the object.
The (3,3) line could show significant excitation in the center of the
object but this emission would be underestimated since the temperature
parameter is largely determined by the strong emission from the other
two lines.  Furthermore, variations in the ortho-to-para ratios could
affect the fits; however we would require substantially more
information to resolve these variations.  The parameter fits represent
emission-weighted averages of the parameters under consideration.

The changes to the parameter estimation have resulted in 
approximately 10\% changes in the \ammonia-based parameters.
For example, the mean \tk\ of the sources in the Gemini OB1 molecular cloud 
have decreased by roughly 2 K from the values presented in D10.  
Because of the changes, we have reanalyzed the Gem OB1
sources as well, and present the results from the newest pipeline in
the comparison in this work.

\section{Results}\label{results}
\begin{figure}
\begin{center}
\epsscale{1.2}
\plotone{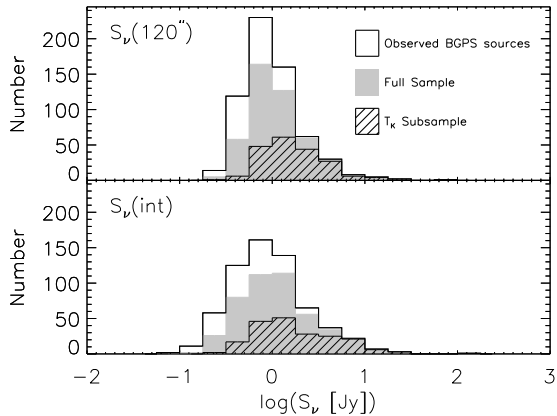}
\figcaption{\label{histsnu}
Distributions of \Snu(120\as) (top panel) and \Snu(int) (bottom panel).  The white histograms include all BGPS sources observed in \ammonia, the gray histograms include the full sample, and the striped histograms include the \tk\ subsample.
}
\end{center}
\end{figure}

Of the 737 \ammonia\ pointings, 722 overlap with BGPS sources
in the v1.0 catalog, while the remaining 15 pointings were 
selected by eye from an earlier BGPS catalog.  The 15 unmatched
pointings are excluded from our analysis because they do not 
have corresponding 1.1 mm data.  Eight of the 15 unmatched sources
do have a detection in the \ammonia(1,1) line, suggesting
that some of the low intensity BGPS emission is likely real.
We have additionally excluded ten of the 737 \ammonia\ pointings
due to radio frequency interference (RFI) in the \ammonia(2,2)
IF.  Six of the ten pointings with RFI were detected in the 
\ammonia(1,1) transition, but we are unable to characterize the
gas properties without a realistic upper limit for the (2,2) line.

To detect a line, we require $W\geq 5\Delta$W and \tpk$\geq 4\sigma$,
where $W$ is the integrated intensity, $\Delta$$W$ is the error in integrated
intensity, and \tpk\ is the peak brightness temperature of the line.  
We have detected the \ammonia(1,1)
line in 514 of the 707 \ammonia\ pointings that were matched with a BGPS
source and were not affected by RFI, for a detection rate of 73\%.
Since the positions of the observations obtained in 2008 were 
determined by eye from early BGPS data products there are multiple 
\ammonia\ pointings per BGPS source in some cases.  
The 514 \ammonia\ pointings detected in the (1,1) line correspond
to 456 unique BGPS sources (72\%\ of the 631 BGPS sources observed).  

Table \ref{nh3_props} lists the observed \ammonia\ properties for the 514 pointings
detected in the (1,1) line,
including BGPS source number (column 1), RA and DEC of the \ammonia\ pointing
(columns 2 and 3), radial velocity ($V_{\rm LSR}$; column 4), velocity
dispersion ($\sigma_{Vlsr}$; column 5), and peak main beam temperature
(\tmb) and integrated intensity ($W$) for the (1,1), (2,2), (3,3), and
(4,4) \ammonia\ transitions (columns 6-13).  Lines that were observed but not
detected are listed as upper-limits based on the observed rms ($\sigma$), and are given by
\tmb$< 4\sigma$ and $W< 5\Delta$$W$, where $\Delta W=\sigma\delta V \sqrt N$ is
the error in the integrated intensity, $\delta V$ is 
the channel width in velocity, and $N$ is the number of channels over 
which the rms was calculated.  These 514 spectra are included in an 
online version of Figure \ref{nh3spectra}.

In the following analysis, we consider
only the \ammonia\ pointings that lie closest to the 1.1 mm peaks,
and refer to the BGPS sources and these pointings as the ``full sample''.
We additionally define two subsamples from the full sample:  the
``\tk\ subsample'' and the ``near subsample''.  The \tk\ subsample 
consists of 199 BGPS sources that were detected in the (2,2) line.
Since sources without a (2,2) detection can only provide an upper limit 
for the gas kinetic temperature, the \tk\ subsample is used throughout
this section and section \ref{analysis} to characterize the properties
of sources excluding sources that only have upper limits.  The near
subsample consists of 199 BGPS sources that were placed at the near
kinematic distance (see section \ref{distances}), and is used in section
\ref{galactictrends} to assess the trends in source properties as a 
function of positon within the Galaxy.

The distribution of 1.1 mm flux densities is shown in Figure 
\ref{histsnu}.  We present flux densities within both a 
well-defined aperture (\Snu(120\as); top panel) and the integrated
flux density, which is dependent on the reduction and source extraction
algorithms (\Snu(int); bottom panel).  The white histograms denote the 
distribution of all observed BGPS sources, the gray histograms denote 
the full sample, and the striped histograms denote the \tk\ subsample.
The high flux density end of the distribution is well represented in
both the full sample and the subsample, while a significant fraction
of the faint BGPS sources are lacking (1,1) and (2,2) detections.

\begin{figure}
\begin{center}
\plotone{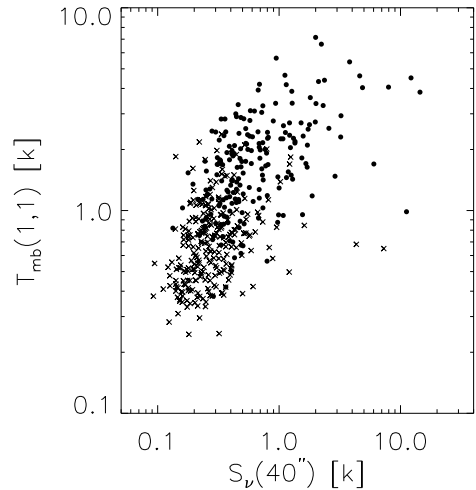}
\figcaption{\label{tmb1122} Observed peak \tmb\ of the \ammonia(1,1) line versus the flux within a beam, \Snu(40\as).  Circles mark sources with a detection in both the (1,1) and (2,2) lines while crosses mark sources with only an upper limit in the (2,2) line.
}
\end{center}
\end{figure}

Figure \ref{tmb1122} plots \tmb(1,1) versus the 1.1 mm flux per beam,
\Snu(40\as), for the full
sample.  The \tk\ subset is shown as black circles, while sources
without a detection in the (2,2) line are shown crosses.
As expected, the (2,2) line was not detected in
pointings toward the weaker 1.1 mm sources with faint (1,1) lines.

\begin{figure}
\begin{center}
\plotone{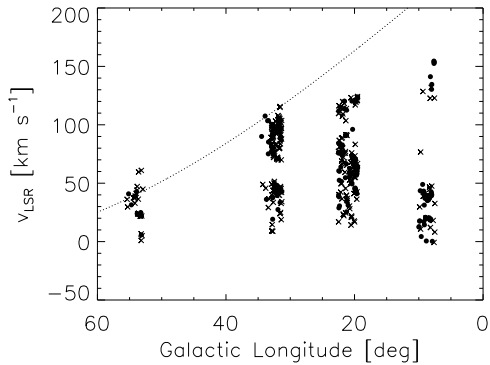}
\figcaption{\label{vlsr}
Observed $v_{lsr}$ for the full sample.  Black circles mark the \tk\ subsample, while crosses mark sources without a (2,2) detection.  The dotted line denotes the maximum radial velocity expected from circular rotation given by $v_{lsr,max} = \Theta(R_o)(1-\mbox{sin}(\ell))$.  Sources with radial velocities in excess of $v_{lsr,max}$ are placed at the tangent distance.
}
\end{center}
\end{figure}

The observed radial velocities for the full sample are shown in
Figure \ref{vlsr}.  The black circles denote the \tk\ subsample
and the crosses mark BGPS sources without a (2,2) detection.  The
dotted line denotes the maximum radial velocity expected from 
Galactic circular rotation as a function of $\ell$ given by
$v_{\rm lsr,max} = \Theta(R_o)(1-sin(\ell))$, where $\Theta(R_o)$
is the orbital velocity of the sun about the Galactic center.
The observed radial velocities follow the distribution of 
$^{13}$CO$(J=1-0)$ seen in the BU-FCRAO Galactic Ring Survey 
(GRS; Jackson et al.~2006).  There are 24 sources that show some evidence
of two velocity components.  These sources are noted in Column 8 of
Table \ref{gas_props}.

\begin{figure}
\begin{center}
\plotone{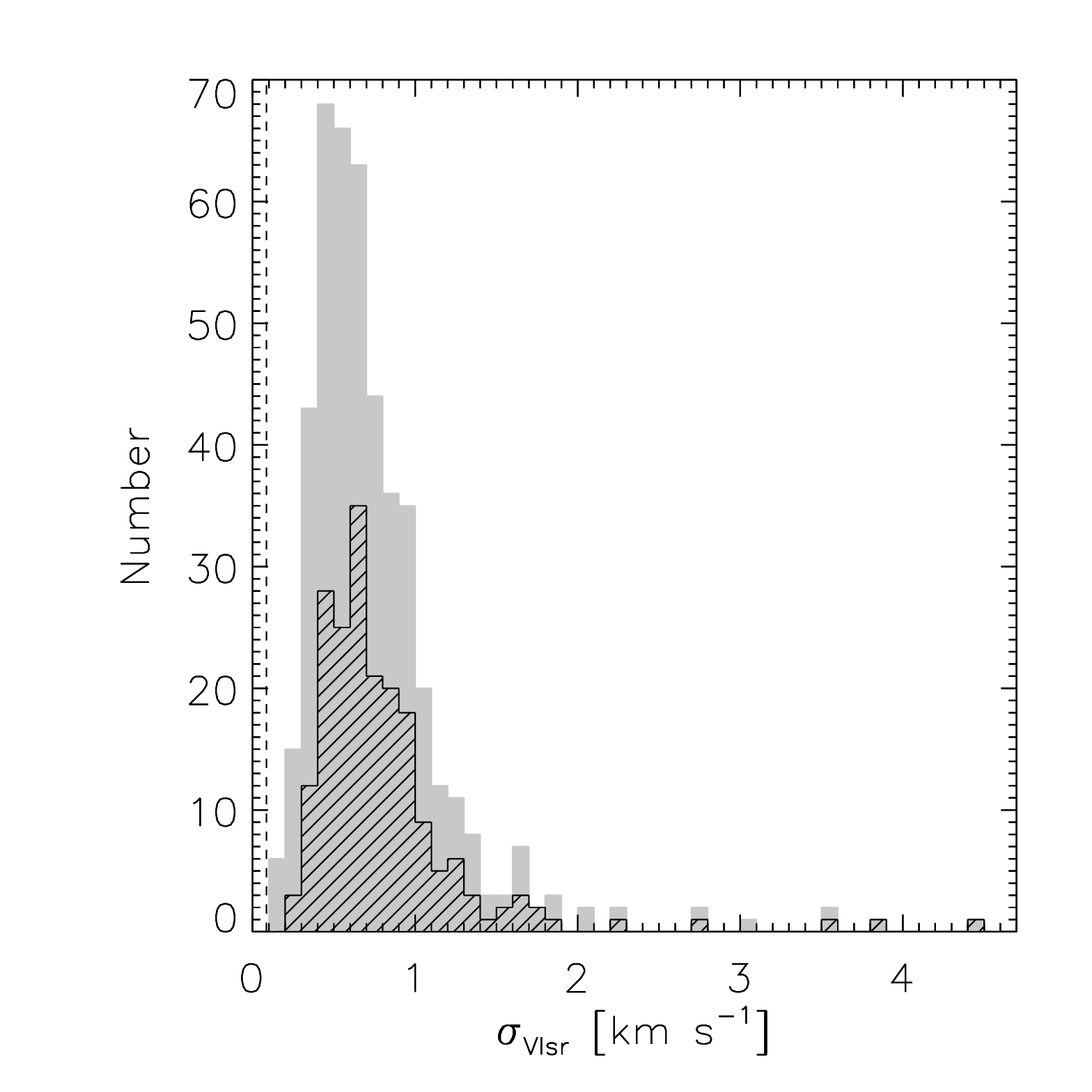}
\figcaption{\label{sigmav}
Distribution of Gaussian line widths for the full sample (gray histogram) and the \tk\ subsample (striped histogram).  Note $\sigma_{Vlsr}=\mbox{FWHM}/\sqrt{8ln2}$.  The dashed line denotes the resolution of the \ammonia\ observations of 0.08 \kms.  Six of the eleven sources with $\sigma_{Vlsr} \ge 2$ \kms\ are weak detections, one has two velocity components that are poorly fit with the current model, and the remaining four sources truly do exhibit line widths greater than 2 \kms.  
}
\end{center}
\end{figure}

We also measure the Gaussian width of the (1,1) line, 
$\sigma_{Vlsr}$, and plot the distribution of the
full sample (gray histogram) and the \tk\ subsample (striped
histogram) in Figure \ref{sigmav}.  Note that $\sigma_{Vlsr}=\mbox{FWHM}/\sqrt{8ln2}$.
The dashed line denotes
the spectral resolution of the \ammonia\ observations of 0.08 \kms.
The full sample can be characterized by 
\mean{\sigma_{Vlsr}}=0.76$\pm$0.49 \kms, and the \tk\ 
subsample by \mean{\sigma_{Vlsr}}=0.81$\pm$0.53 \kms.
We observe line widths up to $\sigma_{Vlsr}\sim$ 4 \kms.  Of the 11 sources with $\sigma_{Vlsr}\ge 2$ \kms,
six are weak detections with poor model fits, one has two velocity components,
and four truly do exhibit line widths greater than 2 \kms.  See Table \ref{stats} for
statistical properties of $\sigma_{Vlsr}$ for both the 
full sample and the \tk\ subsample.

We defer the analysis of the H$_2$O maser observations to a later paper,
but we do note the presence or absence of a water maser in Column 7 of
Table \ref{gas_props}.  We have detected the 22 GHz H$_2$O line towards
182 BGPS sources (40\%).

\section{Resolving the Kinematic Distance Ambiguity}\label{distances}
Without distances to the BGPS sources, the observed properties can not
be translated into physical properties.  The observed \ammonia\ radial 
velocity in conjunction with the Galactic rotation
model of Reid et al.~(2009; $\rm R_o=8.4\pm0.6$ kpc, 
$\Theta_{\rm o}=254\pm16$ \kms) provides kinematic distances to all 
sources with a detection in the \ammonia(1,1) transition.
However, the application of the Galactic rotation model introduces 
a distance ambiguity (known as the kinematic distance ambiguity, or KDA)
for sources within the solar circle.  A single radial velocity
provides two distances (refered to as a near and far distance) found
on either side of the tangent point.  Only sources found at the tangent
velocity do not suffer from the KDA.  We break the KDA based on the
presence of infrared dark clouds (IRDCs) or HI self-absorption (HISA).  

IRDCs are regions seen in 
absorption against the diffuse infrared background, some of which have been shown
to have high 
densities ($>10^5$ \cmv), cold temperatures ($<25$ K), and high 
column densities
($\sim 10^{23-25}$ \cmc; Egan et al.~1998; Carey et al.~1998, 2000;
Simon et al.~2006; Rathborne et al.~2005, 2006, 2007, 2008, 2010).  
Thousands have been cataloged from the MSX (Price et al.~2001) and GLIMPSE 
(Benjamin et al.~2003; Chambers et al.~2009) surveys (Egan et al.~1998; 
Carey et al.~1998; Simon et al.~2006; Peretto \& Fuller 2009; Chambers 
et al.~2009).  Spatial coincidence and structural similarity between a BGPS source and 
an IRDC places the BGPS source at the near kinematic distance where it is
able to absorb the diffuse IR background.
A BGPS source not coincident with an IRDC is assumed to be at the far
kinematic distance.   
We have cross matched the BGPS sources observed in \ammonia\ with the 
catalog of 
IRDCs seen in the \textit{Spitzer} GLIMPSE (Peretto \& Fuller 2009)
and identified 105 BGPS sources coincident with an IRDC.  
We have visually inspected the GLIMPSE 8 \um\ images for spatial 
coincidence and structural similarity between the 1.1 mm emission
and 8 \um\ absorption features 
both identified by Peretto \& Fuller (2009) as well as those not included in
their IRDC catalog.  47 IRDCs from Peretto \& Fuller (2009) that
were matched to a BGPS source
were deemed not to be evidence of the near distance due to a mismatch 
in position or structure of the IRDC compared to the BGPS emission.
Overall, 171 of the 456 BGPS sources in the full sample were found 
to coincide with an 8 \um\ IRDC and were placed at the near kinematic
distance, while 215 were assigned the far distance due to a lack of
correspondence between 8 \um\ absorption and 1.1 mm emission.  
Additionally, 70 BGPS sources had inconclusive evidence of an 8 \um\
IRDC and were flagged as possibly at the near distance.

We note that the assumption that an IRDC is located at the near distance
could be false in some cases.  For example, since the identification of an IRDC relies
on a bright infrared background, a source at the near distance without a
significant infrared background to absorb would be falsely assigned to 
the far kinematic distance.  Similarly, sources exhibiting weak 8\um\ 
absorption could be located at the far distance.  Battersby et al.~(2011)
have identified a sample of 5 high column density sources in the \textit{Herschel}
Hi-GAL data that have weak 8\um\ absorption and exhibit a discrepancy between
the far-IR column density and that calculated from the 8 \um\ 
absorption under the assumption of the near kinematic distance, which suggests
the sources may be located at the far distance.  
While our assumption that a BGPS source associated with an IRDC is at the near distance
is accurate in most cases, it disregards the fact that some IRDCs may be
located at the far distance.  

To minimize false kinematic distance 
assignments, we have also considered HI self-absorption when resolving 
the KDA.
The HI self-absorption technique for resolving the KDA is similar to
the IRDC method in that cold, dense material located at the near 
distance will absorb the emission from warmer material with 
the same radial velocity found at the far kinematic distance 
(e.g., Knapp 1974; Burton et al.~1978).  
The signature of HISA is frequently displayed in molecular clouds (e.g., 
Knapp 1974), and theoretical studies by Flynn et al.~(2004) demonstrate
that all molecular clouds contain enough HI to produce a self-absorption
profile against a warm continuum background.  Indeed, HISA 
has been seen throughout 
the Canadian Galactic Plane Survey of 21 cm emission toward the outer 
Galaxy (Gibson et al.~2000, 2005).  This technique has recently 
been used to resolve the KDA toward star-forming regions and HII regions
(Jackson et al.~2002; Fish et al.~2003; Busfield et al.~2006; 
Pandian et al.~2009; Anderson \& Bania 2009; Roman-Duval et al.~2010).  
There are situations in which this
method for KDA resolution could be incorrect.  For example, a BGPS 
source at the
far distance could have a peculiar velocity that places it at the same
velocity as a warmer source at the far distance and causes it to exhibit
HISA.  Also, an HII region embedded within a cold source at the far 
distance could also be incorrectly assigned to the near distance
(Roman-Duval et al.~2010).
In such cases, the BGPS source would be incorrectly assigned
to the near distance.  
Here we accept the KDA resolution as correct, but note that it may not be 
correct in all cases.

\begin{figure}
\begin{center}
\plotone{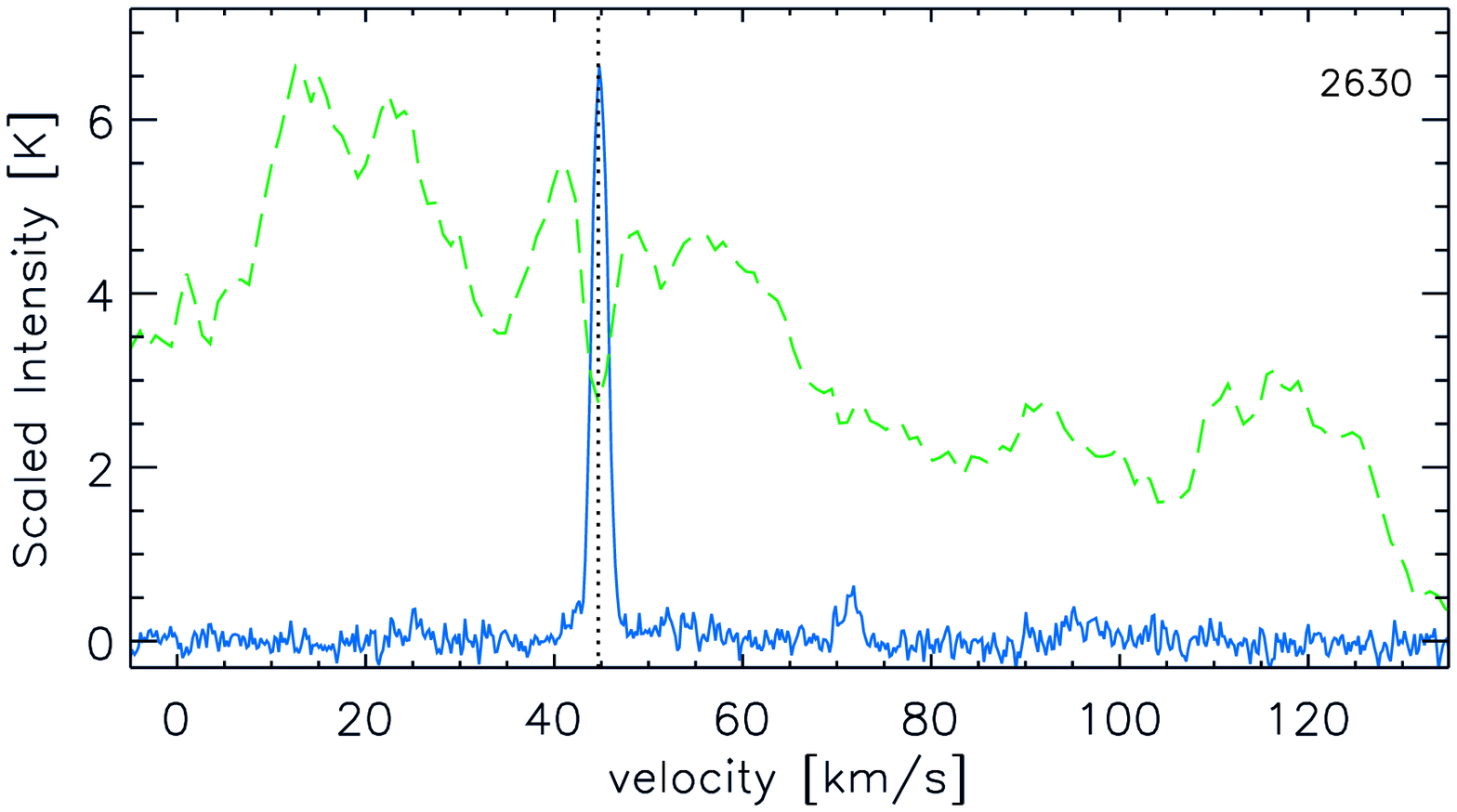} 
\plotone{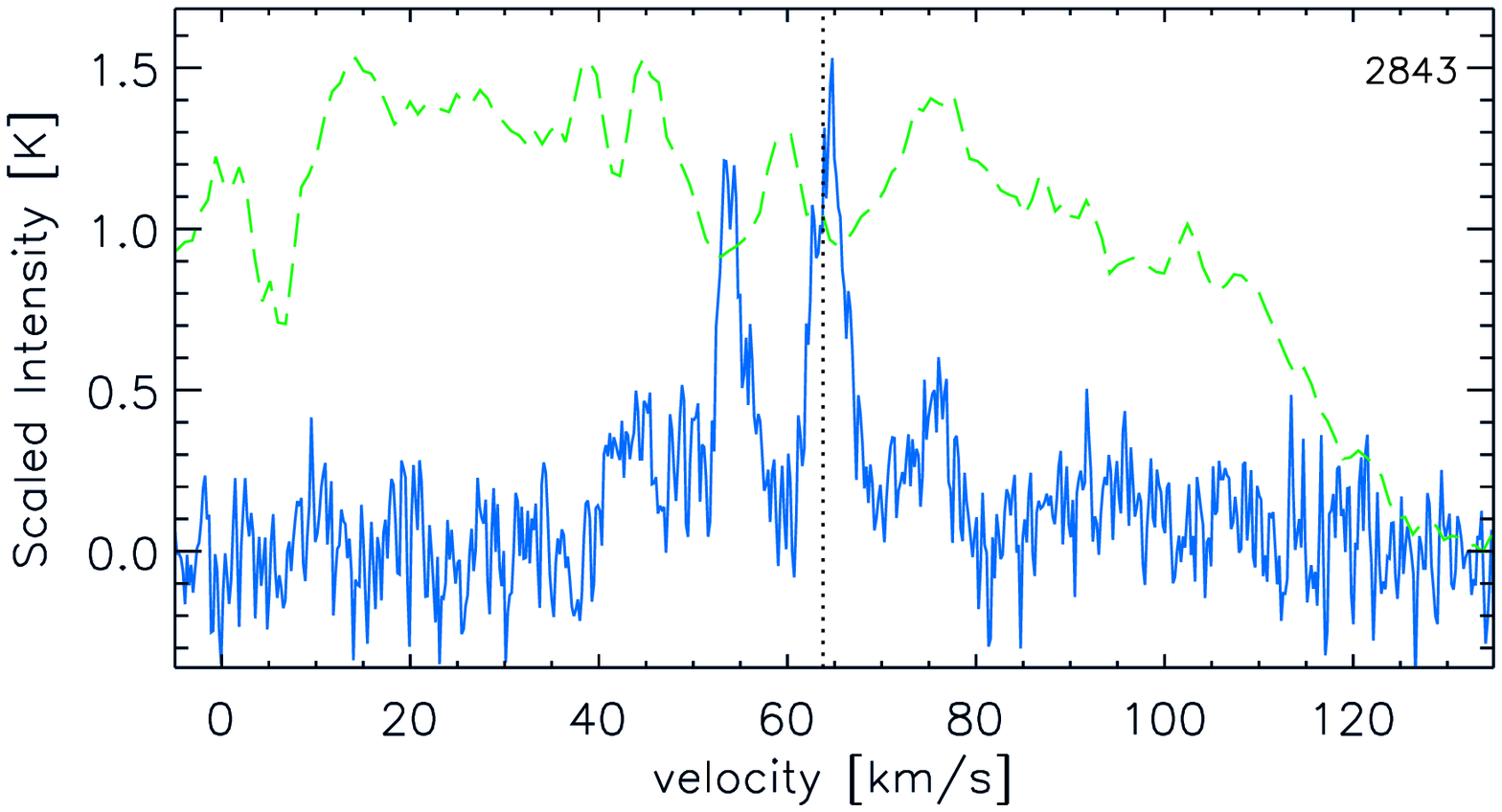} 
\plotone{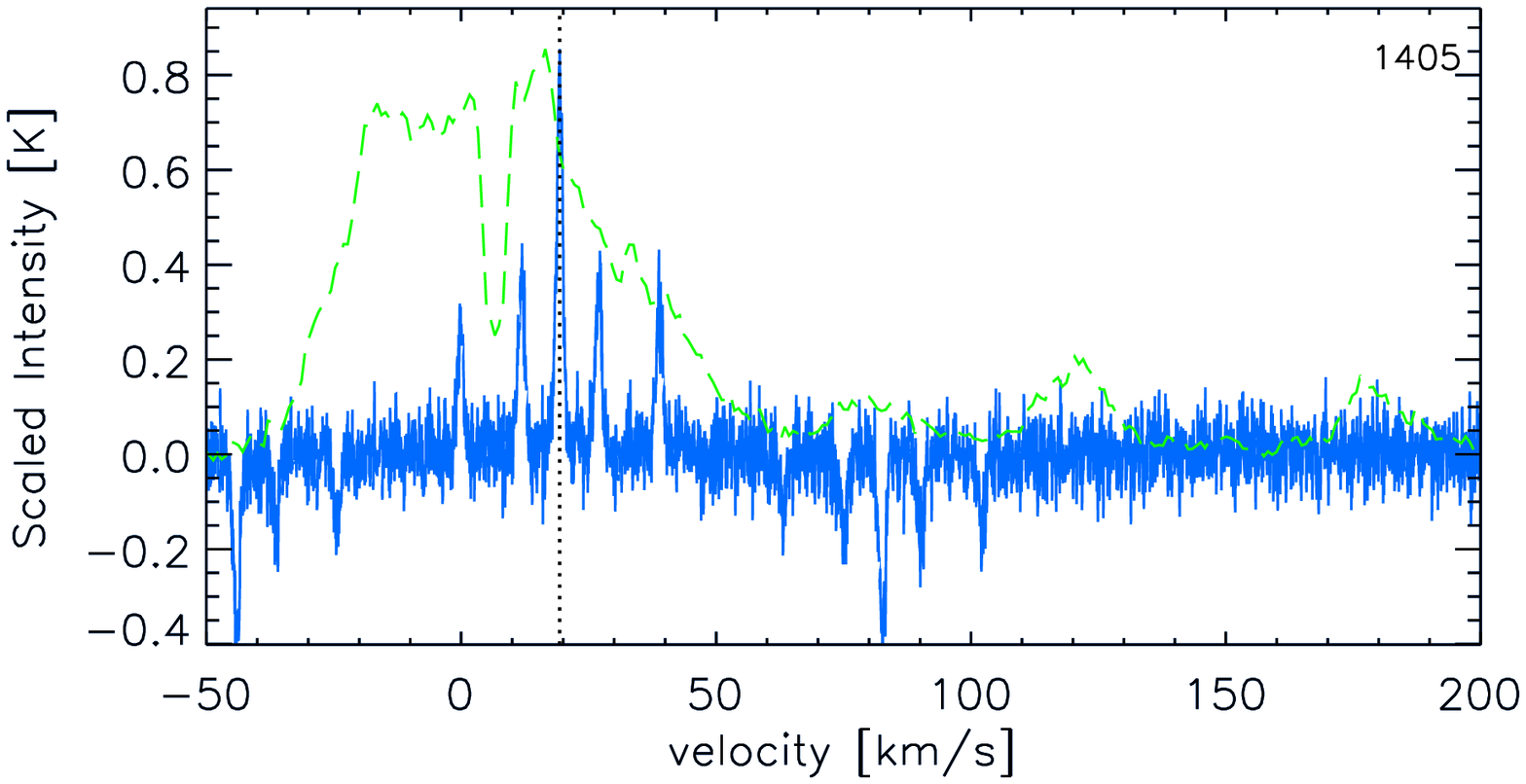} 
\figcaption{\label{hisayes} 
 HISA examples.  The green dashed line shows the HI spectrum, the blue solid line shows the  $^{13}$CO spectrum (top and middle panels) or the observed \ammonia(1,1) spectrum (bottom panel), and the dotted black line marks the observed \ammonia\ velocity.  Top panel: BGPS source 2630.  There is a very clear HISA feature at the same velocity as the $^{13}$CO peak and the observed \ammonia\ velocity.  Middle panel: BGPS source 2843 with potential HISA.  The width of the potential HISA feature and the offset in velocity from the dense gas tracers resulted in a ``m'' flag for HISA.  Bottom panel:  BGPS source 1405 with no HISA.  $^{13}$CO observations were not available in the $7.5\degree \le \ell \le 10.0\degree$ field, we overlay our \ammonia\ spectrum instead.  The two \ammonia\ features seen in absorption on either side of the \ammonia\ emission lines are due to the frequency switching used during observations.  There is a clear lack of HISA at the same velocity as the \ammonia\ spectrum.  
}
\end{center}
\end{figure}

We have used the 21 cm VLA Galactic Plane Survey (VGPS; Stil et al.~2006) 
and Southern
Galactic Plane Survey (SGPS; McClure-Griffiths et al.~2005) to search for 
HI self-absorption toward the 456 BGPS sources in the full sample.
Figure \ref{hisayes} shows examples of possible outcomes of the HISA method.
We have marked the radial velocity observed in our \ammonia\ 
observations, but use $^{13}$CO (1-0) spectra from the BU-FCRAO
Galactic Ring Survey (GRS; Jackson et al.~2006) for comparison since 
$^{13}$CO emission traces the line width and spatial extent of HISA well 
(e.g., Goldsmith \& Li 2003; Li \& Goldsmith 2005). 
The top panel shows a clear example of a HISA for BGPS source 2630.  
The green dashed line is the HI spectrum, the blue solid line is the 
$^{13}$CO spectrum, and the dotted black line marks the observed 
\ammonia\ velocity.  The HISA feature matches both the observed 
\ammonia\ velocity and the $^{13}$CO peak and line width.   
The middle panel shows a spectrum toward a BGPS source with potential HISA.  
The HI spectrum
has an absorption feature slightly offset from the observed \ammonia\
velocity and $^{13}$CO peak and the $^{13}$CO line width does not match
the potential absorption feature.  
The bottom panel shows a clear
example a BGPS source placed at the far distance based on the HISA
method.  Here the blue solid line displays the observed \ammonia(1,1)
spectrum since the GRS did not include our $7.5\degree \le \ell \le 10.0\degree$ field.
The \ammonia\ hyperfine structure can clearly be seen in this spectrum,
and the absorption features seen on either side of the \ammonia\ emission
are a result of the frequency switching performed during observations.  The
HI spectrum is smoothly decreasing at the \ammonia\ velocity, and shows
no absorption feature.  
Of the 456 BGPS sources in the full sample, 
the HISA technique has identified 116 BGPS
sources at the near distance, while 130 displayed questionable HISA and
could not definitively break the KDA
and 210 suggested the far distance due to a lack of HISA.  

When investigating matches between a BGPS source and either an IRDC or
HISA we assign one of three flags to each source for each method:  
``y'' denoting a positive association suggesting the near distance, 
``n'' denoting no association, and ``m'' representing a questionable association.
A BGPS source is assigned the near distance if it receives a single ``y'' 
(143 sources), two ``y'' flags (72 sources),
or two ``m'' flags, one from the IRDC and HISA methods (19 sources).  
Conversely, a BGPS source is assigned the far distance if it receives 
two ``n'' flags (126 sources) or one ``n'' and one ``m'' flag (96 sources).  

\begin{figure}
\begin{center}
\plotone{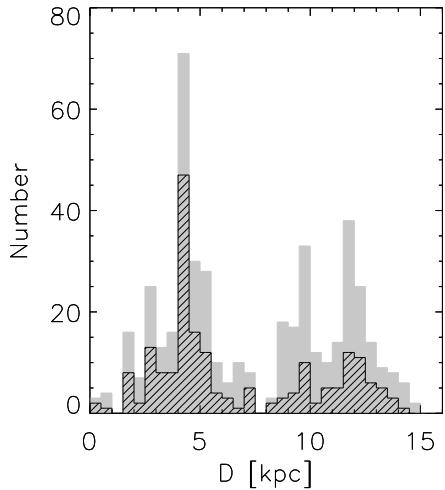}
\figcaption{\label{dhelio}
Distribution of kinematic distances.  The gray histogram marks the full sample of 456 BGPS sources, while the striped histogram denotes the \tk\ subsample.  The peak in the distribution around 4 kpc is a result of the 5 kpc molecular ring.  
}
\end{center}
\end{figure}

\begin{figure}
\begin{center}
\plotone{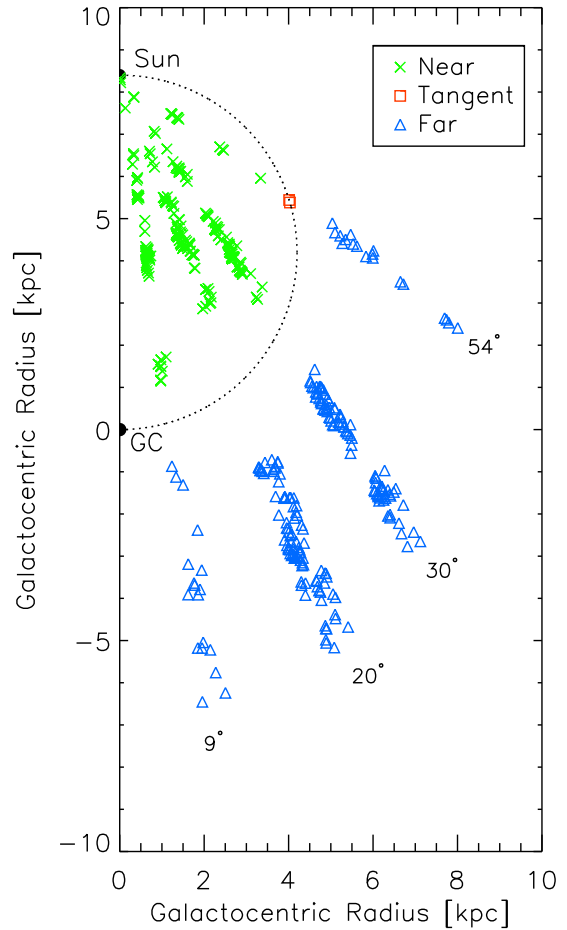}
\figcaption{\label{galaxypic}
Location of the full sample of 456 BGPS sources in the Galaxy.  Green crosses mark BGPS sources placed at the near kinematic distance, red squares mark sources at the tangent distance, and blue triangles mark sources placed at the far kinematic distance.  The dotted line marks the location of the tangent distance as a function of Galactic longitude.
}
\end{center}
\end{figure}

By combining the IRDC and HISA results, we have identified 233 BGPS 
sources at the near distance, 221 at the 
far distance, and two at the tangent distance where the KDA does not 
exist.  Table \ref{distancestable} lists the BGPS source ID (column 1),
HISA, IRDC, and KDA flags 
(columns 2-4, respectively),
as well as the assumed kinematic distance (column 5; asymmetrical errors
listed in parentheses) and resulting \rgal\ (column 6). 
Figure \ref{dhelio} shows the distribution of distances
for the full sample of 456 BGPS sources (gray histogram) and for 
the \tk\ subset (striped histogram).  The peak in the distribution
near 4 kpc is likely a result of  abundant dust and gas
in the 5 kpc molecular ring (e.g.~Burton et al.~1975; Scoville \& Solomon 1975;
Cohen \& Thaddeus 1977; Robinson et al.~1984; Clemens et al.~1988; 
Kolpak et al.~2002; Rathborne et al.~2009).  Figure \ref{galaxypic} 
shows the location of the 456 BGPS sources in the Galaxy.  The green 
crosses mark sources placed at the near distance, red squares mark sources
at the tangent distance, while blue triangles mark sources placed at
the far distance.  The dotted line denotes the location of the tangent 
distance as a function of Galactic longitude.  

As previously discussed, the use of IRDCs and HISA to resolve the KDA may
lead to incorrect assignments of both near and far kinematic distances.
To assess the possible effects of incorrect distance assignments, we 
have explored the worst case scenario, 
in which all sources have been assigned incorrect distances, by reversing 
the KDA resolution for all sources in the full sample.  This test showed 
little effect on the statistical properties of the sample, suggesting that
our results will not be greatly affected by incorrect distance assignments.

The use of a circular rotation curve for the Galaxy necessarily simplifies
Galactic kinematics, and thus introduces errors into our assumed distances.  
For example, spiral arm structure can cause the radial velocities of nearby
sources to deviate from circular rotation and display peculiar velocities.  
Similarly, the radial velocities of nearby sources that are orbiting the 
Galaxy at the angular velocities similar to that of the sun are dominated by
peculiar velocities.  Additionally, sources located within the Galactic
bar exhibit streaming motions that can deviate significantly from circular 
rotation (e.g.~Burton \& Bania 1974).  In these cases, the observed radial
velocities will return inaccurate kinematic distances under the assumption
of simple circular rotation.  Thus, kinematic distances have large inherent 
errors.  Sources located in the $\ell\sim8.5\degree$ field could be strongly 
affected by the streaming motions within the bar.  Sources in this region 
assigned small kinematic distances should be viewed skeptically.  It is 
possible that peculiar velocities or streaming motions have resulted in
a very low radial velocity and the source is truly located within the
Galactic bar or Scutum spiral arm at approximately 4-5 kpc.  These
uncertainities should be noted; however, without a detailed model of the
spiral arm structure, peculiar velocities and streaming motions throughout
the Galaxy, we are unable to correct for these effects.  In this paper,
we accept the kinematic distances as correct but note the potential for 
large errors.  

\begin{figure}
\begin{center}
\plotone{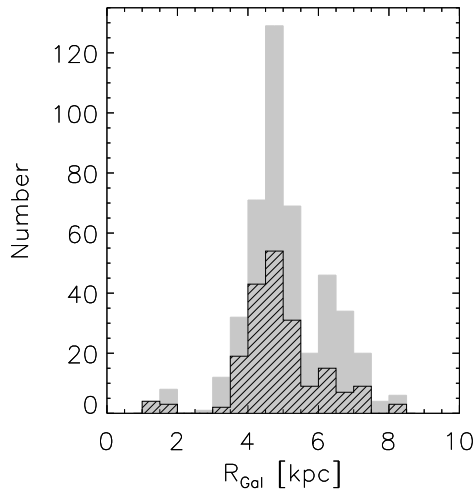}
\figcaption{\label{rgal}
Distribution of Galactocentric radii for the full sample (gray histogram) and the \tk\ subsample (striped histogram).  The peak in number of sources at \rgal$\sim$4.5 kpc corresponds to sources in the 5 kpc molecular ring.
}
\end{center}
\end{figure}

While the kinematic
distance is subject to the near/far distance ambiguity in the inner
Galaxy, each near/far distance pair corresponds to a single distance
from the Galactic Center, the Galactocentric radius.  Figure \ref{rgal}
plots the distribution of Galactocentric radius for the full sample
(gray histogram) and the \tk\ subsample (striped histogram).  The 
peak in the distribution at \rgal$=4-5$ kpc corresponds to the 5 kpc
molecular ring, which corresponds to the end of the Galactic bar and 
beginning of the Scutum spiral arm.  The peak in source number at
\rgal$=$6 kpc corresponds to the Sagittarius arm.  We do not have enough 
sources at large \rgal\ for a definitive detection or non-detection
of the Perseus arm at \rgal$=$8 kpc.  
Section \ref{galactictrends} discusses trends in 
properties as a function of Galactocentric radius.

\section{Analysis}\label{analysis}
With kinematic distances, we are able to calculate physical 
properties such as mass and volume-averaged density.  In this section,
we present each property and describe the biases inherent in each.
Errors are propagated through each derived quantity and are presented 
for individual sources within parentheses in Tables \ref{mm_props}, and
\ref{nh3_props}$-$\ref{derived_masses_densities}.

When discussing mean properties here and throughout,
the uncertainties given are the standard deviation about the mean
rather than the uncertainty in the mean.  Some property 
distributions are non-Gaussian, and the uncertainty given should be 
viewed as a measure of the scatter in the sample rather than the error.

General statistics (minimum, mean, standard deviation, median, and maximum) 
for the properties presented in this section for the full sample and 
the \tk\ subsample can be found in Table \ref{stats}.

\subsection{Physical Size}\label{physicalsize}
\begin{figure*}
\begin{center}
\epsscale{0.75}
\plotone{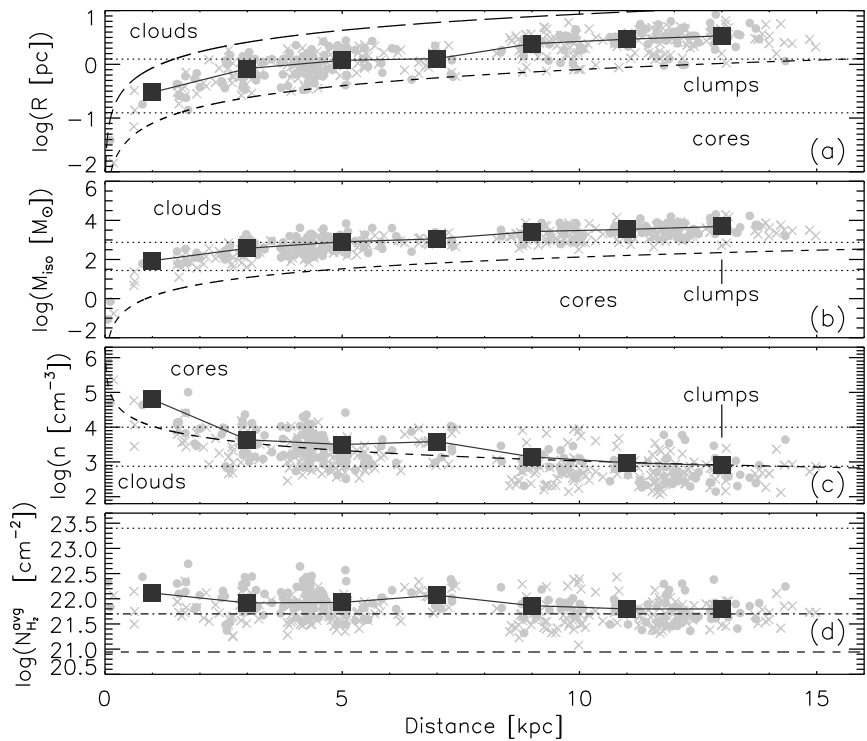}
\figcaption{\label{versusd}
Physical properties versus kinematic distance.  In all panels, the dotted lines delineate typical properties of clouds, clumps, and cores while the short dashed lines mark lower limits of the physical properties within a 33\as\ beam.  The gray circles mark the \tk\ subsample and gray crosses mark sources with no \ammonia(2,2) detection.  The black squares and lines denote the mean of each property within 2 kpc wide bins centered at 1, 3, 5, 7, 9, 11, and 13 kpc.  (a) Physical radius.  The long dashed line denotes the upper limit to source size imposed by the spatial filtering of uniform emission on size scales greater than 5.9\am.  (b) \miso(int).  (c) Volume averaged density.  (d) H$_2$ column density.   The dotted line in panel (d) marks the surface density of 1 g \cmc\ required to prevent fragmentation and allow a massive star to form (Krumholz \& McKee 2008).  The dash-dot line denotes the threshold gas surface density required for efficient star formation as seen in nearby low mass star-forming regions (Lada et al.~2010; Heiderman et al.~2010).
}
\end{center}
\end{figure*}

We calculate the physical radius based on the object radius determined
by the source extraction algorithm (Equation \ref{radiuseqn}) and 
the kinematic distance.  The physical radii for the full sample are given in
column 2 of Table \ref{derived_masses_densities} and are 
shown in Figure \ref{versusd}a.  The gray circles are the \tk\ subsample
and the gray crosses are BGPS sources with no (2,2) detection.  The black
squares and the solid black line mark the mean physical radius within 2 kpc
bins.  The short dashed line marks the physical radii corresponding to the
beam radius of 16.5\as\ as a function of distance, and the long dashed line
marks the physical radii corresponding to 2.95\am\ (half of the spatial
scale at which the BGPS loses sensitivity due to spatial filtering).
BGPS sources with extracted radii smaller than half the beam size
have been set as upper limits with radii equal to half the beam size.
The mean physical radius increases with distance as expected due to 
the upper limit on spatial scales imposed by spatial filtering during 
data reduction and the lower limit due to the beam size.  

Typical radii limits for cores, clumps, and clouds are denoted with the 
dotted lines at 0.125 pc and 1.25 pc (cores $R=0.01-0.1$ pc,
clumps $R=0.15-1.5$ pc, clouds $R=2.0-7.5$ pc; Bergin \& Tafalla 2007)
in Figure \ref{versusd}.  Assuming that BGPS 
sources have a constant density, we can place contraints
on the distances at which each type of source would be included in
the BGPS catalog.  Cores farther
than 1.25 kpc will be unresolved by the 33\as\ beam (for $R=0.1$ pc),  
although unresolved cores can still be detected by the BGPS if they have 
flux densities high enough to counter balance beam-averaging effects
and remain above the beam-averaged detection threshold 
(see Section 7.2 of D10).  The largest ($R=1.5$ pc) uniform density clumps
at distances less than 1.75 kpc would be filtered out but would be resolved
across the Galaxy.  The smallest clumps ($R=0.15$ pc) would become 
unresolved relative to the beam at 1.9 kpc.  Similarly, the smallest clouds 
would be filtered out at distances closer than 1.2 kpc, while the largest
would be filtered out if closer than 8.7 kpc.  The largest clouds 
are resolved even at the far side of the Galaxy.

The mean 1.1 mm source radius increases with distance and sources
are extended compared to the 33\as\ beam.  There are 24 unresolved
BGPS sources for which we will also need to consider mass 
in order to place them in the hierarchical structure.

\subsection{Gas Kinetic Temperature}\label{tk}
The gas kinetic temperature is calculated based on the excitation
temperatures of the observed \ammonia\ lines.  For sources with 
only the (1,1) line
detected, the rms of the (2,2) spectra was used as an upper limit
to calculate the (2,2) excitation temperature and the resulting \tk\ is
only an upper limit.  The uncertainties in individual \tk\ measurements
are small, with $\mean{\sigma_{\tk}/\tk} = 0.06$.  

The derived \tk\ are given in column 3 of Table \ref{gas_props}.  The full sample
is characterized by $\mean{\tk} < 15.6\pm5.0$ K, while the \tk\
subsample is characterized by $\mean{\tk} = 17.4\pm5.5$ K.  The
fractional scatter about the mean (0.32) is five times larger than
the individual uncertainties in the measurements (0.06) so the
differences among the sources are real.  See Table \ref{stats} for
further statistics.  There are 81 sources that likely require multiple
\tk\ components to fit the \ammonia(3,3) emission as well as the
(1,1) and (2,2).  These sources are marked in Column 8 of
Table \ref{gas_props}.  

Without mapping the \ammonia\ emission towards the BGPS sources
we can not characterize kinetic temperature gradients in the gas.
However, since the \ammonia\ pointings correspond to the peak 1.1 mm 
emission, which closely traces the \ammonia\ emission (Friesen
et al.~2009), the measured \tk\ is likely an upper limit as
\tk\ has been shown to increase toward the peak of the \ammonia\
emission (Zinchenko et al.~1997).  Thus the \ammonia\ outside of the
GBT beam are likely to be colder than the \tk\ observed toward the
peak of the 1.1 mm emission.

\subsection{Isothermal and Virial Masses}\label{masses}
Assuming the dust can be characterized by a single temperature,
we calculate the isothermal mass, \miso, as
\begin{eqnarray} \miso= & \frac{\Snu D^{2}}{\kappa_{\nu}B_{\nu}(\td)} \notag \\
= & 13.1\ \mbox{M$_\odot$}\ \left(\frac{\Snu}{1\ Jy}\right)\left(\frac{D}{1\ kpc}\right)^2\left(\frac{e^{13.0\ K / \td}-1}{e^{13.0\ K / 20.0\ K}-1}\right) ,\label{miso}
\end{eqnarray}
where \Snu\ is the 1.1 mm flux density, D is the kinematic distance, 
$\kappa_{\nu}$ is the dust opacity per gram of dust and gas and includes
a gas-to-dust ratio of 100, and $B_{\nu}$ is the Planck function evaluated 
at \td.  We logarithmically interpolate the Ossenkopf \& Henning (1994)
dust opacities (Table 1, Column 5, commonly referred to as OH5 dust)
to 271.1 GHz (the effective central frequency of the Bolocam bandpass
convolved with a 20 K blackbody modified by an opacity varying with 
frequency as $\kappa_{\nu} \propto \nu^{1.8}$; see Aguirre et al.~2010)
and find $\kappa_{\nu} = 0.0114$ cm$^2$ g$^{-1}$.

We make the simplifying
assumption that the dust and gas are collisionally coupled such that $\td=\tk$
and use the derived \tk\ in calculating \miso.  As noted in Section
\ref{tk}, the derived \tk\ are likely an upper limit to the average
\tk\ for an entire BGPS source.  Thus, the assumption that $\td=\tk$ 
characterizes the entire BGPS source will overestimate \td\ and 
underestimate \miso.  As discussed in D10, based on \ammonia\ 
maps of molecular clouds presented in Zinchenko et al.~(1997), 
assuming the peak \tk\ for the entire source may underestimate \miso\ 
by a factor of up to 2.  Additionally, upper limits due to 
non-detections of the \ammonia(2,2) line provide only lower limits 
on \miso.  

\miso\ could also be overestimated if some of the 1.1 mm emission
is radio free-free emission rather than thermal dust emission.
Dunham et al.~(2011) found that the mean contribution of free-free
emission is only 3\%\ based on a comparison of BGPS flux densities
and 6 cm radio emission from Urquhart et al.~(2009).  For most 
BGPS sources, free-free emission will not significantly increase the flux
density, but for a small number of sources approximately half or 
more of the 1.1 mm flux density could be a result of free-free emission.

\begin{figure}
\begin{center}
\epsscale{1.1}
\plotone{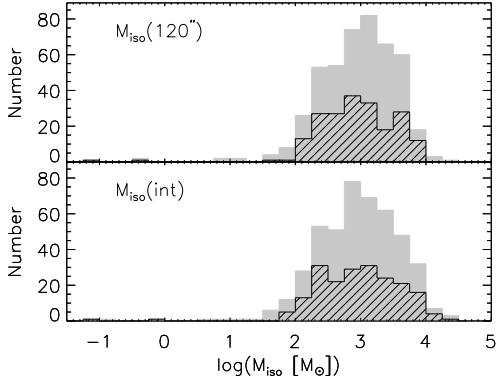}
\figcaption{\label{histmass}
The distribution of isothermal masses \miso(120\as) (top panel) and \miso(int) (bottom panel) for the full sample (gray histograms) and the \tk\ subsample (striped histograms).
}
\end{center}
\end{figure}

\miso(120\as) and \miso(int) are listed in Table \ref{derived_masses_densities}.
The distributions of \miso(120\as) and \miso(int) are shown in Figure
\ref{histmass} for the full sample (gray histograms) and the \tk\
subsample (striped histograms).  The \tk\ subsample is characterized
by $\mean{log(\miso(120\as)/\msun)} = 2.95\pm0.61$ and
$\mean{log(\miso(int)/\msun)} = 2.94\pm0.66$.  See Table \ref{stats}
for additional statistics for these mass distributions.  
The lowest mass sources (\miso$<10$ \msun) at distances less than 
1 kpc, and with velocities near 0 \kms\ have unreliable
kinematic distances as discussed in Section \ref{distances}.  

We also calculate the virial mass for each source assuming a spherical, 
uniform density gas cloud given by
\begin{equation} \mvir=\frac{5\sigma_{Vlsr}^{2}\rm R}{G}, \label{mvireqn}
\end{equation}
where $\sigma_{Vlsr}$ is the Gaussian line width, R is the
physical radius, and G is the gravitational constant.  This equation
assumes that the \ammonia\ line width observed at the peak of the
1.1 mm emission is also valid at the edge of the observed 1.1 mm
emission.  This assumption is unlikely to hold for the larger BGPS
sources at the farthest distances, as \ammonia\ emission is not
likely to trace entire molecular clouds but rather the denser
regions within the clouds.
As discussed in D10, these assumptions likely result
in an overestimate of \mvir.  Zinchenko et al.~(1997) found
that the line width decreases away from the peak of the \ammonia\
emission in approximately half of their sample, which would lead to 
an overestimate of \mvir.  Additionally, the BGPS sources are not likely
spherical or uniform density.  The mean aspect ratio of the full sample
is $1.5\pm0.3$.  We cannot characterize the density distribution of 
the BGPS sources with our current data set, so we consider the mean power-law
density profile found from modeling dust emission from massive 
star-forming regions, $\mean{p}=1.8\pm0.4$ (Mueller et al.~2002).
If we correct for shape and non-uniform density as described in D10, 
$\mean{\mvir/\miso}$ would be decreased by 36\%. Since we do not
have a measure of the density power-law, or \ammonia\ maps to quantify
these effects for each BGPS source, we do not include these corrections
in subsequent analysis but note that the virial masses presented are 
upper limits.  

\begin{figure}
\begin{center}
\plotone{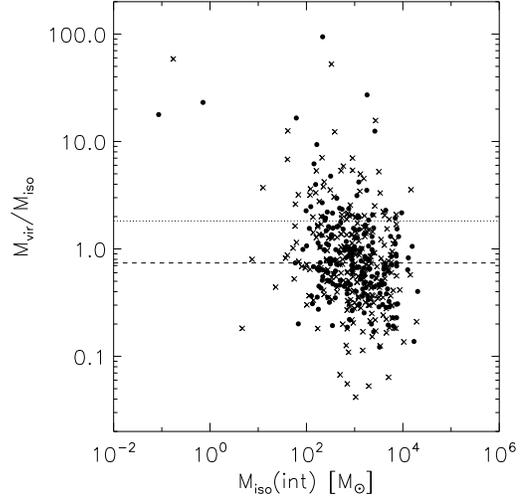}
\figcaption{\label{mvirmiso}
\mvir/\miso(int) ratio versus \miso(int) for the full sample.  The dotted line marks the mean virial parameter of the full sample, $\mean{\mvir/\miso}=1.82$, and the dashed line marks the median virial parameter of 0.74.
}
\end{center}
\end{figure}

\mvir(int) is given in Table \ref{derived_masses_densities}.
The full sample is characterized by $\mean{\mbox{log}(\mvir/\msun)} = 2.85\pm0.66$
while the \tk\ subsample is characterized by  $\mean{\mbox{log}(\mvir/\msun)} = 2.87\pm0.64$.  
Figure \ref{mvirmiso} plots the virial parameter, $\mvir/\miso$,
versus \miso(int).  The full sample is characterized by 
$\mean{\mbox{log}(\mvir/\miso)}=-0.09\pm0.45$, and a median of $10^{-0.13}$, which is more 
representative of the full sample since the mean is skewed by the few
sources with a virial parameter of $\sim$100.  
These BGPS sources with \mvir/\miso$\sim$100 are the local sources 
with unreliable kinematic distances and upper limits for the source 
sizes.  Since the virial parameter is proportional to $D^{-1}$, the 
observed scatter in the virial parameter could be a result of errors 
in the derived kinematic distances.  Additionally, the scatter could 
be due to deviations from a uniform density distribution or varying
source ellipticities.  Alternatively, the assumption that 
the observed line width is applicable at the boundary of the 1.1 mm 
emission could be invalid.

\miso(int) is shown versus kinematic distance in Figure \ref{versusd}b. 
The gray circles mark the \tk\ subsample, and the gray crosses mark
BGPS sources with an upper limit for \tk, and subsequently a lower
limit in \miso.  The dashed black line marks the mass within a beam 
corresponding to a 5$\sigma$ detection, where we assume 
$\sigma = 0.020\pm0.006$ Jy beam$^{-1}$ from the mean and standard
deviation of the rms in each BGPS field, and \td$=20$ K.  The black
squares and black line connecting them mark the mean \miso\ within
2 kpc wide bins.  The dotted lines mark limits between clouds,
clumps, and cores of 27.5 \msun\ and 750 \msun\ based on the
following characteristic values of each type:
clouds have $M=10^3-10^4$ \msun, clumps have $M=50-500$ \msun, and 
cores have $M=0.5-5$ \msun\ (Bergin \& Tafalla 2007).  
The mean \miso\ increases with distance, and all masses are above
the limiting mass per beam because the sources
are significantly extended compared to the beam.  

\begin{figure*}
\begin{center}
\epsscale{0.95}
\plottwo{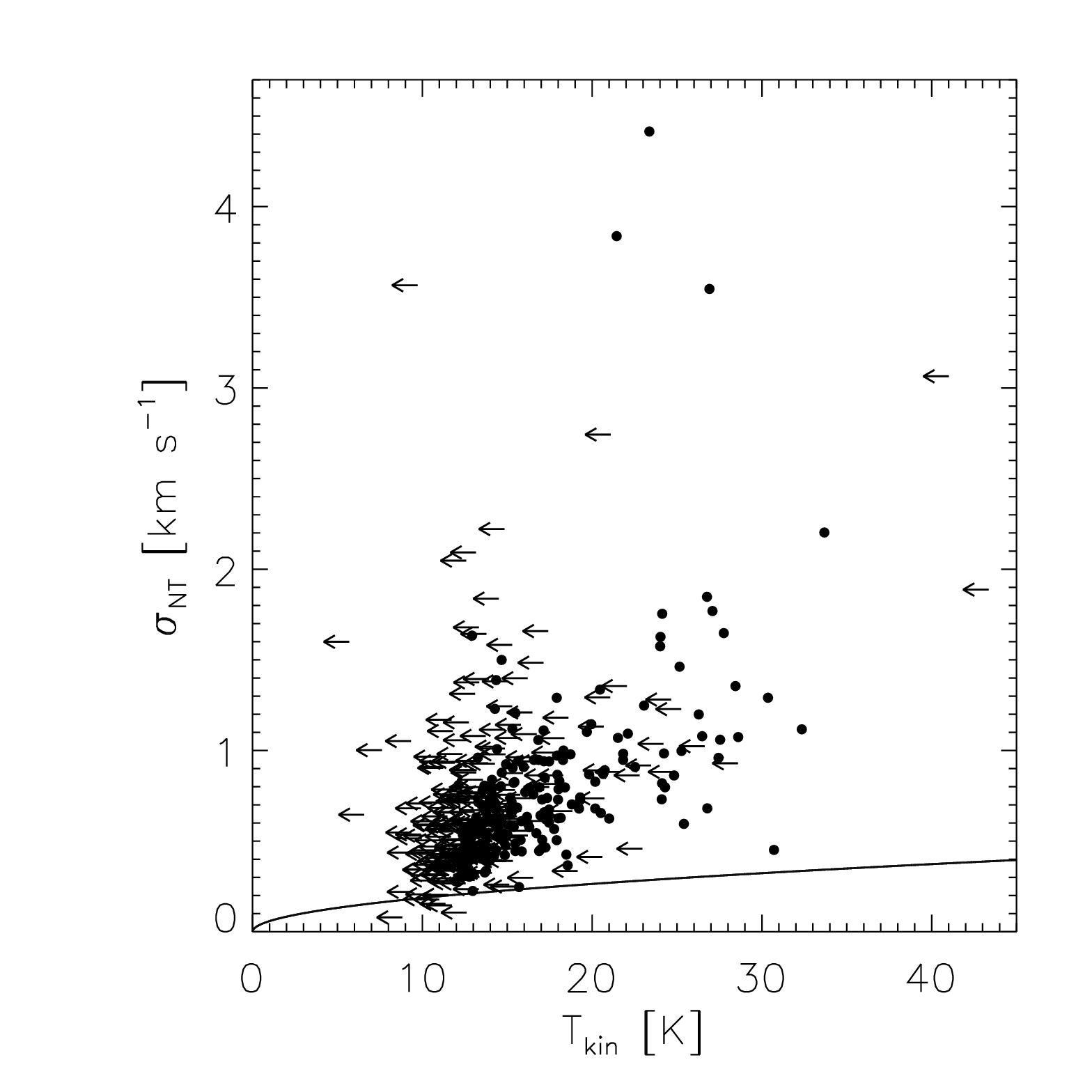}{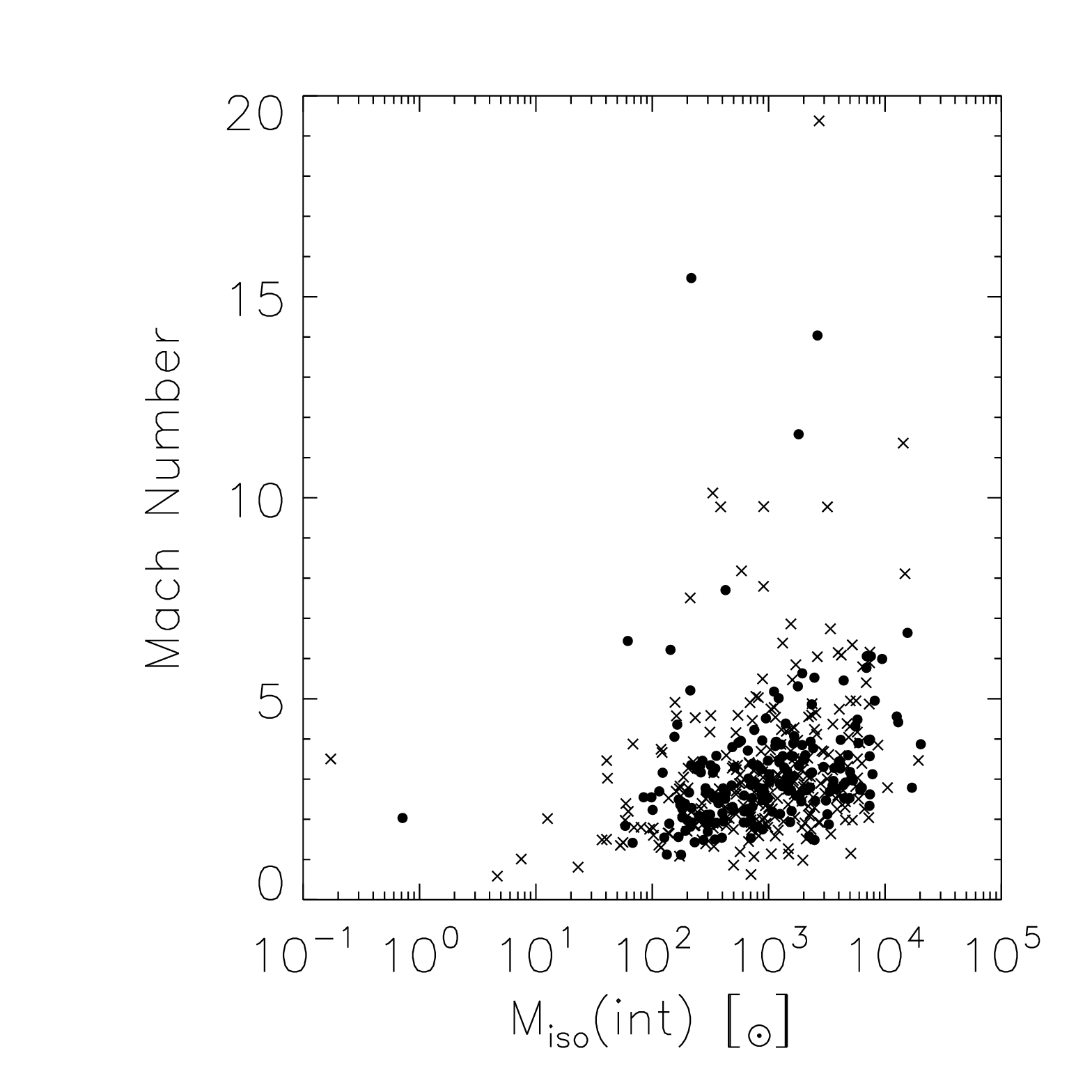}
\figcaption{\label{sigmant}
Left:  The non-thermal line width versus \tk\ for the full sample.  Black circles mark the \tk\ subsample and the arrows mark sources with only an upper limit for \tk\ due to a non-detection in the (2,2) line.  The black line denotes the thermal sound speed given by $a=(k\tk/\mu m_H)^{1/2}$.  The non-thermal component is, on average, three times larger than the thermal sound speed.  Right:  Mach number ($\sigma_{Vlsr}/\sqrt{k\tk/\mu m}$) versus \miso(int).  Here, crosses mark sources without a (2,2) detection.
}
\end{center}
\end{figure*}

\subsection{Sound Speed and Non-thermal Motions}\label{sigmanonthermal}
The observed line widths provide a measure of the internal motions
within each BGPS source.  As discussed in Section \ref{results},
we observe velocity dispersions of up to 4 \kms, with 98\%\ of sources
having $\sigma_{Vlsr} \le 2.0$ \kms, which is similar to
line widths observed toward other samples of massive star-forming regions
(e.g.~$\Delta V(\rm FWHM)\sim1-4$ \kms; Anglada et al.~1996) but
an order of magnitude larger than seen in \ammonia\ observations of 
cores in nearby, low-mass star-forming regions (e.g.~Rosolowsky et al.~2008).

The observed velocity dispersion is a result of both thermal and 
non-thermal motions of the gas.  Given the derived \tk\ we can
calculate the thermal contribution to the line width and remove it
to determine the non-thermal velocity dispersion:
\begin{equation}
\sigma_{\rm NT} = \sqrt{\sigma_{Vlsr}^2-\frac{k\tk}{17 m_H}},
\label{sigmanonthermal}
\end{equation}
where $(k\tk/17 m_H)^{1/2}$ is the thermal broadening due to \ammonia,
and $m_H$ is the mass of a single hydrogen atom.  The full sample
can be characterized by $\mean{\sigma_{\rm NT}} < 0.75\pm0.49$ \kms,
and the \tk\ subsample by $\mean{\sigma_{\rm NT}} = 0.81\pm0.53$ \kms.
The non-thermal velocity dispersions for the full sample are shown 
in Figure \ref{sigmant} versus \tk.  The \tk\ subsample is
shown as black circles, while sources without a (2,2) detection and
an upper limit in \tk\ are plotted as arrows marking the upper-limit in \tk.  
The solid black line 
depicts the thermal sound speed as a function of \tk\ given by 
$a=(k\tk/\mu m_H)^{1/2}$, where $\mu= 2.37$.  
For most sources, the non-thermal line width is significantly larger than 
the thermal component; we find a mean ratio of non-thermal to 
thermal sound speed of $3.2\pm1.8$ for the full sample and 
$3.2\pm1.7$ for the \tk\ subsample.  The sound speed and 
non-thermal velocity dispersion are listed in columns 5 and 6 of
Table \ref{gas_props}.

The right panel of Figure \ref{sigmant} plots the Mach number (given 
as $\sigma_{Vlsr}/a$) versus \miso(int).  The non-thermal
velocity dispersion weakly scales with mass.  As discussed in the previous
section, \miso\ increases with distance, and the increase in $\sigma_{\rm NT}$
seen with mass is likely the result of observing larger objects at
greater distances.  We find a mean Mach number of 3.23$\pm$1.87 for the
full sample, and 3.24$\pm$1.73 for the \tk\ subsample.

\subsection{Volume and Surface Densities}\label{densities}
We calculate the volume-averaged density based on \miso(int) and
$R$ as $n=3\miso(\rm int)/4 \pi \rm R^3$.  This quantity is listed
in Column 6 of Table \ref{derived_masses_densities}, and shown in Figure \ref{versusd}c.
The dashed line marks the volume-averaged density corresponding
to the limiting mass within a beam size shown in panel b and
described in Section \ref{masses}.  In contrast to the limiting 
mass, this measure of density is not a hard limit.  It is possible
to have lower and higher density BGPS sources.  For example, a BGPS
source with an extracted size equal to the beam size will have a 
higher density if the flux is greater than the 5$\sigma$ limit 
required for detection.  A large, extended BGPS source with an
extracted size much larger than the beam size can have a
density lower than this limit.  The observed flux is roughly linear
with source size while the volume is proportional to $R^3$, thereby
allowing a larger source to have a lower density (see Section 7.2 of
D10).  This trend can be seen in Figure \ref{versusd}c
where the volume-averaged density decreases with kinematic distance
while the extracted 1.1 mm source size increases.  The dotted lines
in this figure denote the typical range of clouds, clumps and cores
(clouds, $n=50-500$ \cmv; clumps, $n=10^3-10^4$ \cmv; cores, $n=10^4-10^5$
\cmv; Bergin \& Tafalla 2007).

The full sample has a minimum volume-averaged density of 120 \cmv, 
a maximum of $8.0\times10^5$ \cmv, and $\mean{\log(n/\cmv)}= 3.12\pm0.56$.  
The \tk\ subsample has $\mean{\log(n/\cmv)}=3.27\pm0.58$.
Similarly, we calculate the average surface density as 
$\Sigma = \miso(int)/\pi R^2$.  This quantity is listed in 
column 8 of Table \ref{derived_masses_densities}.  
The distribution of surface density
for the full sample ranges from 0.006 g \cmc\ to 0.24 g \cmc\ with 
$\mean{\Sigma}=0.037\pm0.030$ g \cmc.
The full statistical characteristics of each sample are given in Table
\ref{stats}.  

\begin{figure}
\begin{center}
\plotone{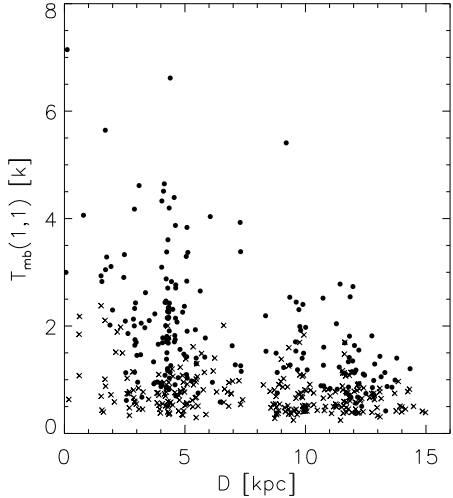}
\figcaption{\label{texd}
Observed \tmb(1,1) versus kinematic distance for the full sample.  Circles mark the \tk\ subsample and crosses mark sources without a (2,2) detection.  The decrease in observed \tmb(1,1) with distance is indicative of beam dilution.
}
\end{center}
\end{figure}

\subsection{Excitation Temperature and Density}\label{excitation}
The observed \tmb(1,1) are shown versus kinematic
distance in Figure \ref{texd}, where circles mark the \tk\ 
subsample and crosses mark sources without a (2,2) detection.
There is a clear decrease in \tmb(1,1) with distance suggesting
that distant sources suffer from beam dilution.  Such beam dilution
causes \tex\ to be underestimated, which in turn underestimates 
the excitation density as discussed below.  Such beam dilution
will not affect the calculated \tk\ since it is based on
the ratio of (1,1) and (2,2) emission, which will suffer from the
same beam dilution.  

The measured excitation temperatures are listed in Table \ref{gas_props}.  
Sources not well fit by the assumed \ammonia\ model have arbitararily been set to have
$\tex=\tk$.  The full sample is characterized by 
$\mean{\tex} = 4.1\pm2.2$ K, and the \tk\ subset by 
$\mean{\tex}=4.5\pm2.1$ K.  The mean difference between the excitation
temperature and kinetic temperature is $\mean{\tk-\tex}=11.5\pm5.1$ K
for the full sample.  This difference can be explained by a combination
of the low volume-averaged densities detected and the effects of beam 
dilution.  The densities are low enough that
the \ammonia\ transitions are not thermalized, as expected for
the mean density we find (see D10 for more detail).  

We can calculate an excitation density, defined to be the density 
which produces the observed level populations described by \tex\ 
for the observed \tk, from \tex, \tk, and $\tau$.  
\begin{equation} n_{ex} = \frac{k(J(\tex)-J(T_{cmb}))}{h \nu_{(1,1)}(1-J(\tex)/J(\tk))}n_{crit}\beta, 
\label{nexeqn}
\end{equation}
where
\begin{equation}
J(T)=\frac{h \nu_{(1,1)}}{k(1-e^{-h\nu_{(1,1)}/kT})},
\label{jeqn}
\end{equation}
$T_{cmb} = 2.73$ K, the escape probability is given by $\beta = (1-
e^{-\tau})/\tau$, where we take $\tau = \tau_{(1,1)}\times 0.233$,
which is the maximum optical depth in a single hyperfine line of the
\ammonia(1,1) transition.  Since \tex\ will be lowered due to the
beam dilution in the more distant sources while \tk\ is unaffected,
the excitation density will be underestimated in sources suffering
from beam dilution.

\begin{figure}
\begin{center}
\plotone{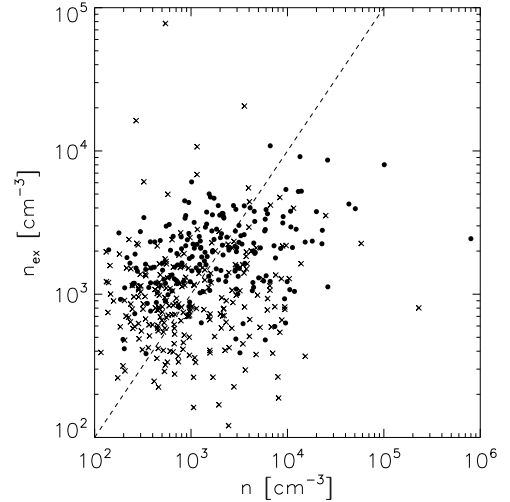}
\figcaption{\label{nexn}
Excitation density versus 1.1 mm volume-averaged density.  Black circles mark the \tk\ subsample while crosses mark sources with only an upper limit for \tk.  The dashed line denotes a one-to-one correlation and is not a fit to the data.
}
\end{center}
\end{figure}

The excitation densities are listed in Column 7 of Table 
\ref{derived_masses_densities}.  
We cannot calculate $n_{ex}$ for the sources where we set $\tex=\tk$.  
The full sample has $\mean{\rm log(n_{ex}/\cmv)}=3.10\pm0.35$.
The excitation density is plotted versus the volume-averaged 1.1 mm
density in Figure \ref{nexn} where black circles mark the \tk\ subsample
and crosses mark BGPS sources without a (2,2) detection.  The dashed line
marks a one-to-one correlation and is not a fit to the data.  
For the full sample, we find $\mean{\mbox{log}(n_{ex}/n)} = -0.02\pm0.58$
and a median of $10^{0.06}$, suggesting that both density estimates are reasonable.  

\begin{figure*}
\begin{center}
\epsscale{0.7}
\plotone{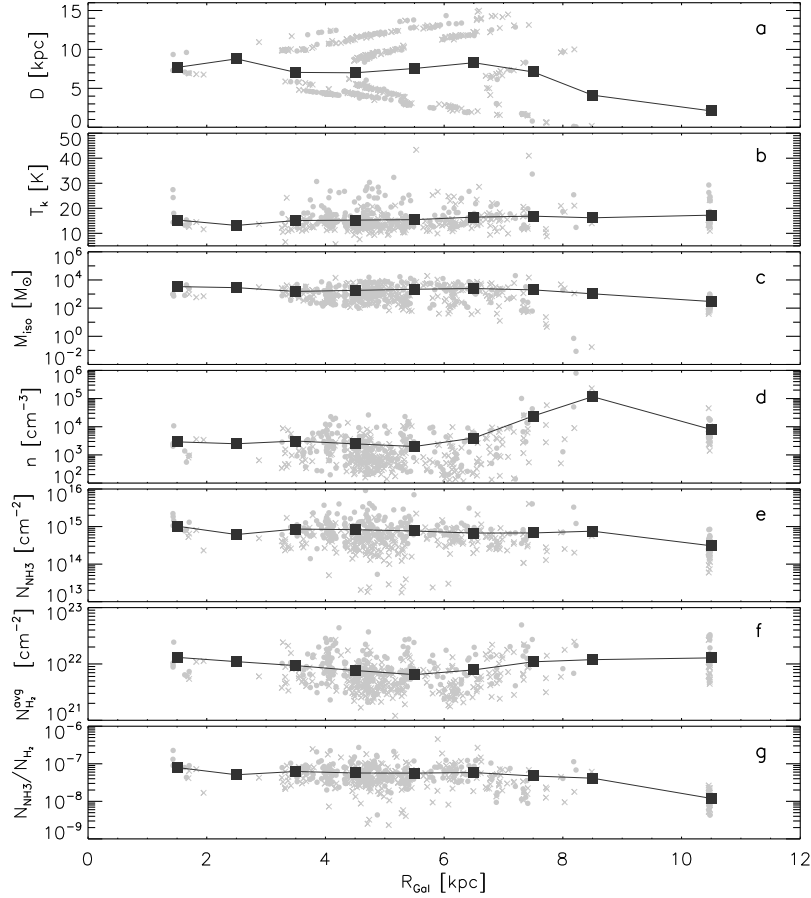}
\figcaption{\label{versusrgal}
Kinematic distance, \tk, \miso, n, $N_{\ammonia}$, $N_{H_2}^{avg}$, and \ammonia\ abundance versus Galactocentric radius.  Gray circles mark the \tk\ subsample, gray crosses denote sources without a (2,2) detection, black squares mark the mean of each property within 2 kpc wide bins.
}
\end{center}
\end{figure*} 

\subsection{Column Densities and \ammonia\ Abundance}\label{coldenabund}
We calculate two measures of the H$_2$ column density, the column density 
in a beam centered on the peak of the 1.1 mm emission ($N_{H_2}^{beam}$)
and the average column density of the entire 1.1 mm source 
($N_{H_2}^{avg}$).  The column density in a beam is calculated as
\begin{equation}
N_{H_2}^{beam} = \frac{\rm \Snu(40\as)}{\Omega(40\as) \mu_{H_2} m_{\rm H} \kappa_{\nu} \rm B_{\nu}(\td)},
\end{equation}
where \Snu(40\as) is the flux density in an aperture of diameter
40\as\ without the aperture correction of 1.46 required 
to correct for point source emission that falls outside of the 
aperture, $\Omega(40\as)=2.95\times10^{-8}$ sr is the solid angle 
of the 40\as\ aperture, $\mu_{H_2}=2.8$, $m_H$ is the mass of a single
hydrogen atom, $\kappa_{\nu}$ is the dust opacity described in
section \ref{masses}, and $B_{\nu}(\td)$ is the Planck function
evaluated at \td.  We use \Snu(40\as) as a measure of the average
1.1 mm surface brightness to measure $N_{H_2}^{beam}$
rather than the single peak 1.1 mm pixel in order to reduce the
affects of pixel-to-pixel variations caused by the noise in
the BGPS images.  The solid angle subtended by the 40\as\ aperture
is very close to the solid angle of the 33\as\ effective beam,
$\Omega(beam)=2.9\times10^{-8}$ sr.  We calculate 
$N_{H_2}^{avg}$ as 
\miso(int)/$\mu_{H_2} m_{H} \pi R^2$ and note
that this quantity is typically smaller than $N_{H_2}^{beam}$
since it is averaged over the entire area of the BGPS source while
$N_{H_2}^{beam}$ is calculated at the peak of the 1.1 mm 
emission.  
$N_{H_2}^{beam}$ and $N_{H_2}^{avg}$ are listed in 
columns 9 and 10 of Table \ref{derived_masses_densities}.

The average H$_2$ column densities, $N_{H_2}^{avg}$, are shown in 
Figure \ref{versusd}d versus
kinematic distance.  The gray circles denote the \tk\ subset, gray crosses
denote sources without a (2,2) detection, and the black squares and black
solid line denote the mean H$_2$ column densities in 2 kpc bins.  
The dashed line denotes the H$_2$ column density corresponding to the
mass per beam limit described in section \ref{masses}.  The dash-dot
line marks a threshold gas surface density of 122.5 \msun\ pc$^{-2}$
above which efficient star formation occurs
(116 \msun\ pc$^{-2}$, Lada et al.~2010; 129 \msun\ pc$^{-2}$, 
Heiderman et al.~2010), and the dotted line
denotes a surface density of 1 g \cmc\ required to prevent excessive 
fragmentation
and allow massive stars to form (Krumholz \& McKee 2008).  See section
\ref{massivestars} for further discussion of these limiting column densities.
The mean
H$_2$ column density decreases slightly with kinematic distance,
but is essentially flat.  The full sample is described by
$\mean{\mbox{log}(N_{H_2}^{avg}/\cmc)}=21.81\pm0.27$, and 180 
\msun\ pc$^{-2}$.  Using $R_{V}=3.1$ and $N_{H_2}^{avg}/A_{V} =
9.4\times10^{20}$ \cmc\ mag$^{-1}$ (Bohlin et al.~1978), the mean
$N_{H_2}^{avg}$ for the full sample corresponds to $8.5\pm6.8$ mag
of visual extinction.  Assuming a $R_{V}=5.5$ extinction model
(Weingartner \& Draine 2001), $N_{H_2}^{avg} /A_{V} =
6.86\times10^{20}$ \cmc\ mag$^{-1}$, the mean of the full sample
corresponds to $11.7\pm9.3$ mag.

We calculate the \ammonia\ column densities as described
in D10 and list them in column 11 of Table \ref{derived_masses_densities}.
Statistical properties are listed in Table \ref{stats}.

The \ammonia\ column densities are described by 
$\mean{\log(N_{NH_3}/\cmc)}=14.75\pm0.37$ and
$\mean{\log(N_{NH_3}/\cmc)}=14.95\pm0.29$ for the
full sample and \tk\ subsample, respectively.  We additionally find
a mean \ammonia\ abundance of 
$\mean{\mbox{log}(N_{NH_3}/N_{H_2}^{beam})}=-7.34\pm0.03$ 
for the full sample.  Our measured \ammonia\ abundance is consistent with
other studies:  $3\times10^{-8}$ (Harju et al.~1993),
$2.8\times10^{-8}$ (Tafalla et al.~2006), and $2\times10^{-8}$ 
(Foster et al.~2009).

\begin{figure*}
\begin{center}
\epsscale{0.9}
\plotone{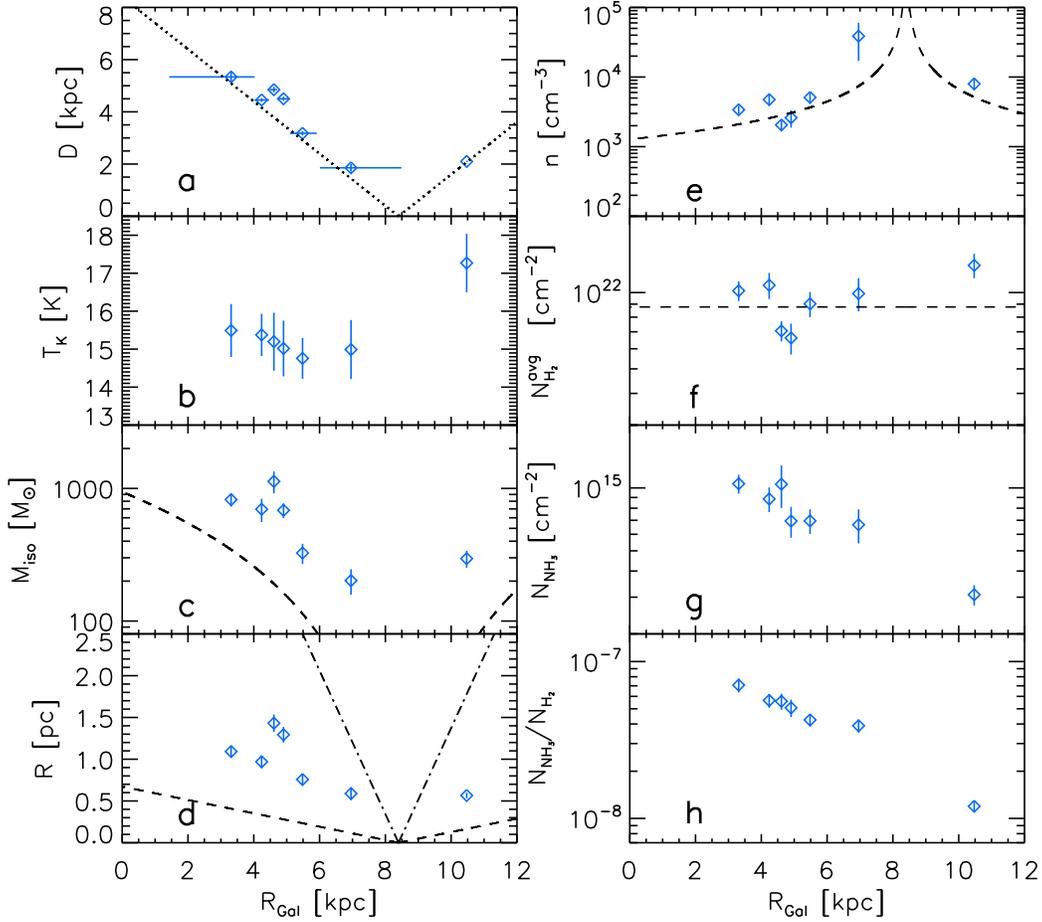}
\figcaption{\label{meanversusrgalnear} \footnotesize
Mean properties of near BGPS sources versus Galactocentric radius for (a) kinematic distance, (b) \tk, (c) \miso, (d) $R$, (e) $n$, (f) $N_{H_2}^{avg}$, (g)$N_{NH_3}$, and (h) $N_{NH_3}/N_{H_2}^{beam}$.  Each point represents the mean over 39 BGPS sources, except for the last two points, which represent 38 and 33 sources, respectively.  The points are centered at the mean \rgal\ of 3.3, 4.2, 4.6, 4.9, 5.5, 7.0, and 10.5 kpc.  The blue diamonds and vertical lines denote the mean and standard error in the mean (given by $\sigma/\sqrt N$ where $\sigma$ is the standard deviation and $N$ is the number of sources per bin) for each property.  The blue horizontal lines in panel (a) represent the range of \rgal\ included in each point.  The dotted line in panel (a) represents the distance to the nearest source at a given \rgal, which correspnds to the $\ell=0\degree$ line of sight.  The dashed lines represent:  (c) the 50$\sigma$ mass per beam, (d) beam size, which represents a lower limit on extracted source size, and the scale above which we filter out uniform emission representing an upper limit to extracted source size (5.9\am, dash-dot line), (e) the 5$\sigma$ volume-averaged density per beam, and (g) the 50$\sigma$ average column density per beam.
}
\end{center}
\end{figure*}

\section{Galactic Trends in Physical Properties}\label{galactictrends}
As discussed in section \ref{distances}, each BGPS source has a unique
\rgal\ given by the observed radial velocity and assumed Galactic
rotation curve, allowing us to study
the trends in physical properties as a function of distance
from the Galactic center.  In this section we include the Gemini
OB1 molecular cloud (Gem OB1) sources from D10 as an outer Galaxy data point.
Since the BGPS did not perform a blind survey of
the outer Galaxy, but rather targeted known star-forming regions (Aguirre et al.~2011),  
the Gem OB1 sources may be biased in terms of evolutionary stage.
We discuss possible effects of this bias in Section \ref{biasesandimplications}.
We also note that the Gem OB1 sources may not be representative of all outer 
Galaxy star formation.

Figure \ref{versusrgal} plots the physical properties as a function of
Galactocentric radius for the \tk\ subsample (gray circles) and
the BGPS sources without a (2,2) detection (gray crosses).  The
black squares denote the mean properties within 2 kpc bins centered
at 1, 3, 5, 7, 9, and 11 kpc.  We observe BGPS sources
at both the near and far kinematic distances for each
Galactocentric radius.  The concentration
of BGPS sources in lines in Figure \ref{versusrgal}a is due to the
range of Galactic longitudes included in the full sample.  Each 
sideways V formation corresponds to one of the four ranges in 
Galactic longitude, where the vertex of the V created by the  
$\ell \sim 9\degree$ sources is at the smallest \rgal\ and the 
$\ell \sim 54\degree$ sources are found at \rgal$\sim 7$ kpc.

The observed dichotomy of kinematic distance as a function of 
Galactocentric radius causes a spread in physical properties 
such as \miso, n, and N$_{H_2}$ as a function of \rgal\ because 
of their dependence on the kinematic distance (see Figure 
\ref{versusd}).  This spread in properties will serve to mask 
any trends in properties as a funciton of \rgal.  In this
section, we 
consider the near subsample (233 BGPS sources placed at the near
kinematic distance) in order to remove the ambiguity introduced by the
kinematic distance dichotomy.   

The mean and standard error in the mean
are presented for each property  for the near subsample
in Figure \ref{meanversusrgalnear}.  The 233 BGPS sources were sorted
by \rgal\ and split evenly into six bins with 39 sources in the first
five bins and 38 sources in the sixth bin.  The 33
Gem OB1 sources are presented separately in the seventh bin.
This binning method provides more measurements of the mean 
properties where there are more BGPS sources.  The 7 bins
are centered at \rgal\ of 3.3, 4.2, 4.6, 4.9, 5.5, 7.0, and 10.5 kpc.
Bins 2$-$4 centered at 4.2, 4.6, and 4.9 kpc represent 
sources in the 5 kpc molecular ring.
The blue diamonds represent the mean properties within each bin 
and the vertical blue lines represent the standard error
in the mean, which is given by $\sigma/\sqrt N$ where $\sigma$ is the
standard deviation and $N$ is the number of sources in each bin.  

Figure \ref{meanversusrgalnear}a plots the mean kinematic distance
versus \rgal.  The horizontal blue lines depict the range of \rgal\
values included in each bin.  The dotted lines correspond to the distance to the 
nearest source at a given \rgal, which lies along $\ell=0\degree$ between
the sun and the Galactic center.   The near subsample 
decreases in kinematic distance as \rgal\ increases such that 
sources at higher \rgal\ are at smaller kinematic distances.
The kinematic distances are similar to the nearest source 
distance shown by the dotted lines allowing us to apply our
understanding of the biases as described in Sections \ref{analysis}
and \ref{whatisbgps} to the near subsample as a function of \rgal.

\subsection{Observed Trends}\label{observedtrends}
Here we highlight the observed trends in physical properties
with \rgal, and discuss the possible biases in the following section.

The sources within the molecular ring are located at larger kinematic
distances than expected given the smooth trend of decreasing kinematic
distance with increasing \rgal\ seen in the inner Galaxy sources located
outside of the molecular ring (see Figure \ref{meanversusrgalnear}a).
We will explore the possible effects of the larger kinematic distance toward
the molecular ring sources in the following section.

Figure \ref{meanversusrgalnear}b shows the kinetic gas temperature
derived from the \ammonia\ observations.  There is no real trend in
\tk\ of sources in the inner Galaxy, but there is a 3 K difference 
between the mean \tk\ of the inner Galaxy and Gem OB1 sources.

\miso\ is shown in Figure \ref{meanversusrgalnear}c.  The dashed lines
depict a 50$\sigma$ mass per beam at distances corresponding to the
dotted lines in panel (a).  \miso\ steadily decreases with \rgal\
with the trend roughly following the shape of the curve representing
a 50$\sigma$ mass per beam.  The 
three bins closest to the molecular ring at 4.2, 4.6, and 4.9 kpc
deviate from the general trend in decreasing \miso\ with \rgal\ and
are roughly a factor of a few more massive than described by the general
decreasing trend.  

Figure \ref{meanversusrgalnear}d displays the mean observed physical
radius in each bin.  The dashed lines mark the beam size at distances
corresponding to the dotted lines in panel (a), which represent a 
minimum source size, and the dash-dot lines mark the upper limit in
source size above which the data reduction filters out emission.
The mean radius decreases with \rgal, but two of the three bins representing
the molecular ring have radii approximately 1.5 times that
expected from the general decreasing trend.

The volume-averaged density is shown in Figure \ref{meanversusrgalnear}e
as a function of \rgal.  The dashed lines mark the 5$\sigma$ density
per beam required for a detection in the BGPS.  Two of the bins representing
the molecular ring fall below this line while the other points lie above.

Figure \ref{meanversusrgalnear}f
plots N$_{H_2}^{avg}$ as a function of \rgal.  The average
H$_2$ column density is roughly constant with \rgal, with the lowest
column densities seen in the molecular ring and the highest in
Gem OB1.  The dashed line denotes a 50$\sigma$ column density
per beam.  The observed \ammonia\ column density is shown in Figure 
\ref{meanversusrgalnear}g as a function of \rgal.  The column 
density decreases with \rgal\, and only one of the molecular ring
bins falls below the general trend.  Figure 
\ref{meanversusrgalnear}h plots the observed 
\ammonia\ abundance versus \rgal.  The abundance, given by 
$N_{NH_3}/N_{H_2}^{beam}$, decreases with \rgal\
and is approximately a factor of 7 higher in the inner Galaxy
than in the Gem OB1 sources.

\subsection{Biases and Implications}\label{biasesandimplications}
As mentioned above, the Gem OB1 sources belong to a molecular cloud known 
to be forming stars.  This could bias the physical properties thereby 
providing an inappropriate comparison point for the inner Galaxy sources.
The Gem OB1 sources differ dramatically from the inner Galaxy sources in
only two properties:  \tk\ and \ammonia\ abundance.  The warmer mean \tk\
observed in the Gem OB1 sources could be due to a bias toward later 
evolutionary stages, which typically have warmer dust temperatures
(Battersby et al.~2010).  Similarly, 
\tk\ of BGPS sources increases with the number of associated mid-IR sources
(see section \ref{starformationactivity}).
By targeting a region of known star formation,
we have potentially excluded the earliest evolutionary stages that have
colder temperatures, therefore resulting in a warmer mean \tk.  

Foster et al.~(2009) found no significant difference between the \ammonia\ 
abundance of starless and protostellar cores in Perseus, suggesting that
the potential bias of the Gem OB1 sources to later evolutionary states
is not likely to be the cause of the lower \ammonia\ abundance in Gem OB1 sources. 
The general decrease in \ammonia\ abundance (Figure \ref{versusrgal}h) could be the
result of changes in dust properties and/or metallicity with \rgal.  Changes in the
gas-to-dust ratio, grain size distribution, and temperature of dust grains
across the Galaxy could all result in the observed decrease in \ammonia\ abundance
with \rgal.  Similarly, a decrease with \rgal\ in the amount of nitrogen 
available to form \ammonia\ would result in a lower abundance at larger \rgal.
The decrease we observe mimics the decreasing nitrogen 
abundance seen as a function of \rgal\ (e.g.~Gummersbach et al.~1998;
Shaver et al.~1983; Smartt et al.~2001; Rolleston et al.~2000).
We fit a line to the \ammonia\ abundance as a function of \rgal\ and find
\begin{equation}
\label{abundtrend}
X_{\ammonia}=10^{-6.90\pm0.05}10^{-0.096\pm0.008\rgal}.
\end{equation}
We observe a gradient of -0.096 dex/kpc in the \ammonia\ abundance, which
agrees within the error bars with the observed nitrogen gradient ($-0.09\pm0.015$
dex/kpc; Shaver et al.~1983).  Thus, the \ammonia\ abundance decreases at least
as quickly as the nitrogen abundance but possibly faster due to possible changes in
dust properties to which our observations are not sensitive.  
This trend needs to be taken 
into account when calculating H$_2$ column densities and masses based on
observed \ammonia\ column densities.

The effects of distance must also be taken into account when intrepreting
trends with \rgal.  The Malmquist bias together with the spatial filtering 
in the BGPS data has resulted in the detection of different types of structures 
at different distances
(i.e.~cores, clumps, clouds; see section \ref{whatisbgps}), which complicates
the study of BGPS sources as a function of position in the Galaxy.  This 
complication, which will affect all continuum studies, needs to be addressed 
carefully.  

\begin{figure*}
\begin{center}
\epsscale{0.4}
\plotone{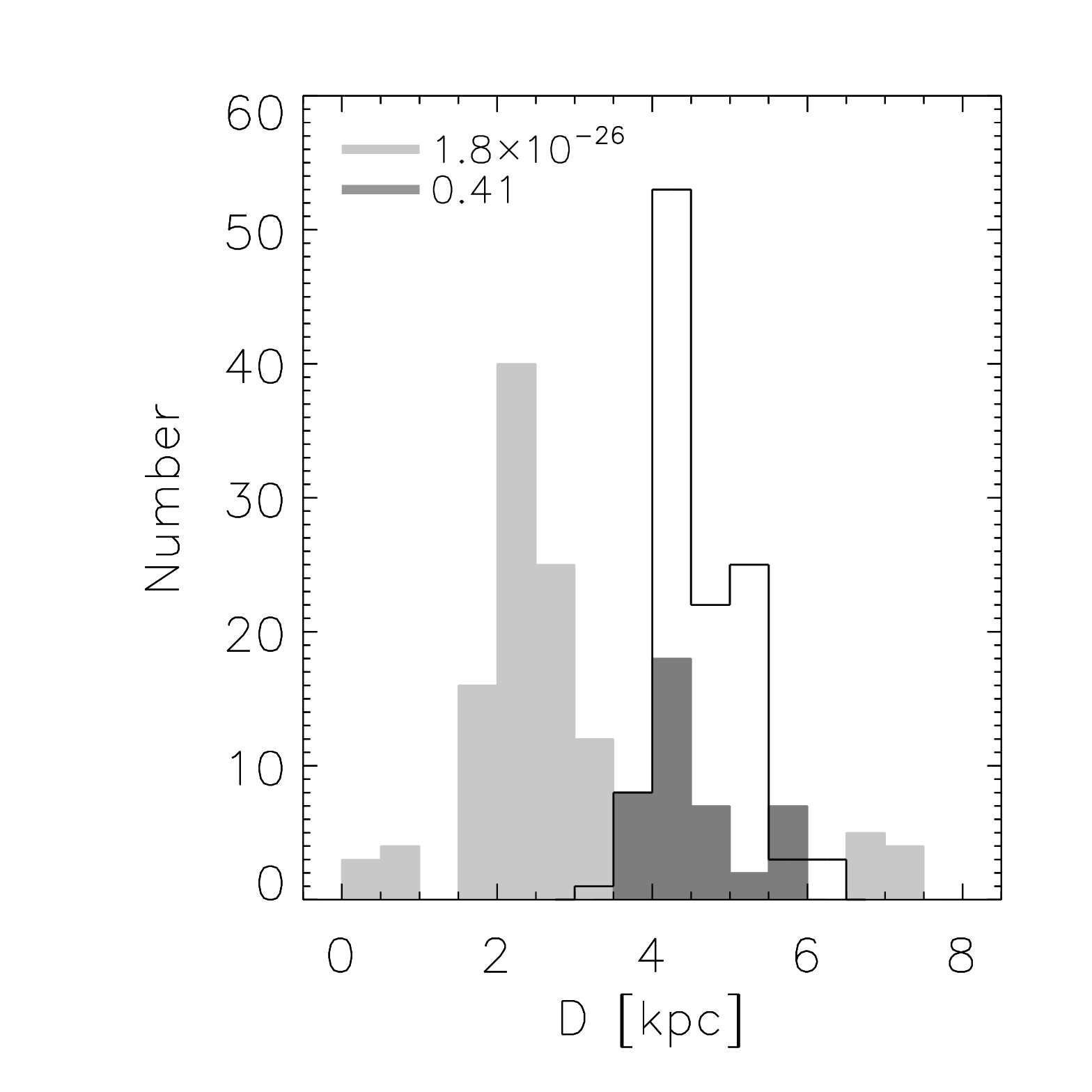}\\
\epsscale{0.84}
\plottwo{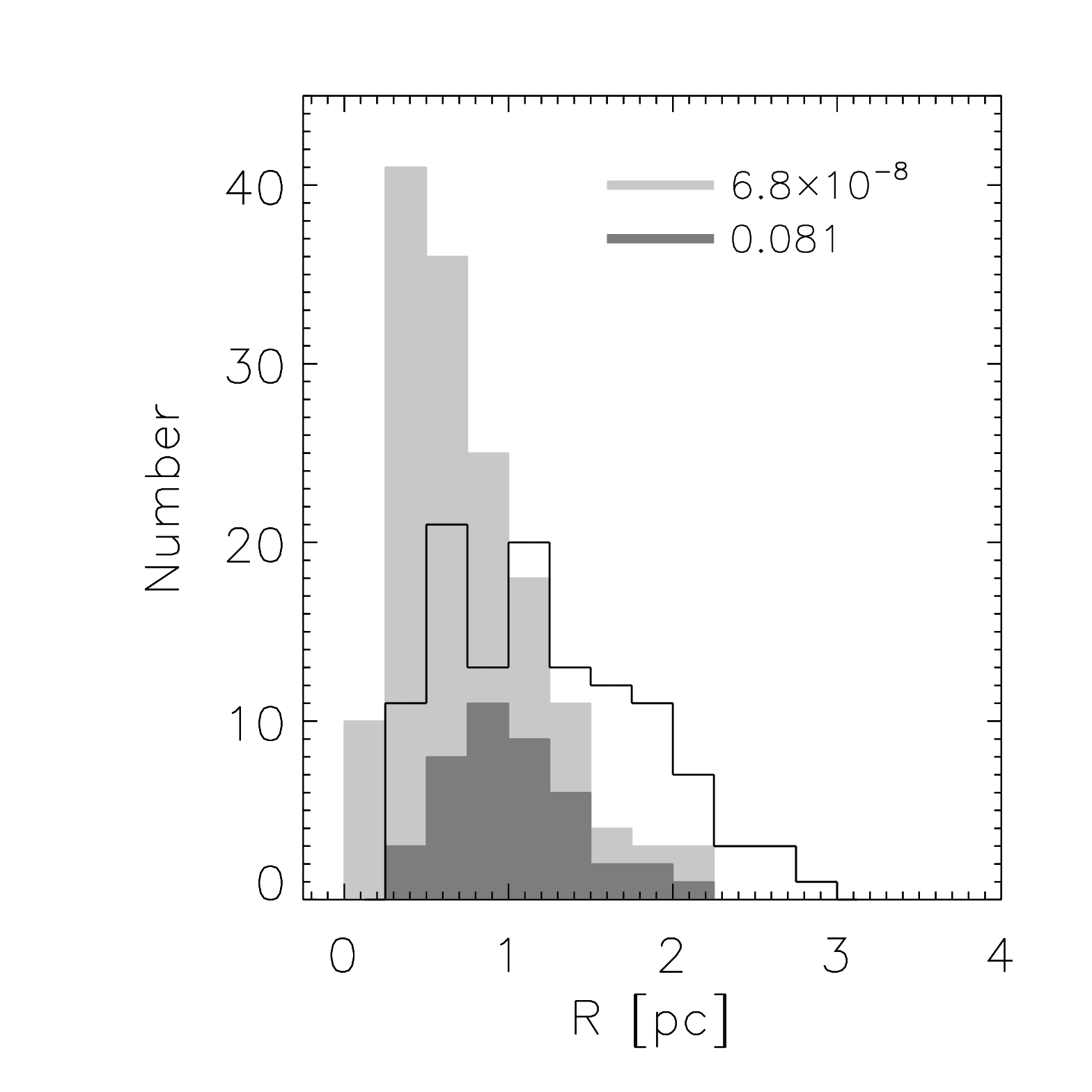}{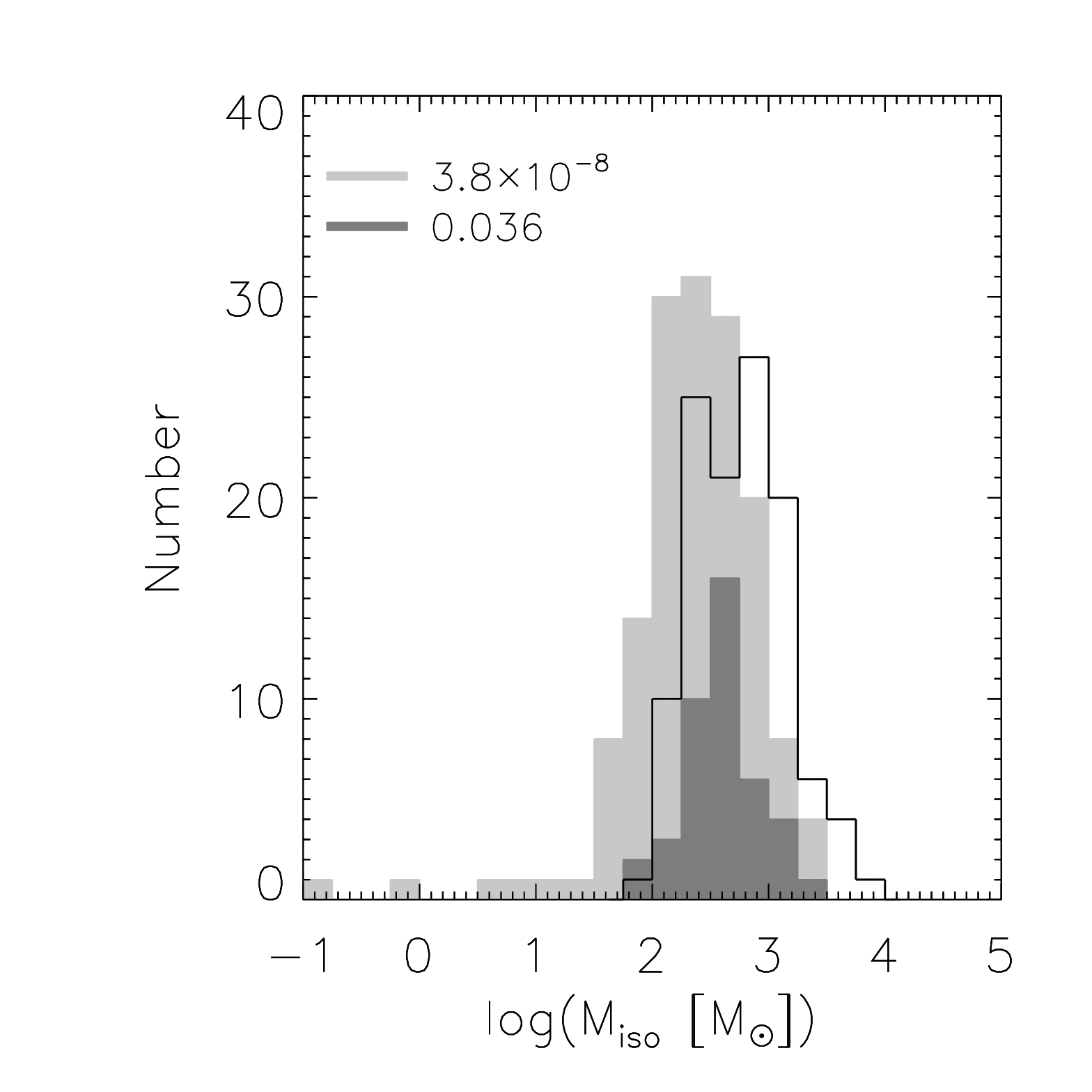}
\figcaption{\label{molecring_hists}
Distributions of kinematic distance (upper panel), physical radius (lower left panel), and \miso(int) (lower right panel) for the near kinematic distance sources located within the molecular ring (solid line), outside the molecular ring (light gray), and outside the molecular ring with 3.5 kpc$\leq$ D $\leq$ 6.5 kpc (dark gray).  The legend gives the KS probabilities comparing the full sample of sources inside and outside of the molecular ring (light gray) and for the kinematic distance selected samples of sources inside and outside of the molecular ring (dark gray).  
}
\end{center}
\end{figure*}

Both the mass and radius of BGPS sources exhibit
the same general trend of decreasing values with increasing \rgal\ with
peaks within the molecular ring.  For the near subsample, \rgal\ and 
$D$ are inversely proportional
so that large values of \rgal\ correspond to small kinematic distances
and small \rgal\ corresponds to large kinematic distances.  Therefore, 
the larger masses and radii seen at low \rgal\ are likely the result of
the Malmquist bias.  Indeed, the slope of the mean mass versus \rgal\ 
roughly follows the slope of the 50$\sigma$ mass per beam line suggesting
that trend is caused by distance effects.

The general trends of decreasing masses and radii with increasing \rgal\ are 
due to distance effects, but are the larger masses and radii seen toward 
molecular ring sources real?  The difference in mean masses and radii plotted in Figure 
\ref{meanversusrgalnear} suggest the molecular ring sources are truly larger and 
more massive, but the sources with \rgal$=4-5$ kpc are also located at larger 
distances than described by the general trend of decreasing kinematic distance
with \rgal\ (Figure \ref{meanversusrgalnear}a).
We explore the possible effects of distance by 
comparing properties of sources inside ($4\leq \rgal \leq 5$ kpc; 115 BGPS sources)
and outside the molecular ring (151 BGPS sources) using Kolomogorov-Smirnov (KS) tests.
The two-sided KS test compares the cumulative distribution functions of two
samples and measures the probability that the sources are drawn from the same
parent distribution, but high KS probabilities do not necessarily indicate the
same parent population.  Figure \ref{molecring_hists} plots the distributions
of kinematic distance (upper panel), 
physical radius (lower left panel), and \miso\ (lower right panel)
for sources within the molecular ring (solid line histogram) and
sources outside the molecular ring (light gray histogram).  
The KS probability for the kinematic distance ($10^{-26}$) shows that we
are not comparing similar distance distributions by splitting the sample
into sources found inside the molecular ring and those found outside the
molecular ring.  The sources located outside the molecular ring span a large range of 
kinematic distances, as seen in the upper panel of Figure \ref{molecring_hists}.
We have attempted to further untangle the effects of distance by considering only the 
sources inside and outside the molecular ring
within the same range of kinematic distances, specifically $3.5 \leq D \leq 6.5$ kpc.  
This distance range includes 114 of the 115 sources inside the molecular ring,
and 42 of the 151 sources outside the molecular ring (dark gray histogram
in Figure \ref{molecring_hists}).
We performed similar KS tests on these kinematic distance selected samples
and found KS probabilities of 0.41, 0.081 and 0.036
for D, R, and \miso, respectively.  The KS probabilities
have increased compared to the non-kinematic distance selected sample
and no longer suggest measurable differences for any of these properties.

Therefore, the larger mean masses and radii of sources within the 
molecular ring are likely caused by distance effects.  When we compare sources at
similar kinematic distances, we see that there is little variation in physical
properties regardless of environment (inside versus outside the molecular ring, 
for example).
This is in contrast to the molecular clouds seen in the $^{13}$CO
GRS survey.  The molecular clouds within the molecular ring typically 
have warmer temperatures, larger areas, larger masses, higher peak H$_2$ column densities, 
and typically contain more clumps than molecular clouds located outside the molecular ring 
(Rathborne et al.~2009).  This suggests that environment may affect the large scale 
molecular clouds while the regions of dense gas ($n\sim10^3$ \cmv) within the molecular
clouds exhibit similar properties regardless of environment.

\section{Discussion}\label{discussion}
\subsection{Comparison to Other Surveys}\label{othersurveys}
Schlingman et al.~(2011) conducted a survey of BGPS sources in
HCO$^+$(3-2) and N$_2$H$^+$(3-2) and were able to break the kinematic
distance ambiguity for 648 sources that they placed at the near distance.
They find a median radius of 0.75 pc, median isothermal mass assuming \td$=20$ 
K of 330 \msun, and a median volume-averaged density of $2.5\times10^3$ \cmv.
For the 233 BGPS sources placed at the near kinematic distance in our 
sample, we find a median
radius of 0.95 pc, median mass of 350 \msun, and a median volume-averaged density
of $2.2\times10^3$ \cmv, very similar to the properties of the Schlingman et 
al.~sample.  In contrast to our sample, where we find a median virial parameter of 0.74, 
Schlingman et al.~find typical virial parameters of a few and only a small number of 
sources have virial parameters less than one.  This difference is likely due to 
the larger line widths detected in HCO$^+$ (median FWHM$=2.98$ \kms) compared to 
the line widths detected in \ammonia\ (median FWHM $=0.64$ \kms\ $\sqrt{8 \mbox{ln}2}=1.5$ 
\kms).  

The APEX Telescope Large Area Survey of the Galaxy (ATLASGAL; Schuller et al.~2009)
has surveyed the Galactic plane at 870 \um.  Early results from this survey and
\ammonia\ follow-up observations reveal similar physical properties as those presented in 
this work.  For 24 ATLASGAL sources near $\ell\sim19\degree$ that were detected
in \ammonia, Schuller et al.~find \tk\ ranging from 14 to 33 K, masses ranging from
roughly $10^2$ to $10^4$ \msun, and H$_2$ column densities from $7\times10^{21}$ \cmc\
to $1.3\times10^{22}$ \cmc.  The range of properties of ATLASGAL sources are consistent
with the range of properties of BGPS sources presented here.

Compared to the GRS clouds studied by Roman-Duval et al.~(2009),
the BGPS sources are smaller (BGPS $R=0.01 - 8.5$ pc; GRS $R=0.1 - 40$ pc),
less massive (BGPS \miso$=0.1 - 2.4\times10^4$ \msun; GRS $M=10 - 10^6$ \msun),
higher density (BGPS \mean{n}$=5\times10^3$ \cmv; GRS \mean{n}$=230$ \cmv),
and have higher surface densities (BGPS \mean{\Sigma}$=180$ \msun\ pc$^{-2}$;
GRS \mean{\Sigma}$=144$ \msun\ pc$^{-2}$).  This is consistent with the paradigm
that the BGPS sources are typically the smaller, denser regions located within
molecular clouds.  

\begin{figure}
\begin{center}
\plotone{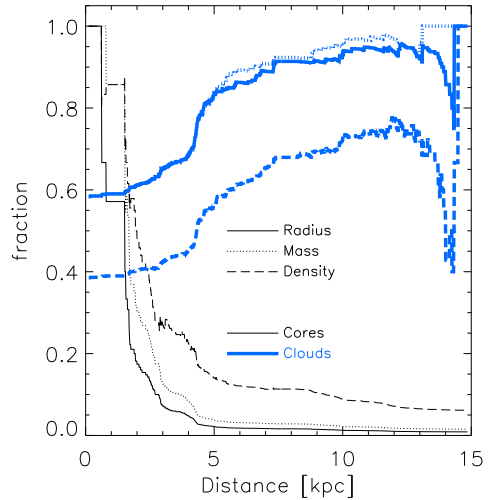}
\figcaption{\label{ccclimits}
Fraction of sources in the full sample of 456 BGPS sources detected in \ammonia\ nearer that are cores (thin black lines) and farther that are clouds (thick blue lines) as a function of kinematic distance determined by radius (solid lines), mass (dotted lines), and volume-averaged density (dashed lines).  Both ends of the distance range suffer from small number statistics. See section \ref{whatisbgps} for further details.
}
\end{center}
\end{figure}

\subsection{What is a BGPS Source?}\label{whatisbgps}
D10 addressed the question ``what is a BGPS source''
in terms of expected masses and densities
as a function of distance based on sources at a single, well-known 
distance.  With a large sample spanning a large range of kinematic 
distances, we can address what type of source the BGPS is detecting 
at each distance empirically.  
As discussed throughout section \ref{analysis}, BGPS sources can be 
classified as clouds, clumps, or cores based on their size, mass, and 
volume-averaged density.  While there can be significant overlap in
the physical properties of clouds, clumps, and cores, they can be
approximately separated by the dotted lines shown in Figure 
\ref{versusd}(a$-$c).  

Figure \ref{ccclimits}
shows the fraction of sources closer than a given distance that
are described as cores (thin black lines; $R\le 0.125$ pc, 
\miso\ $\le 27.5$ \msun, and $n\ge 10^4$ \cmv) based on radius 
(solid), mass (dotted), and volume-averaged density
(dashed).  We find that greater than 90\%\ of the BGPS
sources out to a distance of 0.68 kpc are well-described as cores.  
However, we suffer from small number statistics at distances
less than 2 kpc (see Figure \ref{dhelio}), and the distance out to 
which we detect cores could be greater.  Since no sources are detected 
between 1 and 
approximately 1.75 kpc, the fraction of BGPS sources that are cores 
can not be characterized in this distance range.  Thus, this result is 
consistent with D10, which determined that cores are 
detected out to approximately 2 kpc.  

\begin{figure*}
\begin{center}
\epsscale{0.9}
\plottwo{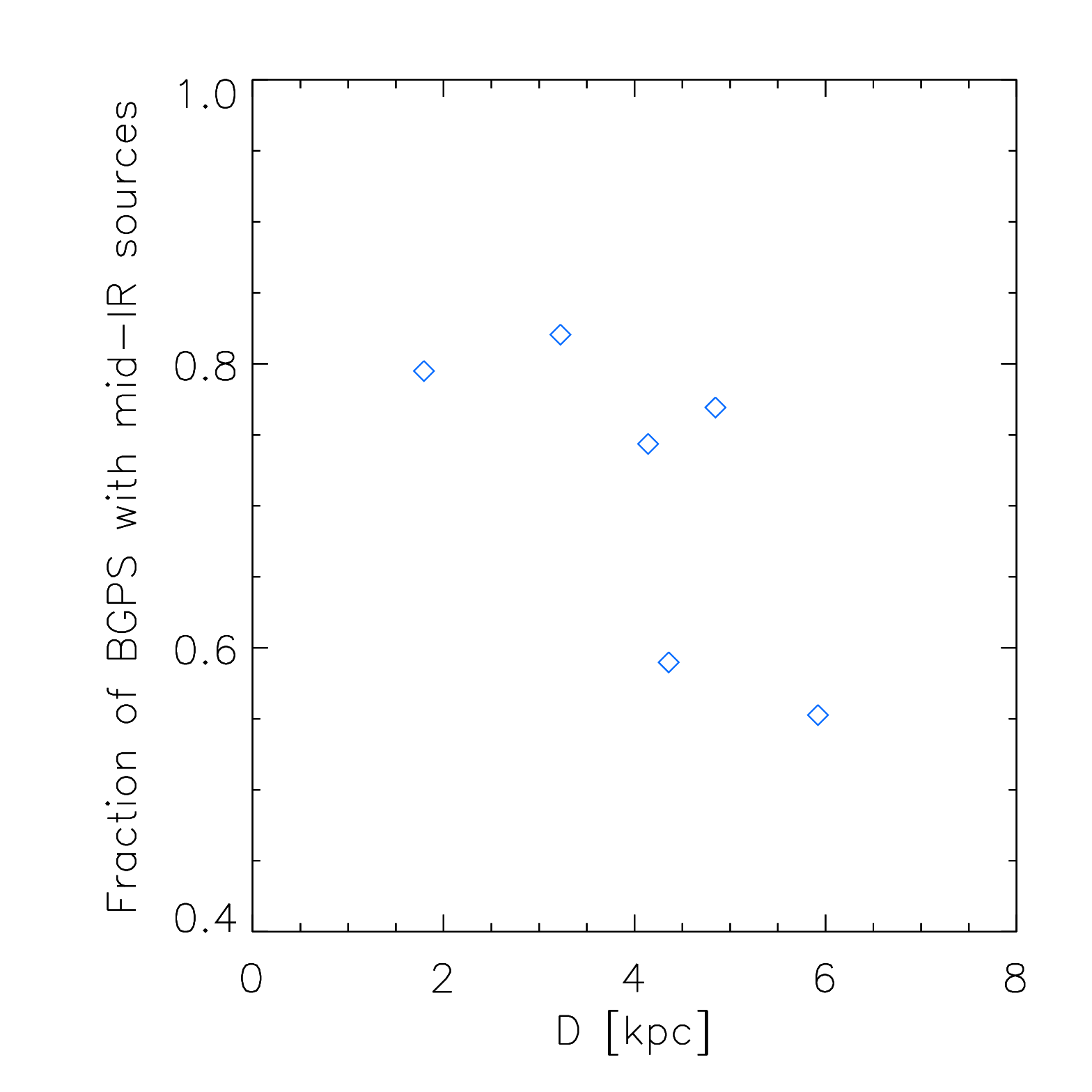}{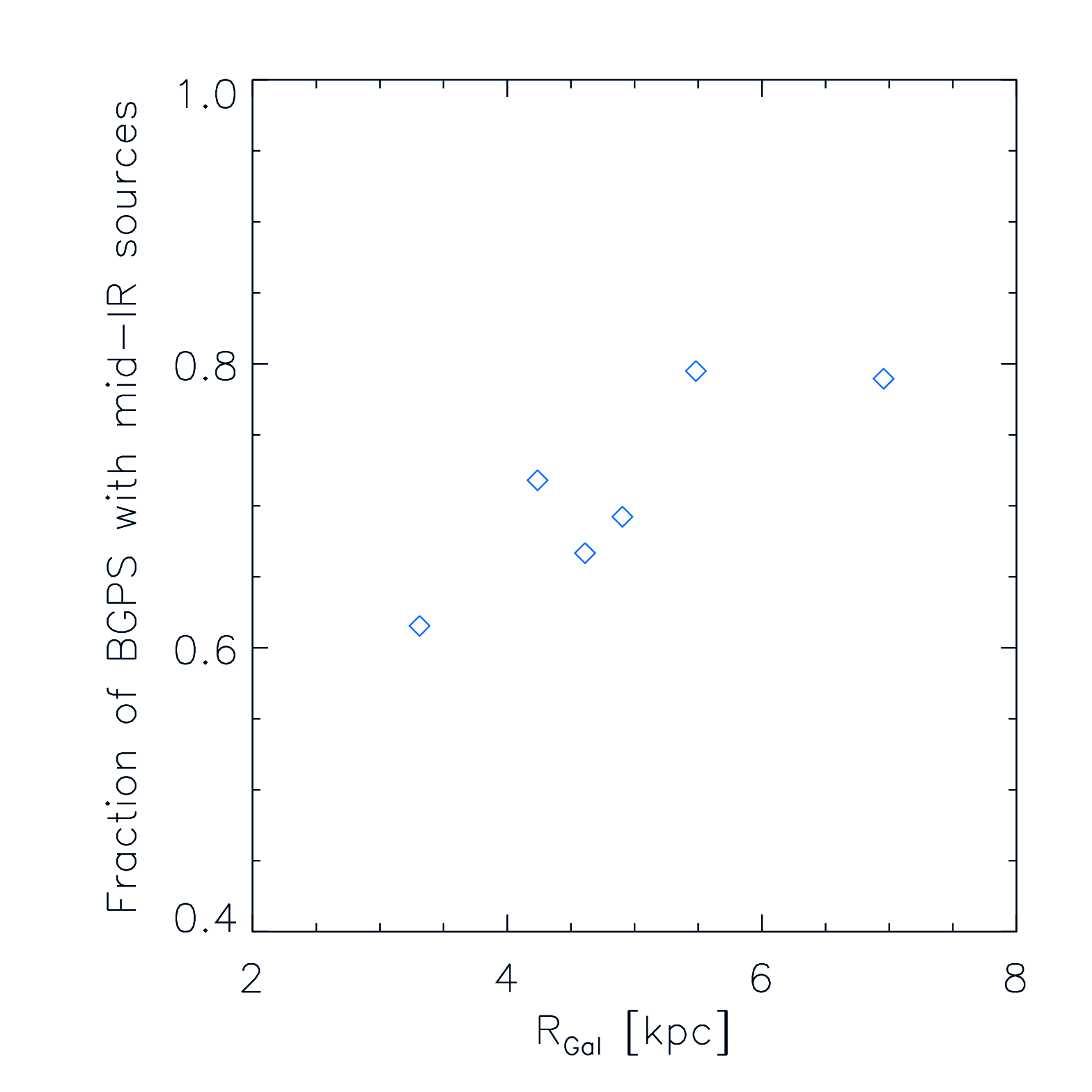}
\figcaption{\label{sffracvsd}
Fraction of BGPS sources associated with at least one mid-IR source (Dunham et al.~2011) as a function of kinematic distance (left) and \rgal\ (right) for the near subsample of 233 BGPS sources placed at the near kinematic distance.  Each diamond represents includes 39 BGPS sources, except for the bin at \rgal$\sim$7 kpc, which represents 38 BGPS sources.  
}
\end{center}
\end{figure*}

Similarly, we determine the fraction of sources located farther than 
a given distance that are characterized as clouds ($R\ge 1.25$ pc, 
\miso\ $\ge 750$ \msun, and $n\le 750$ \cmv).  These fractions are
shown in Figure \ref{ccclimits} as the thick blue lines.  The 
fraction of sources with volume-averaged densities below the limit
for clouds never reaches 0.9, but does follow the general trends seen
in the fractions based on radius and mass.  The density curve never reaches
90 percent because it is suffering from effects of volume-averaging.  
Thus, we do not consider the volume-averaged density when determining
the distance at which we detect $\ge$90\%\ clouds.  We find 7.3 kpc
based on the fraction with radii $\ge 1.25$ kpc, and 6.6 kpc from
the fraction with \miso(int) $\ge 750$ \msun.  We take the average of
these two distances, 7.0 kpc, as the distance above which the BGPS detects 
$\ge$90\%\ clouds.

At distances up to approximately 1 kpc, the BGPS detects cores.  
At distances greater than approximately 7 kpc, the BGPS detects
clouds.  At intermediate distances, the BGPS will detect predominately
clumps, but also cores and clouds.  While the high column 
density features seen in the BGPS do identify regions of dense gas, 
they represent very different scales of structures when looking 
through the Galaxy.  Similar continuum based studies (e.g.~ATLASGAL, 
Schuller et al.~2009; Hi-Gal, Molinari et al.~2010) will also be
subject to these effects.  Kinematic data is crucial in 
interpreting the results of such studies and determining what
structures are actually observed.

\subsection{Star Formation Activity}\label{starformationactivity}
Dunham et al.~(2011) characterized the star formation activity
of the BGPS sources based on the presence of mid-infrared (mid-IR)
sources along a line of sight coincident with each BGPS source.
They placed each BGPS source in one of four groups representing
increasing confidence in the type of associated mid-IR source reliably
indicating star formation activity (as opposed to a chance alignment).  
Group 3 represents the highest confidence in star formation activity
and includes BGPS sources matched with an \textit{Extended
Green Object} (EGO; Cyganowski et al.~2008) or Red MSX Survey (RMS; 
Hoare et al.~2004; Urquhart et al.~2008) source.
Groups 1 and 2 include BGPS sources matched with GLIMPSE
red sources cataloged by Robitaille et al.~(2008) or a deeper 
GLIMPSE red source catalog
created by Dunham et al.~(2011).  Group 0 includes BGPS 
sources that were not matched with any mid-IR sources, and we 
refer to them as ``starless'' although they are not necessarily truly
starless (see Dunham et al.~2011 for more details).  

Dunham et al.~(2011) find that 49\% of BGPS sources in the inner 
Galaxy where all surveys overlap contain at least one mid-IR source. 
We find that 67\% of the full sample contains at least one 
mid-IR source, while 33\% are ``starless.''  68\% of the BGPS sources
without a detection in the \ammonia(1,1) line were ``starless,''
and groups 1 through 3 comprise roughly 21\%, 9\%, and 3\%, respectively.
The BGPS sources without \ammonia\ or any corresponding mid-IR 
sources may be false detections in the BGPS catalog. 

Figure \ref{sffracvsd} plots the fraction of BGPS sources in the
near subsample that contain at least one mid-IR source as a function 
of kinematic distance (left panel).  The first five data points represent 39 BGPS sources, 
and the sixth represents 38.  By placing the same number of BGPS sources
in each bin, we obtain more data points at distances where more BGPS
sources are located.
The fraction peaks near 85\% at a distance
of approximately 3.0 kpc, and reaches a minimum of 55\% at 6 kpc.
The decrease in fraction of BGPS sources with mid-IR sources is
likely a result of the increasing distances.  At the farthest distances
shown here, the BGPS is dectecting the large scale objects, clumps
or clouds, and the GLIMPSE and RMS surveys would only detect the 
most luminuous YSOs, which become increasingly rare with
increasing luminosity and mass.  Thus, the decreasing fraction 
we see with distance may simply be a reflection of a decreasing
number of detected YSOs in the mid-IR catalogs.
The fraction of BGPS sources with mid-IR sources also reaches a minimum at 
kinematic distances of roughly 4.5 kpc.  Sources in this kinematic
distance bin correspond to sources within the 5 kpc molecular ring, and we 
further discuss the drop in fraction with star formation activity
in the following section.

\begin{figure}
\begin{center}
\plotone{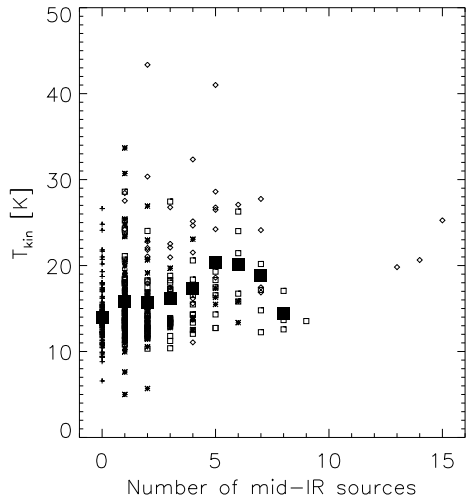}
\figcaption{\label{tkvsmidir}
\tk\ versus number of mid-IR sources associated with each BGPS source from Dunham et al.~(2011).  Plus signs are group 0 sources, asterisks are group 1 sources, open squares are group 2, diamonds are group 3, and the large, filled squares mark the mean \tk\ for each number of mid-IR sources.  The \mean{\tk} increases with number of mid-IR sources up to 6 mid-IR sources, above which we suffer from small number statistics.
}
\end{center}
\end{figure}

The additional gas properties provided by the \ammonia\ survey
presented in this work allow for further exploration of physical
properties as a function of the star formation activity groups
defined by Dunham et al.~(2011).   We find that the mean \tk, 
$\sigma_{Vlsr}$,
and $N_{\ammonia}$ increase with star formation activity 
group such that BGPS sources associated with an 
EGO or RMS source have higher gas kinetic temperatures,
larger observed line widths, and higher \ammonia\ column 
densities.  The mean \tk\ increases from 13.9 K
in group 0 to 22.7 K for group 3, while \mean{\sigma_{Vlsr}}
increases from 0.70 \kms\  to 1.10 \kms\
from group 0 to 3.  Similarly, \mean{N_{\ammonia}} increases
from $5.6\times10^{14}$ \cmc\ for group 0 to
$1.6\times10^{15}$ \cmc.  See Table \ref{sfactivitytrends}
for further statistical characterization of each property distribution.

We see a trend of increasing \mean{\tk} with increasing number
of associated mid-IR sources such that BGPS sources with a larger
number of mid-IR sources are typically warmer than BGPS sources
with fewer mid-IR sources (see Figure \ref{tkvsmidir}).  
There is considerable spread in \tk\ with number of mid-IR sources,
and the observed trend is a result of increase in the minimum
\tk\ detected with number of mid-IR sources.
The warmest temperatures in our sample are seen in BGPS sources 
across the entire range of number of mid-IR sources.  

We also explore the fraction of BGPS sources with star formation
activity as function of \rgal.  The right panel of Figure 
\ref{sffracvsd} plots the fraction of BGPS sources with mid-IR signs 
of star formation versus Galactocentric radius for the 233 sources 
at the near kinematic distances.  The binning is the same as in Figure \ref{meanversusrgalnear} 
excluding the sources in Gemini OB1 since the GLIMPSE catalogs
do not cover the outer Galaxy.  The fraction with 
mid-IR sources is lowest at the smallest \rgal\ and highest
at the largest \rgal\ and exhibits a 10\% decrease towards
the 5 kpc molecular ring.  

\begin{figure}
\begin{center}
\plotone{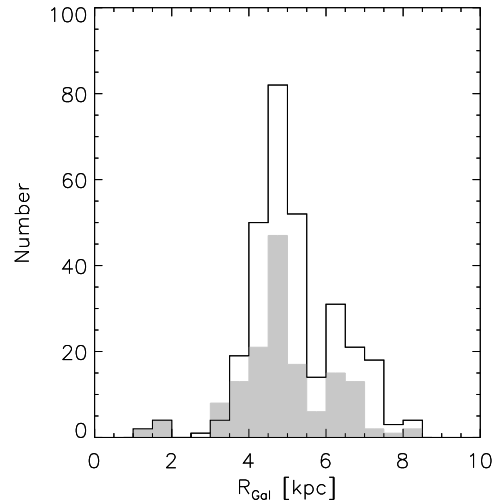}
\figcaption{\label{histsfactivity}
Distributions of BGPS sources with (solid line) and without (shaded gray) star formation activity as described in Dunham et al.~(2011).
}
\end{center}
\end{figure}

The decrease in the fraction of BGPS sources associated with a 
mid-IR source seen in the molecular ring (Figure \ref{sffracvsd})  
is not indicative of less star formation activity within the 
ring.  We see a clear peak in sources toward R$_{\rm Gal}= 4-5$ kpc
(Figure \ref{histsfactivity}) in sources that are associated 
as well as those that are not associated with a mid-IR source.  
Although the gas and 
dust needed to form stars is more abundant in the molecular
ring, the fraction of BGPS sources with mid-IR sources is lower.
The decrease in the fraction of BGPS sources with mid-IR sources 
could be a result of a few different effects. The first possibility 
is the increased mean kinematic distances to the sources within the molecular ring.  
The limiting luminosity (and mass) of the mid-IR catalogs considered by
Dunham et al.~(2011) increases with distance such that only the most luminuous
(highest mass) YSOs will be detected towards the most distant BGPS
sources.  The overall increase in the fraction of BGPS sources with a mid-IR 
source as a function of \rgal\ is therefore a result of the decreasing kinematic distance
with increasing \rgal\ and the rarity of high-mass stars and YSOs.
Similarly, since the BGPS sources within the molecular ring are located at
larger kinematic distances, the lower
fraction of mid-IR sources could simply be a result of the higher mass detection limit
of the mid-IR catalogs at the larger kinematic distances.

A second possible cause of the lower fraction of sources with mid-IR counterparts
is the abundant dense gas within the molecular ring (e.g.~Scoville \& Solomon 1975;
Burton et al.~1975; Clemens et al.~1988; Kolpak et al.~2002; Rathborne et al.~2009;
Roman-Duval et al.~2010), which would provide higher
extinction toward the mid-IR sources in this region.  Greater extinction would
effectively raise the mass detection limit of the YSOs at similar distances.  
This could reduce the fraction of BGPS sources with a mid-IR counterpart 
simply because the most massive YSOs are more rare.  The third possibility
is the increased background of PAH emission toward the longitudes corresponding
to the molecular ring.  Point sources are easily lost in the increased background 
8\um\ emission, and are therefore unlikely to be included in the point source 
catalogs.  Robitaille et al.~(2008) employed a flux cut to avoid this issue 
and ensure the reliability and completeness of the GLIMPSE Red Source Catalog.
However, the mid-IR sources considered by D11 are dominated by an additional
GLIMPSE catalog created expressly for that work, which did not include a flux cut
and will certainly be sensitive to the background emission.  It is likely that
the reduced fraction of BGPS sources that coincide with a mid-IR source
is the combined result of all three of the effects discussed here.

\subsection{Comparisons to Criteria for Efficient or Massive Star Formation}\label{massivestars}

\begin{figure}
\begin{center}
\epsscale{1.25}
\plotone{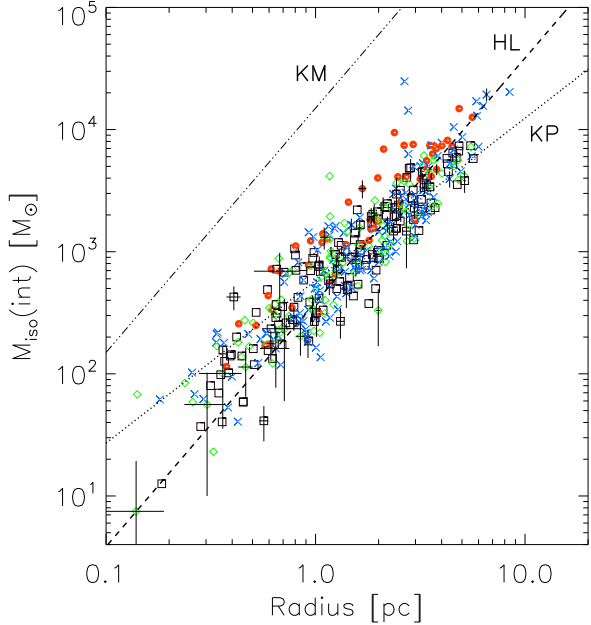}
\figcaption{\label{massradiussf} \miso(int) versus radius for the full sample of 456 BGPS sources.  Symbols and colors denote the star formation activity within each BGPS source as defined by D11.  Black squares represent group 0 sources (no corresponding mid-IR sources), blue crosses mark group 1 sources, green diamonds denote group 2 sources, and red circles depict group 3 sources (young RMS sources and EGOs).  Error bars are shown for every tenth BGPS source.  The dash-dotted line marks the Krumholz \&\ McKee (2008) surface density threshold, the dotted line marks the Kauffmann \&\ Pillai (2010) relation, and the dashed line denotes the Heiderman et al.~(2010) and Lada et al.~(2010) threshold for efficient star formation.
}
\end{center}
\end{figure}

In this section, we compare various criteria for efficient or massive
star formation with our BGPS sample.  We plot mass (\miso(int), Equation \ref{miso})
versus radius ($R$, Equation \ref{radiuseqn}) for the full sample of BGPS sources
with distance estimates in Figure \ref{massradiussf}. Error bars are plotted for
every tenth data point to avoid clutter. Points are coded by shape
and color according to the four groups discussed in Section \ref{starformationactivity}.
Note that uncertainties in distance will tend to create a correlation
of the form $M \propto r^2$.

Heiderman et al.~(2010) compared star formation rate surface surface densities
($\Sigma(SF)$) and gas surface densities ($\Sigma(gas)$)
for 20 nearby clouds that are forming primarily
low-mass stars and a group of more distant dense clumps that are forming
massive stars. They found that $\Sigma(SF)$ was a strong non-linear
function of $\Sigma(gas)$ up to about $129\pm14$ \msun\ pc$^{-2}$.
Above that ``threshold", $\Sigma(SF)$ was roughly linear with $\Sigma(gas)$.
Independently, Lada et al.~(2010) found that the star formation rate in
a sample of clouds was linearly proportional to the cloud mass above
$\Sigma(gas) \sim 116$ \msun\ pc$^{-2}$, while comparisons to total
cloud mass showed no regularity.

In Figure \ref{massradiussf}, a dotted line of $\Sigma(gas) = 122.5 \msun {\rm pc^{-2}}$, or
$N_{H_2}^{avg} \ge 6.49\times10^{21}$ \cmc,
represents an average of the criteria found by Heiderman et al.~(2010)
and Lada et al.~(2010). We refer to this as the HL criterion for
``efficient" star formation.
Of the 456 sources in the full sample, 211 (46.3\%) satisfy the HL criterion
when we use the surface density of the entire source.
However, since the BGPS sources have higher surface
density structures on smaller scales, BGPS sources that do not satisfy
the HL criterion overall may still have subregions that do satisfy it.
Thus, 46.3\%\ should be viewed as a lower limit. If, for example,
we consider the peak column density, $N_{H_2}^{beam}$,
86.8\%\ of the full sample of BGPS sources would have some part that
satisfies the HL criterion. Of the BGPS sources with the most certain
star formation (Group 3), 70.5\%\ (36 of 51) have average surface densities
above the HL criterion, and 98\%\ (50 of 51) have peak surface densities
above the HL criterion.

A dash-dotted line in Figure \ref{massradiussf} shows the criterion
to avoid excessive fragmentation of $\Sigma(gas) = 1$
g cm$^{-2}$ $ = 4787$ \msun\ pc$^{-2}$ advocated by
Krumholz \&\ McKee (2008), and presented earlier by McKee \&\ Tan (2003). 
None of the BGPS sources in our full sample come close to satisfying
this KM criterion. As with the HL criterion, parts of these sources
may satisfy this criterion. The surveys by Plume et al. (1992) toward
regions known to be forming massive stars found quite a few sources satisfying
this criterion, though Wu et al.~(2010) found that most exceeded the
KM criterion only as measured in the highest density tracers. Since
the Plume et al.~sources are certainly included in our BGPS survey,
they must form the densest parts of the BGPS sources.

Kauffmann et al.~(2010) and Kauffmann \& Pillai (2010) have recently
suggested a much less stringent criterion for massive star formation
than the KM criterion discussed above.
They argued that clouds known to be forming massive ($M_* \sim 10$ \msun)
stars have structural properties  described by
$m(r) > 870\ \msun\ (r/\rm pc)^{1.33}$, where $m(r)$ is the mass within
radius $r$ about a peak surface density. Clouds below this criterion
are generally not forming massive stars. This ``KP" criterion
corresponds to a line of different slope in Figure \ref{massradiussf}, and
a criterion for surface density that decreases with radius:
$\Sigma(gas) \propto r^{-0.67}$. While we have not characterized
the internal structure of our BGPS sources, we can compare their
total mass and radius to the KP relation.
When determining their relationship, KP reduced Ossenkopf \&\
Henning (1994) dust opacities by a factor of 1.5.
Since we have assumed Ossenkopf \&\ Henning
dust opacities without the factor of 1.5 reduction, we scale the mass-radius
relation to our assumed dust opacities and use
$m(r) > 580\ \msun\ (r/\rm pc)^{1.33}$ for the KP relation appropriate for
our mass estimates.
In the full sample of BGPS sources, 47.6\%\ have masses above the
KP relation. Among the Group 3 sources, those most certain to be forming
stars, and likely to be forming massive stars,
80\% (41 of 51) lie above the KP relation.

Since the BGPS includes sources ranging from cores to clouds, we 
consider all three star formation criteria discussed above in order 
to identify all BGPS sources that have high enough densities to be 
forming stars on vastly different size scales.  We estimate that about 
half the full sample lies above both the
HL and KP criteria and should be forming stars, including massive
stars, with high efficiency. Smaller parts of a larger fraction
(86.8\%) of BGPS sources may be forming stars efficiently, and
a still smaller part may satisfy the KM criterion for massive star
formation.
As discussed in section \ref{starformationactivity}, we observe mid-IR
sources toward 67\%\ of the full sample.  Since the mid-IR
catalogs considered by Dunham et al.~(2011) include high-mass YSOs across
the Galaxy, we are capable of seeing all massive YSO counterparts of
BGPS sources.  Conversely, the mid-IR catalogs used in
Dunham et al.~(2011) include only the nearby low-mass YSOs.
Therefore, we should not expect to detect mid-IR counterparts toward the
full 86.8\%\ BGPS sources with some surface densities high enough
for efficient star formation. Roughly speaking, the statistics of
mid-IR source detection are consistent with both the HL and
KP relations.

The 47.6\%\ of BGPS sources that satisfy the KP relation
contain 79.3\%\ of the total mass in the full sample.
Thus the majority of the mass in the BGPS full sample
would be predicted by the KP relation to be forming massive stars.
These sources are characterised by
the following mean properties:  \mean{\log(\miso(120\as)/\msun)}$=3.29\pm0.42$,
\mean{\log(\miso(int)/\msun)}$=3.30\pm0.49$, \mean{R}$=2.30\pm1.51$ pc,
\mean{\log(n(120\as)/\cmv)}$=3.29\pm0.49$,
\mean{\log(n(int)/\cmv)}$=3.16\pm0.56$, \mean{\Sigma(int)}$=0.051\pm0.036$
\cmc, and \mean{\log(N_{H_2}^{beam}/\cmc)}$=22.23\pm0.28$, where
the standard deviation of the distribution is given as an error representing
the spread in each property.  The BGPS sources satisfying the KP relation
are typically larger, more massive, and less dense than
the previous samples of massive star-forming regions chosen based on
tracers of star formation that were discussed in section \ref{intro}.

\section{Summary}\label{summary}
We present the results of a survey of \ammonia(1,1), (2,2), and (3,3)
toward a sample of 631 1.1 mm continuum 
sources from the Bolocam Galactic Plane Survey (BGPS).  We have detected 
the \ammonia(1,1) line towards 456 BGPS sources (72\%), demonstrating that
the high column density features identified in the BGPS and other continuum
studies are excellent predictors of the presence of dense gas.  We have 
determined kinematic distances and 
resolved the kinematic distance ambiguity (KDA) to all 456 BGPS sources via
association with an 8 \um\ IRDC seen in the GLIMPSE mosaics or the 
presence of HISA in the VGPS or SGPS.  We have placed 233 BGPS sources
at the near distance, 221 at the far distance, and two at the tangent 
distance.  The kinematic distances range from 0.1 to 15.0 kpc.  

With the KDA resolved, we calculate physical properties based on both the
1.1 mm continuum emission (e.g.\ $R$, \miso, $n$, $\Sigma$, and $N_{H_2}$) and the 
\ammonia\ line emission (e.g. \tk, \tex, \mvir, $\sigma_{Vlsr}$, 
$\sigma_{NT}$, N$_{\ammonia}$).  The large range in kinematic distances
results in a large range in most physical properties (see Table \ref{stats}
for a full statistical description of the distribtuion of each physical 
quantity).  Different scale
structures are detected at different distances due to the Malmquist bias
and the spatial filtering in the BGPS data, which is a complication
that will affect all continuum based studies.
Based on the characteristic mass, radius, and density of cores, 
clumps, and clouds presented by Bergin \& Tafalla (2007), we find that 
the BGPS predominately detects cores at distances of 1 to 2 kpc, clumps
at distances of 2-7 kpc, and clouds at distances greater than 7 kpc.
Therefore, the effects of distance must be carefully accounted for.

We have studied the Galactic trends in each physical property as a
function of Galactocentric radius, \rgal.  We see a strong peak in the 
number of BGPS sources within the 5 kpc molecular ring, which corresponds
to the end of the Galactic bar and beginning of the Scutum spiral arm.
We also see a peak in sources at \rgal$\sim$6 kpc, which corresponds 
to the Sagittarius Arm.
We have found the following trends in physical properties with \rgal:
\begin{itemize}
\item There are no obvious trends among the inner Galaxy 
sources.  Sources in Gem OB1 have a higher $\mean{\tk}$ than sources
within the inner Galaxy, although this could be caused by a bias of
the Gem OB1 sources toward later evolutionary stages.  
\item \ammonia\ column density decreases with \rgal\ with a factor 
of approximately 8 difference between the sources closest to the 
Galactic center and those in Gem OB1.
\item Similarly, \ammonia\ abundance also decreases smoothly with \rgal\ with
a factor of 7 difference between the inner and outer most BGPS sources.
The \ammonia\ abundance changes with \rgal\ at the same rate as the nitrogen 
abundance, suggesting the decrease in \ammonia\ abundance is caused by less 
nitrogen to form \ammonia\ toward the
outer Galaxy, although changes in dust properties (e.g.~gas-to-dust ratio, grain
size, grain temperature) likely also affect the measured \ammonia\ abundance.
\item When comparing sources at similar kinematic distances,
there is little variation in physical properties 
across the Galaxy regardless of 
enviornment (inside versus outside the molecular ring, for example).
This is in contrast to the molecular clouds seen in the GRS, which were
found to be larger, more massive, warmer, and have higher column densities
within the molecular ring (Rathborne et al.~2009).  This suggests that environment 
may affect the large scale molecular clouds while interior regions of dense gas 
exhibit similar properties regardless of environment.
\end{itemize}

We have characterized the star formation activity of the BGPS sources
based on the presence of an associated mid-IR source as described in
Dunham et al.~(2011).  We find that 67\%\ of the BGPS sources in this
study have an associated mid-IR source.  BGPS sources associated with
either an Extended Green Object (EGO; Cyganowski et al.~2008) or a Red
MSX Source (RMS; Hoare et al.~2004; Urquhart et al.~2008) have a
higher mean \tk, $\sigma_{Vlsr}$, and $N_{\ammonia}$ than sources without
any mid-IR indication of star formation activity.

We also compared the BGPS sources to criteria suggested for both
efficient star formation and for massive star formation.  We find
that 46.3\%\ have average surface densities
above the threshold for efficient star formation (Heiderman et al.~2010;
Lada et al.~2010).  If we consider the peak column density,
we find 86.8\%\ of the full sample have some part with surface
densities above the threshold for efficient star formation.  Similarly,
we have compared the BGPS sources to the criterion for massive star 
formation presented by Kauffmann et al.~(2010), and found that 47.6\%\
of the full sample have masses above their criterion.  Thus, roughly half
of the full sample satisfies both criteria and should be forming stars
with high efficiency, while smaller parts of 86.8\%\ 
of the full sample may be forming stars efficiently.  This is consistent
with the 67\%\ of the full sample that has corresponding mid-IR sources.  
The BGPS sources that satisfy the Kauffman et al.~(2010) criterion for 
massive star formation are, on average, larger,
more massive, and less dense than seen in previous large-scale studies of 
massive star-forming regions that were based on tracers of massive star formation 
(i.e.~Plume et al.~1992, 1997; Shirley et al.~2003; Wu et al.~2010).

\acknowledgements
The authors would like to thank the anonymous referee for constructive comments, C.~Figura and L.~Morgan for the \ammonia\ observations they have provided for this work, as well as T.~Robitaille and M.~M.~Dunham for many enlightening conversations.  The BGPS is supported by the National Science 
Foundation through NSF grant AST-0708403.  M.~K.~Dunham and N.~J.~Evans 
were supported by NSF grant AST-0607793 to the University of Texas at 
Austin.  M.~K.~Dunham was additionally supported by a grant from the 
National Radio Astronomy Observatory (NRAO) Student Observing Support
Program, award number GSSP09-0004.  The NRAO is a facility of the 
National Science Foundation, operated under cooperative agreement by 
Associated Universities, Inc.  E.~Rosolowsky is supported by a Discovery Grant from
the Natural Sciences and Engineering Research Council Canada.  C.~J.~C.~is supported by a NSF
Astronomy and Astrophysics Postdoctoral Fellowship under award
AST-1003134, and also acknowledges support from NSF 
grant AST-0808119.  J.~S.~U.~is supported by a CSIRO OCE postdoctoral grant.
This paper made use of information from the
Red MSX Source (RMS) survey database at www.ast.leeds.ac.uk/RMS, which
was constructed with support from the Science and Technology Facilities
Council of the UK.
This publication makes use of molecular line data from the Boston University-FCRAO Galactic Ring Survey (GRS). The GRS is a joint project of Boston University and Five College Radio Astronomy Observatory, funded by the National Science Foundation under grants AST-9800334, AST-0098562, \& AST-0100793.  
This research used the facilities of the Canadian Astronomy Data Centre operated by the National Research Council of Canada with the support of the Canadian Space Agency.


\clearpage

\clearpage

\clearpage

\clearpage
\clearpage
\begin{landscape}
\begin{deluxetable*}{rcccccccccccc}
\tabletypesize{\scriptsize}
\tablewidth{0pt}
\tablecaption{\label{nh3_props}Observed NH$_{3}$ Properties}
\tablehead{
\colhead{ID} & \colhead{RA} & \colhead{Dec} & \colhead{$V_{LSR}$} & \colhead{$\sigma_{Vlsr}$} & \colhead{\tmb(1,1)} & \colhead{$W$(1,1)} & \colhead{\tmb(2,2)\tablenotemark{b}} & \colhead{$W$(2,2)\tablenotemark{b,c}} & \colhead{\tmb(3,3)\tablenotemark{b,c}} & \colhead{$W$(3,3)\tablenotemark{b,c}} & \colhead{\tmb(4,4)\tablenotemark{b,c}} & \colhead{$W$(4,4)\tablenotemark{b,c}} \\                    
\colhead{Number\tablenotemark{a}} & \colhead{(J2000)} & \colhead{(J2000)} & \colhead{(km s$^{-1}$)} & \colhead{(km s$^{-1}$)} & \colhead{(K)} & \colhead{(K km s$^{-1}$)} & \colhead{(K)} & \colhead{(K km s$^{-1}$)} & \colhead{(K)} & \colhead{(K km s$^{-1}$)} & \colhead{(K)} & \colhead{(K km s$^{-1}$)} 
}
\startdata
1307 & 18 02 30.0 & -22 28 06.9 &  32.63 (0.03) & 0.94 (0.03) & 0.85 (0.10) &  6.08 (0.20) & $<$ 0.35 & $<$ 0.97 & $<$  0.39 & $<$ 0.95 & \nodata & \nodata \\ 
1309 & 18 01 00.5 & -22 15 48.5 &   8.16 (0.06) & 0.94 (0.06) & 0.39 (0.09) &  3.21 (0.18) & $<$ 0.35 & $<$ 0.97 & $<$  0.39 & $<$ 0.98 & \nodata & \nodata \\ 
1310 & 18 02 47.8 & -22 26 03.3 & 122.87 (0.03) & 0.60 (0.02) & 0.75 (0.09) &  4.08 (0.18) & $<$ 0.35 & $<$ 0.86 & $<$  0.38 & $<$ 0.83 & \nodata & \nodata \\ 
1311 & 18 03 15.0 & -22 27 09.7 & 153.30 (0.01) & 0.80 (0.01) & 2.54 (0.10) & 19.31 (0.21) & 1.38 (0.09) &  5.88 (0.19) & 1.02 (0.10) &  7.09 (0.19) & \nodata & \nodata \\
1313 & 18 02 45.3 & -22 21 46.8 &  -0.71 (0.04) & 0.77 (0.04) & 0.63 (0.09) &  3.94 (0.19) & $<$ 0.35 & $<$ 0.92 & $<$  0.37 & $<$ 0.88 & \nodata & \nodata \\ 
1314 & 18 03 11.9 & -22 24 32.5 & 154.02 (0.01) & 0.97 (0.01) & 3.38 (0.11) & 28.53 (0.23) & 2.07 (0.09) & 10.59 (0.20) & 1.60 (0.10) & 12.32 (0.20) & \nodata & \nodata \\
1316 & 18 03 31.5 & -22 26 48.4 & 152.52 (0.00) & 0.63 (0.00) & 3.93 (0.10) & 24.58 (0.20) & 2.09 (0.09) &  7.47 (0.18) & 1.19 (0.09) &  6.53 (0.17) & \nodata & \nodata \\
1317 & 18 03 26.8 & -22 25 16.1 & 154.55 (0.02) & 1.01 (0.02) & 1.26 (0.10) & 10.45 (0.21) & 0.48 (0.09) &  2.49 (0.20) & $<$ 0.38 & $<$ 0.96& \nodata & \nodata \\
1320 & 18 02 46.6 & -22 12 17.7 &  18.11 (0.03) & 0.76 (0.03) & 0.75 (0.09) &  4.24 (0.18) & $<$ 0.34 & $<$ 0.90 & $<$  0.38 & $<$ 0.89 & \nodata & \nodata \\ 
1322 & 18 03 23.7 & -22 07 45.9 &   0.26 (0.00) & 0.42 (0.00) & 3.00 (0.09) & 15.52 (0.16) & 1.08 (0.09) &  2.12 (0.15) & $<$ 0.38 & $<$ 0.74& \nodata & \nodata
\enddata \\
\tablecomments{Errors are given in parentheses.  The full table is available in the online journal.}
\tablenotetext{a}{B, C, D, and E denote multiple ammonia pointings that fall within a single 1.1 mm source.}
\tablenotetext{b}{Upper-limits are $T_{mb} < 4\sigma$ and $W < 5\sigma \Delta v  \sqrt{N}$, where $\sigma$ is the rms noise, $\Delta v$ is the width of a single channel in velocity, and $N$ is the number of pixels over which the average RMS was calculated.}
\tablenotetext{c}{\nodata denotes sources that were not observed in the given transition.}
\end{deluxetable*}
\clearpage
\end{landscape}

\begin{deluxetable}{rccccccc}
\tabletypesize{\scriptsize}
\tablewidth{0pt}
\tablecaption{\label{gas_props}Derived Gas Properties}
\tablehead{
\colhead{ID} & \colhead{} & \colhead{\tk\tablenotemark{b}} & \colhead{\tex} & \colhead{$a$} & \colhead{$\sigma_{NT}$} & \colhead{H$_2$O}  & \colhead{Multiple}  \\                    
\colhead{Number\tablenotemark{a}} & \colhead{$\tau$(1,1)} & \colhead{(K)} & \colhead{(K)} & \colhead{(\kms)} & \colhead{(\kms)} & \colhead{maser?\tablenotemark{c}}  & \colhead{Comp?\tablenotemark{d}}   
}
\startdata
1307 &  4.31 & $<$ 11.90 (0.47) &  3.29 (0.13) &  0.20(0.002) &  0.93 (0.06) & Y &   \\
1309 &  2.70 & $<$ 28.59 (1.17) &  3.01 (0.13) &  0.32(0.004) &  0.93 (0.11) & Y &   \\
1310 &  4.31 & $<$ 12.61 (0.46) &  3.27 (0.11) &  0.21(0.002) &  0.59 (0.05) & N  &   \\
1311 &  6.06 & 24.29 (0.19) &  4.03 (0.04) &  0.29(0.001) &  0.80 (0.02) & Y & $v_{lsr}$ \\
1313 &  2.79 & $<$ 14.00 (0.75) &  3.24 (0.17) &  0.22(0.003) &  0.77 (0.08) & N  &   \\
1314 &  5.12 & 27.44 (0.13) &  4.60 (0.03) &  0.31(0.000) &  0.96 (0.01) & Y & \tk\ \\
1316 &  5.15 & 18.01 (0.10) &  5.51 (0.04) &  0.25(0.000) &  0.63 (0.01) & Y & \tk\ \\
1317 &  3.62 & 14.40 (0.23) &  3.72 (0.06) &  0.22(0.001) &  1.01 (0.04) & N &   \\
1320 &  2.14 & $<$ 15.76 (0.68) &  3.42 (0.15) &  0.23(0.002) &  0.76 (0.07) & Y &   \\
1322 &  8.34 & 12.38 (0.12) &  4.79 (0.05) &  0.21(0.000) &  0.41 (0.01) & N & \tk\ 
\enddata
\tablecomments{Errors are given in parentheses.  The full table is available in the online journal.}
\tablenotetext{a}{B, C, D, and E denote multiple ammonia pointings that fall within a single 1.1 mm source.}
\tablenotetext{b}{Upper-limits to T$_{\mbox{kin}}$ are due to a non-detection of the \ammonia(2,2) transition.}
\tablenotetext{c}{Y denotes detected, N denotes not detected.}
\tablenotetext{d}{\tk\ denotes additional warm temperature component might provide a better fit, and $v_{lsr}$ denotes multiple velocity components required.}
\end{deluxetable}

\begin{deluxetable}{cccccc}
\tabletypesize{\scriptsize}
\tablewidth{0pt}
\tablecaption{\label{distancestable}Kinematic Distances}
\tablehead{
\colhead{ID} & \colhead{IRDC} & \colhead{HISA} & \colhead{KDA} & \colhead{kinematic} & \colhead{\rgal} \\
\colhead{Number} & \colhead{flag\tablenotemark{1}} & \colhead{flag\tablenotemark{1}} & \colhead{flag\tablenotemark{2}} & \colhead{distance (kpc)} & \colhead{(kpc)}
}
\startdata
1307 & n & n & f &  12.4($^{+  0.52}_{ -0.42}$) &   4.1 \\
1309 & n & m & f &  15.0($^{+  1.42}_{ -1.00}$) &   6.6 \\
1310 & m & m & n &   6.9($^{+  0.10}_{ -0.10}$) &   1.7 \\
1311 & n & m & f &   9.3($^{+  0.09}_{ -0.09}$) &   1.4 \\
1313 & y & y & n &   0.2($^{+  2.45}_{-15.26}$) &   8.5 \\
1314 & y & n & n &   7.3($^{+  0.09}_{ -0.09}$) &   1.4 \\
1316 & y & m & n &   7.3($^{+  0.09}_{ -0.09}$) &   1.4 \\
1317 & y & m & n &   7.3($^{+  0.09}_{ -0.09}$) &   1.4 \\
1320 & n & n & f &  13.7($^{+  0.87}_{ -0.66}$) &   5.3 \\
1322 & y & y & n &   0.1($^{+ 18.47}_{ -1.44}$) &   8.2
\enddata
\tablecomments{Errors are given in parentheses.  The full table is available in the online journal.}
\tablenotetext{1}{y denotes positive association, n denotes no association, and m denotes questionable association.}
\tablenotetext{2}{n denotes near kinematic distance, f denotes far kinematic distance, and t denotes tangent distance.}
\end{deluxetable}

\clearpage
\clearpage
\begin{landscape}
\begin{deluxetable*}{lccccccccccc}
\tabletypesize{\scriptsize}
\tablewidth{0pt}
\tablecaption{\label{derived_masses_densities}Derived Masses and Densities}
\tablehead{
\colhead{ID} & \colhead{Radius} & \colhead{\miso(120$^{\prime\prime}$)} & \colhead{\miso(int)} & \colhead{\mvir(int)} & \colhead{$n_{p}$(int)} & \colhead{$n_{ex}$\tablenotemark{a}} & \colhead{$\Sigma$(int)} & \colhead{$N_{NH_3}$} & \colhead{$N_{H_2}^{avg}$} & \colhead{$N_{H_2}^{beam}$} & \colhead{$X_{\ammonia}$} \\
\colhead{Number} & \colhead{(pc)} & \colhead{(M$_{\odot}$)} & \colhead{(M$_{\odot}$)} & \colhead{(M$_{\odot}$)} & \colhead{($10^{3}$cm$^{-3}$)} & \colhead{($10^{3}$cm$^{-3}$)} & \colhead{($10^{-2}$ g cm$^{-2}$)} & \colhead{($10^{14}$ cm$^{-2}$)} & \colhead{($10^{21}$ cm$^{-2}$)} & \colhead{($10^{22}$ cm$^{-2}$)} & \colhead{($10^{-8}$)} 
}
\startdata
1307 &  2.1 &  2621(690) &  2340(580) &     2169(100) &       0.99(0.33) &  \nodata               &  3.44(1.01) & 10.07(0.93) &  7.35(2.15) &  1.19(0.18) & 13.69(4.20) \\
1309 &  2.1 &  1526(580) &  1632(590) &     2131(210) &       0.74(0.45) &       0.37(0.07) &  2.50(1.22) &  8.04(2.56) &  5.35(2.60) &  0.62(0.07) & 15.01(8.72) \\
1310 &  1.6 &  1163(240) &   985(230) &      665(28) &       0.97(0.24) &       0.76(0.15) &  2.54(0.62) &  6.47(0.76) &  5.43(1.32) &  1.00(0.23) & 11.93(3.23) \\
1311 &  1.2 &   952(150) &   904(120) &      879(11) &       2.33(0.34) &       1.33(0.27) &  4.42(0.64) & 18.41(0.50) &  9.46(1.33) &  0.81(0.09) & 19.45(2.78) \\
1313 &  0.0 &     0(100) &     0(34) &       10(490) &     230.61(81966.01) &       0.80(0.16) &  5.46(0.74) &  5.56(1.07) & 11.67(3278.20) &  0.90(0.21) &  4.76(1336.30) \\
1314 &  1.2 &   932(100) &  1013(90) &     1280(17) &       2.54(0.29) &       2.11(0.42) &  4.87(0.51) & 22.15(0.40) & 10.42(1.06) &  1.68(0.09) & 21.26(2.20) \\
1316 &  0.7 &   745(130) &   719(87) &      302(4) &      10.77(1.53) &       3.79(0.76) & 11.39(1.61) & 14.40(0.22) & 24.37(3.19) &  1.54(0.12) &  5.91(0.78) \\
1317 &  1.2 &  1238(190) &   947(150) &     1474(32) &       2.04(0.37) &       1.46(0.29) &  4.12(0.71) & 10.62(0.61) &  8.81(1.50) &  1.18(0.17) & 12.05(2.17) \\
1320 &  3.0 &  3019(980) &  2923(930) &     2002(140) &       0.45(0.21) &       1.13(0.23) &  2.20(0.86) &  4.53(0.91) &  4.71(1.84) &  0.82(0.18) &  9.61(4.22) \\
1322 &  0.0 &     0(63) &     0(56) &        1(240) &     799.73(929799.74) &       2.44(0.49) &  9.91(9102.62) & 12.08(0.22) & 21.19(19466.69) &  1.77(0.24) &  5.70(5235.07) 
\enddata
\tablecomments{Errors are given in parentheses.  The full table is available in the online journal.}
\tablenotetext{a}{n$_{ex}$ cannot be calculated for sources where we have set \tk=\tex.}
\end{deluxetable*}

\clearpage
\end{landscape}

\clearpage
\clearpage
\begin{landscape}
\begin{deluxetable*}{lcccccccccl}
\tabletypesize{\scriptsize}
\tablewidth{0pt}
\tablecaption{\label{stats}Statistical Summary}
\tablehead{
\colhead{} & \colhead{} & \multicolumn{4}{c}{Full Sample} & \colhead{} & \multicolumn{4}{c}{\tk\ subsample}  \\ 
\cline{3-6} \cline{8-11} 
\colhead{Property}& \colhead{units} & \colhead{minimum} & \colhead{mean\tablenotemark{a}} & \colhead{median} & \colhead{maximum}  & \colhead{} & \colhead{minimum} & \colhead{mean\tablenotemark{a}} & \colhead{median} & \colhead{maximum} 
}
\startdata
$D$ & kpc &  0.1 &  7.4(3.8) &  6.5 & 15.0 & &  0.1 &  6.6(3.7) &  4.7 & 14.3 \\
\rgal & kpc &  1.4 &  5.1(1.2) &  4.8 &  8.5 & &  1.4 &  4.9(1.2) &  4.7 &  8.2 \\
$R_{maj}$ & pc &0.01 & 1.20(0.77) & 1.04 & 4.79 & & 0.01 & 1.28(0.77) & 0.93 & 5.65 \\
$R_{min}$ & pc & 0.00 & 0.79(0.49) & 0.68 & 2.75 & & 0.01 & 0.84(0.49) & 0.67 & 2.88  \\
$R$ & pc & 0.01 & 1.87(1.36) & 1.51 & 8.41 & & 0.01 & 2.02(1.36) & 1.40 & 8.55  \\
log(\Snu(120\as)) & Jy & -0.55 & 0.06(0.32) & 0.00 &  1.35 & & -0.50 & 0.22(0.34) & 0.15 &  1.35  \\
log(\Snu(int)) & Jy &-0.88 & 0.06(0.40) &  0.01 &    1.48 & & -0.54 & 0.24(0.41) &  0.16 &    1.48  \\
\tmb(1,1)  & K &0.25 & 1.27(0.98) &  0.92 &  7.14 & & 0.38 & 1.93(1.12) &  1.70 &  7.14 \\
\tmb(2,2)  & K & 0.17 & 0.59(0.54) &  0.41 &  4.73 & & 0.24 & 0.91(0.68) &  0.73 &  4.73 \\
\tex(1,1)  & K &  2.82 &   4.07(2.15) &  3.58 & 24.81 & &   2.99 &   4.45(2.11) &  4.12 & 24.12 \\
\tk  & K &   5.00 &  15.57(5.02) & 14.24 & 59.41 & &  11.95 &  17.39(5.49) & 15.70 & 59.41 \\
(\tk$-$\tex)  & K &  0.00 &  11.50(5.05) & 10.38 & 56.17 & &   0.00 &  12.94(5.59) & 11.55 & 56.17 \\
$V_{LSR}$  & \kms\ & -0.71 &   63.7(  32.9) &   60.6 & 154.55 & &   0.26 &   59.6(  33.3) &   49.0 &  154.5 \\
$\sigma_{v}$  & \kms\ &0.10 & 0.76(0.49) &  0.64 &  4.42 & & 0.24 & 0.81(0.53) &  0.69 &  4.42 \\
$a$  & \kms\ &  0.13 &   0.23(0.03) &  0.22 &  0.46 & &   0.20 &   0.24(0.03) &  0.23 &  0.46 \\
$\sigma_{NT}$  & \kms\ &  0.08 &   0.75(0.49) &  0.63 &  4.41 & &   0.23 &   0.81(0.53) &  0.68 &  4.41 \\
Mach ($\sigma_{Vlsr}/a$) & \nodata &  0.59 &   3.23(1.87) &  2.83 & 19.38 & &   1.12 &   3.24(1.73) &  2.85 & 15.47 \\
$\tau$(1,1)  & \nodata\ &  0.01 &   3.44(1.73) &  3.37 & 12.22 & &   0.02 &   4.07(1.79) &  3.87 & 12.22 \\
log(\miso(120\as))  & \msun\ &-1.02 & 2.95(0.61) & 3.02 & 4.40 & & -1.02 & 2.93(0.62) & 2.94 & 4.00 \\
log(\miso(int))  & \msun\ &-1.07 & 2.94(0.66) & 2.98 & 4.39 & & -1.07 & 2.95(0.67) & 2.98 & 4.31 \\
log(\mvir(int))  & \msun\ &-0.07 & 2.85(0.66) & 2.88 & 4.72 & &  0.18 & 2.87(0.64) & 2.90 & 4.69 \\
log(\mvir/\miso(int))  & \nodata\ & -1.38 &  -0.09(0.45) & -0.13 &  1.97 & &  -0.91 &  -0.08(0.42) & -0.13 &  1.97 \\
log($n$(int))  &  \cmv\ &  2.07 &   3.12(  0.56) &  3.03 &     5.90 & &   2.15 &   3.27( 0.58) &  3.22 &     5.90 \\
$\Sigma$(int)  & g \cmc\ &  0.006 &   0.037(0.030) &   0.029 &   0.236 & &   0.010 &   0.046(0.033) &   0.036 &   0.231 \\
log($n_{ex}$)  &  \cmv\ &   2.08 &    3.10( 0.35) &    3.09 &    4.88 & &    2.58 &    3.28( 0.26) &    3.29 &    4.03 \\
log($n_{ex}$/n(int))  & \nodata\ &  -2.51 &   -0.02(0.58) &    0.06 &    2.15 & &   -2.51 &    0.01(0.55) &    0.09 &    1.17 \\
log($N_{NH_3}$)  &  \cmc\ &  13.24 &   14.75( 0.37) &   14.76 &   15.96 & &   13.73 &   14.95(  0.29) &   14.92 &   15.96 \\
log($N_{H_2}^{avg}$)  & \cmc\ &  21.08 &   21.81( 0.27) &   21.79 &   22.70 & &   21.32 &   21.91( 0.27) &   21.89 &   22.69 \\
log($N_{H_2}^{beam}$)  &  \cmc\ &  21.30 &   22.09( 0.27) &   22.06 &   23.29 & &   21.55 &   22.22( 0.28) &   22.18 &   23.29 \\
log($N_{NH_3}/N_{H_2}^{beam}$)  & \nodata &  -8.63 &   -7.34( 0.30) &   -7.33 &   -6.34 & &   -8.36 &   -7.27( 0.26) &   -7.27 &   -6.56 \\
\enddata \\
\tablenotetext{a}{Standard deviation is given in parentheses.}
\end{deluxetable*}

\clearpage
\end{landscape}

\begin{deluxetable}{lcccccc}
\tabletypesize{\scriptsize}
\tablewidth{0pt}
\tablecaption{\label{sfactivitytrends}Trends with Star Formation Activity}
\tablehead{
\colhead{Property} & \colhead{Probability} & \colhead{} & \colhead{} & \colhead{standard} & \colhead{} & \colhead{} \\
\colhead{(units)} & \colhead{Group} & \colhead{minimum} & \colhead{mean} & \colhead{deviation} & \colhead{median} & \colhead{maximum}
}
\startdata
$\sigma_{Vlsr}$ & 0 &  0.1 &  0.7 &  0.4 &  0.6 &  3.6 \\
(\kms) & 1 &  0.2 &  0.7 &  0.5 &  0.6 &  4.4 \\
 & 2 &  0.2 &  0.7 &  0.5 &  0.7 &  3.8 \\
 & 3 &  0.3 &  1.1 &  0.5 &  1.0 &  3.1 \\
 & & & & & & \\
\tk\ & 0 &  6.6 & 13.9 &  3.0 & 13.5 & 26.6 \\
(K) & 1 &  5.0 & 14.6 &  3.8 & 13.8 & 33.7 \\
 & 2 & 10.3 & 15.8 &  3.9 & 14.6 & 28.6 \\
 & 3 & 11.1 & 22.7 &  8.2 & 21.1 & 59.4 \\
 & & & & & & \\
$N_{NH_3}$ & 0 &  0.2 &  5.6 &  3.8 &  4.6 & 21.1 \\
($10^{14}$ \cmc) & 1 &  0.2 &  7.1 &  5.5 &  5.8 & 40.2 \\
 & 2 &  0.2 &  8.0 &  6.6 &  6.1 & 40.0 \\
 & 3 &  2.0 & 16.3 & 16.9 & 10.4 & 91.2 \\
\enddata
\end{deluxetable}


\begin{thebibliography}{}

\bibitem[Aguirre et al.(2011)]{2011ApJS..192....4A} Aguirre, J.~E., Ginsburg, A.~G., Dunham, M.~K., et al.\ 2011, \apjs, 192, 4
\bibitem[Anderson \& Bania(2009)]{2009ApJ...690..706A} Anderson, L.~D., \& Bania, T.~M.\ 2009, \apj, 690, 706 
\bibitem[Battersby et al.(2010)]{2010ApJ...721..222B} Battersby, C., Bally, J., Jackson, J.~M., et al.\ 2010, \apj, 721, 222
\bibitem[Battersby et al.(2011)]{2011arXiv1101.4654B} Battersby, C., Bally, J., Gingsburg, A.~G., et al.\ 2011, A\&A, in press, arXiv:1101.4654
\bibitem[Benjamin et al.(2003)]{Benjamin:03} Benjamin, R.~A., Churchwell, E., Babler, B.~L., et al.\ 2003, \pasp, 115, 953
\bibitem[Bergin \& Tafalla(2007)]{2007ARA&A..45..339B} Bergin, E.~A., \& Tafalla, M.\ 2007, \araa, 45, 339  
\bibitem[Beuther et al.(2002a)]{2002A&A...390..289B} Beuther, H., Walsh, A., Schilke, P., et al.\ 2002a, \aap, 390, 289
\bibitem[Beuther et al.(2002b)]{2002ApJ...566..945B} Beuther, H., Schilke, P., Menten, K.~M., et al.\ 2002b, \apj, 566, 945 
\bibitem[Bohlin et al.(1978)]{1978ApJ...224..132B} Bohlin, R.~C., Savage, B.~D., \& Drake, J.~F.\ 1978, \apj, 224, 132
\bibitem[Burton \& Bania(1974)]{1974A&A....33..425B} Burton, W.~B., \& Bania, T.~M.\ 1974, \aap, 33, 425
\bibitem[Burton et al.(1975)]{1975ApJ...202...30B} Burton, W.~B., Gordon, M.~A., Bania, T.~M., \& Lockman, F.~J.\ 1975, \apj, 202, 30
\bibitem[Burton et al.(1978)]{1978ApJ...219L..67B} Burton, W.~B., Liszt, H.~S., \& Baker, P.~L.\ 1978, \apjl, 219, L67 
\bibitem[Busfield et al.(2006)]{2006MNRAS.366.1096B} Busfield, A.~L., Purcell, C.~R., Hoare, M.~G., et al.\ 2006, \mnras, 366, 1096
\bibitem[Carey et al.(1998)]{1998ApJ...508..721C} Carey, S.~J., Clark, F.~O., Egan, M.~P., et al.\ 1998, \apj, 508, 721
\bibitem[Carey et al.(2000)]{2000AAS...197.0516C} Carey, S.~J., Egan, M.~P., Kuchar, T.~A., et al.\ 2000, Bulletin of the American Astronomical Society, 197, 516
\bibitem[Cesaroni et al.(1988)]{1988A&AS...76..445C} Cesaroni, R., Palagi, F., Felli, M., et al.\ 1988, \aaps, 76, 445
\bibitem[Chambers et al.(2009)]{2009ApJS..181..360C} Chambers, E.~T., Jackson, J.~M., Rathborne, J.~M., \& Simon, R.\ 2009, \apjs, 181, 360 
\bibitem[Clemens et al.(1988)]{1988ApJ...327..139C} Clemens, D.~P., Sanders, D.~B., \& Scoville, N.~Z.\ 1988, \apj, 327, 139 
\bibitem[Cohen \& Thaddeus(1977)]{1977ApJ...217L.155C} Cohen, R.~S., \& Thaddeus, P.\ 1977, \apjl, 217, L155
\bibitem[Cyganowski et al.(2008)]{Cyganowsky:08} Cyganowski, C.~J., Whitney, B.~A., Holden, E., et al.\ 2008, \aj, 136, 2391
\bibitem[Dunham et al.(2010a)]{2010ApJ...717.1157D} Dunham, M.~K., Rosolowsky, E., Evans, N.~J., II, et al.\ 2010a, \apj, 717, 1157 
\bibitem[Dunham et al.(2011)]{2011ApJ...731...90D} Dunham, M.~K., Robitaille, T.~P., Evans, N.~J., II, et al.\ 2011, \apj, 731, 90
\bibitem[Egan et al.(1998)]{1998ApJ...494L.199E} Egan, M.~P., Shipman, R.~F., Price, S.~D., et al.\ 1998, \apjl, 494, L199 
\bibitem[Fish et al.(2003)]{2003ApJ...587..701F} Fish, V.~L., Reid, M.~J., Wilner, D.~J., \& Churchwell, E.\ 2003, \apj, 587, 701
\bibitem[Flynn et al.(2004)]{2004ASPC..317...44F} Flynn, E.~S., Jackson, J.~M., Simon, R., et al.\ 2004, in ASP Conf. Ser. 317, Milky Way Surveys: The Structure and Evolution of our Galaxy, ed.~D.~Clemens, R.~Shah, \& T.~Brainerd (San Francisco, CA:ASP), 44
\bibitem[Foster et al.(2009)]{2009ApJ...696..298F} Foster, J.~B., Rosolowsky, E.~W., Kauffmann, J., et al.\ 2009, \apj, 696, 298
\bibitem[Friesen et al.(2009)]{2009ApJ...697.1457F} Friesen, R.~K., Di Francesco, J., Shirley, Y.~L., \& Myers, P.~C.\ 2009, \apj, 697, 1457 
\bibitem[Gibson et al.(2000)]{2000ApJ...540..851G} Gibson, S.~J., Taylor, A.~R., Higgs, L.~A., \& Dewdney, P.~E.\ 2000, \apj, 540, 851
\bibitem[Gibson et al.(2005)]{2005ApJ...626..195G} Gibson, S.~J., Taylor, A.~R., Higgs, L.~A., Brunt, C.~M., \& Dewdney, P.~E.\ 2005, \apj, 626, 195
\bibitem[Goldsmith \& Li(2005)]{2005ApJ...622..938G} Goldsmith, P.~F., \& Li, D.\ 2005, \apj, 622, 938
\bibitem[Gummersbach et al.(1998)]{1998A&A...338..881G} Gummersbach, C.~A., Kaufer, A., Schaefer, D.~R., Szeifert, T., \& Wolf, B.\ 1998, \aap, 338, 881 
\bibitem[Harju et al.(1993)]{1993A&AS...98...51H} Harju, J., Walmsley, C.~M., \& Wouterloot, J.~G.~A.\ 1993, \aaps, 98, 51 
\bibitem[Heiderman et al.(2010)]{2010ApJ...723.1019H} Heiderman, A., Evans, N.~J., II, Allen, L.~E., Huard, T., \& Heyer, M.\ 2010, \apj, 723, 1019 
\bibitem[Hoare et al.(2004)]{Hoare:04} Hoare, M.~G., Lumsden, S.~L., Oudmaijer, R.~D., et al.\ 2004, in ASP Conf. Ser. 317, Milky Way Surveys: The Structure and Evolution of our Galaxy, ed. D.~Clemens, R.~Shah, \& T.~Brainerd (San Francisco, CA:ASP), 156
\bibitem[Jackson et al.(2002)]{2002ApJ...566L..81J} Jackson, J.~M., Bania, T.~M., Simon, R., et al.\ 2002, \apjl, 566, L81
\bibitem[Jackson et al.(2006)]{2006ApJS..163..145J} Jackson, J.~M., Rathborne, J.~M., Shah, R.~Y., et al.\ 2006, \apjs, 163, 145  
\bibitem[Kennicutt(2005)]{2005IAUS..227....3K} Kennicutt, R.~C.\ 2005, in IAU Symposium Proceedings of the International Astronomical Union 227, Massive Star Birth: A Crossroads of Astrophysics, ed.~R.~Cesaroni, M.~Felli, E.~Churchwell, \& M.~Walmsley (Cambridge: Cambridge University Press), 3
\bibitem[Kauffmann et al.(2010)]{2010ApJ...716..433K} Kauffmann, J., Pillai, T., Shetty, R., Myers, P.~C., \& Goodman, A.~A.\ 2010, \apj, 716, 433
\bibitem[Kauffmann \& Pillai(2010)]{2010ApJ...723L...7K} Kauffmann, J., \& Pillai, T.\ 2010, \apjl, 723, L7
\bibitem[Knapp(1974)]{1974AJ.....79..527K} Knapp, G.~R.\ 1974, \aj, 79, 527
\bibitem[Kolpak et al.(2002)]{2002ApJ...578..868K} Kolpak, M.~A., Jackson, J.~M., Bania, T.~M., \& Dickey, J.~M.\ 2002, \apj, 578, 868  
\bibitem[Krumholz \& McKee(2008)]{2008Natur.451.1082K} Krumholz, M.~R., \& McKee, C.~F.\ 2008, \nat, 451, 1082 
\bibitem[Lada et al.(2010)]{2010ApJ...724..687L} Lada, C.~J., Lombardi, M., \& Alves, J.~F.\ 2010, \apj, 724, 687
\bibitem[Li \& Goldsmith(2003)]{2003ApJ...585..823L} Li, D., \& Goldsmith, P.~F.\ 2003, \apj, 585, 823
\bibitem[Lumsden et al.(2002)]{2002MNRAS.336..621L} Lumsden, S.~L., Hoare, M.~G., Oudmaijer, R.~D., \& Richards, D.\ 2002, \mnras, 336, 621 
\bibitem[Markwardt(2009)]{2009ASPC..411..251M} Markwardt, C.~B.\ 2009, in ASP Conf.~Ser. 411, Astronomical Data Analysis Software and Systems XVIII, ed. D.~A.~Bohlender, D.~Durand, \& P.~Dowler (San Francisco, CA:ASP), 251
\bibitem[Matthews et al.(2009)]{2009AJ....138.1380M} Matthews, H., Kirk, H., Johnstone, D., et al.\ 2009, \aj, 138, 1380 
\bibitem[McClure-Griffiths et al.(2005)]{2005ApJS..158..178M} McClure-Griffiths, N.~M., Dickey, J.~M., Gaensler, B.~M., et al.\ 2005, \apjs, 158, 178 
\bibitem[McKee \& Tan(2003)]{2003ApJ...585..850M} McKee, C.~F., \& Tan, J.~C.\ 2003, \apj, 585, 850
\bibitem[Molinari et al.(2010)]{2010PASP..122..314M} Molinari, S., Swinyard, B., Bally, J., et al.\ 2010, \pasp, 122, 314 
\bibitem[Motte et al.(2003)]{2003ApJ...582..277M} Motte, F., Schilke, P., \& Lis, D.~C.\ 2003, \apj, 582, 277
\bibitem[Motte et al.(2007)]{2007A&A...476.1243M} Motte, F., Bontemps, S., Schilke, P., et al.\ 2007, \aap, 476, 1243 
\bibitem[Mueller et al.(2002)]{2002ApJS..143..469M} Mueller, K.~E., Shirley, Y.~L., Evans, N.~J., II, \& Jacobson, H.~R.\ 2002, \apjs, 143, 469 
\bibitem[Pandian et al.(2009)]{2009ApJ...706.1609P} Pandian, J.~D., Menten, K.~M., \& Goldsmith, P.~F.\ 2009, \apj, 706, 1609 
\bibitem[Peretto \& Fuller(2009)]{2009A&A...505..405P} Peretto, N., \& Fuller, G.~A.\ 2009, \aap, 505, 405 
\bibitem[Plume et al.(1992)]{1992ApJS...78..505P} Plume, R., Jaffe, D.~T., \& Evans, N.~J., II 1992, \apjs, 78, 505
\bibitem[Plume et al.(1997)]{1997ApJ...476..730P} Plume, R., Jaffe, D.~T., Evans, N.~J., II, Martin-Pintado, J., \& Gomez-Gonzalez, J.\ 1997, \apj, 476, 730 
\bibitem[Price et al.(2001)]{Price:01} Price, S.~D., Egan, M.~P., Carey, S.~J., Mizuno, D.~R., \& Kuchar, T.~A.\ 2001, \aj, 121, 2819 
\bibitem[Rathborne et al.(2005)]{2005ApJ...630L.181R} Rathborne, J.~M., Jackson, J.~M., Chambers, E.~T., et al.\ 2005, \apjl, 630, L181  
\bibitem[Rathborne et al.(2006)]{2006ApJ...641..389R} Rathborne, J.~M., Jackson, J.~M., \& Simon, R.\ 2006, \apj, 641, 389 
\bibitem[Rathborne et al.(2007)]{2007ApJ...662.1082R} Rathborne, J.~M., Simon, R., \& Jackson, J.~M.\ 2007, \apj, 662, 1082 
\bibitem[Rathborne et al.(2008)]{2008ApJ...689.1141R} Rathborne, J.~M., Jackson, J.~M., Zhang, Q., \& Simon, R.\ 2008, \apj, 689, 1141 
\bibitem[Rathborne et al.(2009)]{2009ApJS..182..131R} Rathborne, J.~M., Johnson, A.~M., Jackson, J.~M., Shah, R.~Y., \& Simon, R.\ 2009, \apjs, 182, 131 
\bibitem[Rathborne et al.(2010)]{2010ApJ...715..310R} Rathborne, J.~M., Jackson, J.~M., Chambers, E.~T., et al.\ 2010, \apj, 715, 310 
\bibitem[Robinson et al.(1984)]{1984ApJ...283L..31R} Robinson, B.~J., Manchester, R.~N., Whiteoak, J.~B., et al.\ 1984, \apjl, 283, L31 
\bibitem[Robitaille et al.(2008)]{Robitaille:08} Robitaille, T.~P., Meade, M.~R., Babler, B.~L., et al.\ 2008, \aj, 136, 2413 
\bibitem[Rolleston et al.(2000)]{2000A&A...363..537R} Rolleston, W.~R.~J., Smartt, S.~J., Dufton, P.~L., \& Ryans, R.~S.~I.\ 2000, \aap, 363, 537 
\bibitem[Roman-Duval et al.(2009)]{2009ApJ...699.1153R} Roman-Duval, J., Jackson, J.~M., Heyer, M., et al.\ 2009, \apj, 699, 1153 
\bibitem[Rosolowsky et al.(2008)]{2008ApJS..175..509R} Rosolowsky, E.~W., Pineda, J.~E., Foster, J.~B., et al.\ 2008, \apjs, 175, 509 
\bibitem[Rosolowsky et al.(2010)]{2010ApJS..188..123R} Rosolowsky, E., Dunham, M.~K., Ginsburg, A., et al.\ 2010, \apjs, 188, 123
\bibitem[Schlingman et al.(2011)]{2011ApJS..195...14S} Schlingman, W.~M., Shirley, Y.~L., Schenk, D.~E., Rosolowsky, E., Bally, J., et al.\ 2011, \apjs, 195, 14
\bibitem[Schuller et al.(2009)]{2009A&A...504..415S} Schuller, F., Menten, K.~M., Contreras, Y., et al.\ 2009, \aap, 504, 415 
\bibitem[Scoville \& Solomon(1975)]{1975ApJ...199L.105S} Scoville, N.~Z., \& Solomon, P.~M.\ 1975, \apjl, 199, L105 
\bibitem[Shaver et al.(1983)]{1983MNRAS.204...53S} Shaver, P.~A., McGee, R.~X., Newton, L.~M., Danks, A.~C., \& Pottasch, S.~R.\ 1983, \mnras, 204, 53 
\bibitem[Shirley et al.(2003)]{2003ApJS..149..375S} Shirley, Y.~L., Evans, N.~J., II, Young, K.~E., Knez, C., \& Jaffe, D.~T.\ 2003, \apjs, 149, 375
\bibitem[Simon et al.(2006)]{2006ApJ...653.1325S} Simon, R., Rathborne, J.~M., Shah, R.~Y., Jackson, J.~M., \& Chambers, E.~T.\ 2006, \apj, 653, 1325 
\bibitem[Smartt et al.(2001)]{2001A&A...367...86S} Smartt, S.~J., Venn, K.~A., Dufton, P.~L., et al.\ 2001, \aap, 367, 86 
\bibitem[Sridharan et al.(2002)]{2002ApJ...566..931S} Sridharan, T.~K., Beuther, H., Schilke, P., Menten, K.~M., \& Wyrowski, F.\ 2002, \apj, 566, 931 
\bibitem[Stil et al.(2006)]{2006AJ....132.1158S} Stil, J.~M., Taylor, A.~R., Dickey, J.~M., et al.\ 2006, \aj, 132, 1158 
\bibitem[Tafalla et al.(2006)]{2006A&A...455..577T} Tafalla, M., Santiago-Garc{\'{\i}}a, J., Myers, P.~C., et al.\ 2006, \aap, 455, 577 
\bibitem[Takano et al.(2002)]{2002PASJ...54..195T} Takano, S., Nakai, N., \& Kawaguchi, K.\ 2002, \pasj, 54, 195 
\bibitem[Urquhart et al.(2008)]{2008ASPC..387..381U} Urquhart, J.~S., Hoare, M.~G., Lumsden, S.~L., Oudmaijer, R.~D., \& Moore, T.~J.~T.\ 2008, in ASP Conf. Ser. 387, Massive Star Formation: Observations Confront Theory, ed. H.~Beuther, H.~Linz, \& T. Henning (San Francisco, CA:ASP), 381 
\bibitem[Urquhart et al.(2009)]{2009A&A...501..539U} Urquhart, J.~S., Hoare, M.~G., Purcell, C.~R., et al.\ 2009, \aap, 501, 539 
\bibitem[Urquhart et al.(2011)]{2011MNRAS.410.1237U} Urquhart, J.~S., Moore, T.~J.~T., Hoare, M.~G., et al.\ 2011, \mnras, 410, 1237
\bibitem[Weingartner \& Draine(2001)]{2001ApJ...548..296W} Weingartner, J.~C., \& Draine, B.~T.\ 2001, \apj, 548, 296 
\bibitem[Wood \& Churchwell(1989a)]{1989ApJS...69..831W} Wood, D.~O.~S., \& Churchwell, E.\ 1989a, \apjs, 69, 831 
\bibitem[Wood \& Churchwell(1989b)]{1989ApJ...340..265W} Wood, D.~O.~S., \& Churchwell, E.\ 1989b, \apj, 340, 265
\bibitem[Wu et al.(2010)]{2010ApJS..188..313W} Wu, J., Evans, N.~J., Shirley, Y.~L., \& Knez, C.\ 2010, \apjs, 188, 313 
\bibitem[Zinchenko et al.(1997)]{1997A&AS..124..385Z} Zinchenko, I., Henning, T., \& Schreyer, K.\ 1997, \aaps, 124, 385



\end{thebibliography}
\end{document}